\documentclass[aps,prx,twocolumn,superscriptaddress,
groupedaddress,10pt]{revtex4}
\usepackage{amsmath}
\usepackage{amssymb}
\usepackage{graphicx}
\usepackage{epsfig}
\usepackage{dcolumn}
\usepackage{bm}
\usepackage{bm}
\usepackage{blindtext}
\usepackage{natbib}
\usepackage{siunitx} 

\usepackage[titletoc]{appendix}

\usepackage{simplewick}

\usepackage{bbm}
\makeatletter
\newcommand{\mathleft}{\@fleqntrue\@mathmargin0pt}
\newcommand{\mathcenter}{\@fleqnfalse}
\makeatother

\usepackage{afterpage}

\usepackage[usenames,dvipsnames]{xcolor}
\usepackage{amsthm}
\usepackage{booktabs}
\usepackage{hyperref}
\usepackage{tikz}
\usetikzlibrary{calc}
\usepackage[printwatermark]{xwatermark}
\usepackage{mathtools}

\def\nn{\nonumber}

\newcommand{\ket}[1]{| #1 \rangle}

\begin{document}
\title{
Comprehensive study of the phase diagram of the spin-1/2 Kitaev-Heisenberg-Gamma chain
}

\author{Wang Yang}
\affiliation{Department of Physics and Astronomy and Stewart Blusson Quantum Matter Institute,
University of British Columbia, Vancouver, B.C., Canada, V6T 1Z1}

\author{Alberto Nocera}
\affiliation{Department of Physics and Astronomy and Stewart Blusson Quantum Matter Institute, 
University of British Columbia, Vancouver, B.C., Canada, V6T 1Z1}


\author{Ian Affleck}
\affiliation{Department of Physics and Astronomy and Stewart Blusson Quantum Matter Institute, 
University of British Columbia, Vancouver, B.C., Canada, V6T 1Z1}

\begin{abstract}

A central question on Kitaev materials is the effects of additional couplings on the Kitaev model which is proposed to be a candidate for realizing topological quantum computations.
However, two  spatial dimension typically suffers the difficulty of lacking controllable approaches.
In this work, using a combination of powerful analytical and numerical methods available in one dimension,
 we perform a comprehensive study on the phase diagram of a one-dimensional version of the spin-1/2 Kitaev-Heisenberg-Gamma model in its full parameter space.
A strikingly rich phase diagram is found with nine distinct phases,
including  four Luttinger liquid phases, a ferromagnetic phase, a N\'eel ordered phase, 
an ordered phase of distorted-spiral spin alignments, and two ordered phase which both break a $D_3$ symmetry albeit in different ways,
where $D_3$ is the dihedral group of order six.
Our work paves the way for studying one-dimensional Kitaev materials and may provide hints to the physics in higher dimensional situations.

\end{abstract}
\maketitle
\tableofcontents

\section{Introduction}

The Kitaev spin-1/2 model on the honeycomb lattice \cite{Kitaev2006} with anisotropic bond-dependent Ising interactions is proposed to host
exotic quasiparticle excitations, including Majorana fermions, and non-abelian anyons under applied magnetic fields. 
A remarkable feature of these excitations 
is that their braiding and fusion operations can be used to realize topological quantum computations \cite{Nayak2008}.
For this reason, the model has stimulated intense research interest in the past decade \cite{Nayak2008,Witczak-Krempa2014,Rau2016}.

It was first proposed that the Mott insulating A$_2$IrO$_4$ (A=Li, Na) compounds on the honeycomb lattice provide a platform for realizing the Kitaev spin-1/2 model \cite{Jackeli2009}.
The $d^5$-configuration of the magnetic Ir$^{4+}$ ion is subject to strong cubic crystal fields 
due to the surrounding octahedral environment formed by O$^{2-}$ ions.
The Ir$^{4+}$ ion is in its low spin configuration with all five electrons residing in the three $t_{2g}$ orbitals,
and the spin and effective orbital angular momenta are $S=1/2$ and $L=1$, respectively.
Strong spin-orbit coupling plays a crucial role in transferring the oxygen-mediated orbital-dependent
superexchange Hamiltonian to an anisotropic effective spin-1/2 model.
Indeed, due to strong spin-orbit couplings, the ground state of a single ion is a Kramers doublet
with a total angular momentum $j=1/2$,
and the projection of the superexchange model to the $j=1/2$ subspace is the desired 
bond-dependent Kitaev spin-1/2 model \cite{Jackeli2009}.

However, the direct overlaps between Iridium $d$-orbitals introduce
additional interactions, including the Heisenberg coupling \cite{Chaloupka2010} 
and the off-diagonal symmetric Gamma coupling \cite{Rau2014}.
The generalized Kitaev models including these additional terms were proposed \cite{Catuneanu2018,Gohlke2018,Ran2017,Wang2017},
and have been supported by many theoretical and experimental studies to be good descriptions for real materials
\cite{Singh2010,Choi2012,Singh2012,Modic2014,Johnson2015,Sandilands2015,Sears2015,Banerjee2016,Baek2017,Banerjee2017,Zheng2017,Jansa2018,Yu2018,Hentrich2018}.
In real materials, the signs of the couplings \cite{Gordon2019} are determined  to be ferromagnetic (FM) for the Kitaev coupling,
antiferromagnetic (AFM) for the Heisenberg coupling, and AFM for the Gamma coupling.
Besides Iridium oxides, other 4d transition metal materials including $\alpha$-RuCl$_3$ are also candidates for the Kitaev model although 
the spin-orbit coupling strength is significantly weaker \cite{Plumb2014,Kim2015}.
Recently there has been experimental evidence for Majorana excitations in the $\alpha$-RuCl$_3$ material \cite{Kasahara2018}.
We also note that there have been proposals for realizing Kitaev materials in $f$-electron systems which have an AFM Kitaev coupling \cite{Motome2020}.

Many theoretical efforts have been devoted to studying the phase diagram and fractional excitations of the Kitaev model on the honeycomb lattice
augmented with Heisenberg and Gamma couplings \cite{Chaloupka2010,Chaloupka2013,Reuther2011,Jiang2011,Price2012,Yadav2016,Janssen2016,Janssen2017,Liu2018,Gordon2019}. 
However, most of the studies are based on a classical analysis, mean field theories, or exact diagonalization (ED) on a small system, 
and a controllable understanding is usually lacking, which is a typical difficulty in two dimension (2D).
On the other hand, in one dimension (1D), there are more reliable analytical and large-scale numerical methods
to study the low energy properties,
including bosonization \cite{Haldane1981,Haldane1981a}, conformal field theory (CFT) \cite{Belavin1984,Knizhnik1984,Affleck1985,Affleck1988,Affleck1995a}, 
and the density matrix renormalization group (DMRG) methods \cite{White1992,White1993,Schollwock2011}. 
Hence, a detailed investigation of a 1D version of the Kitaev model may shed light on the physics in 2D.
It also provides a starting point for an extrapolation to 2D by coupling the 1D chains together and tuning the interchain coupling strength from weak to strong.
In addition, a study of the 1D generalized Kitaev model also has its own merit, 
since it can be realized in systems of Ruthenium stripes within a- or b-oriented superlattices of RuCl$_3$ \cite{Gruenewald2017}.

\begin{figure*}[htbp]
\centering
\includegraphics[width=18cm]{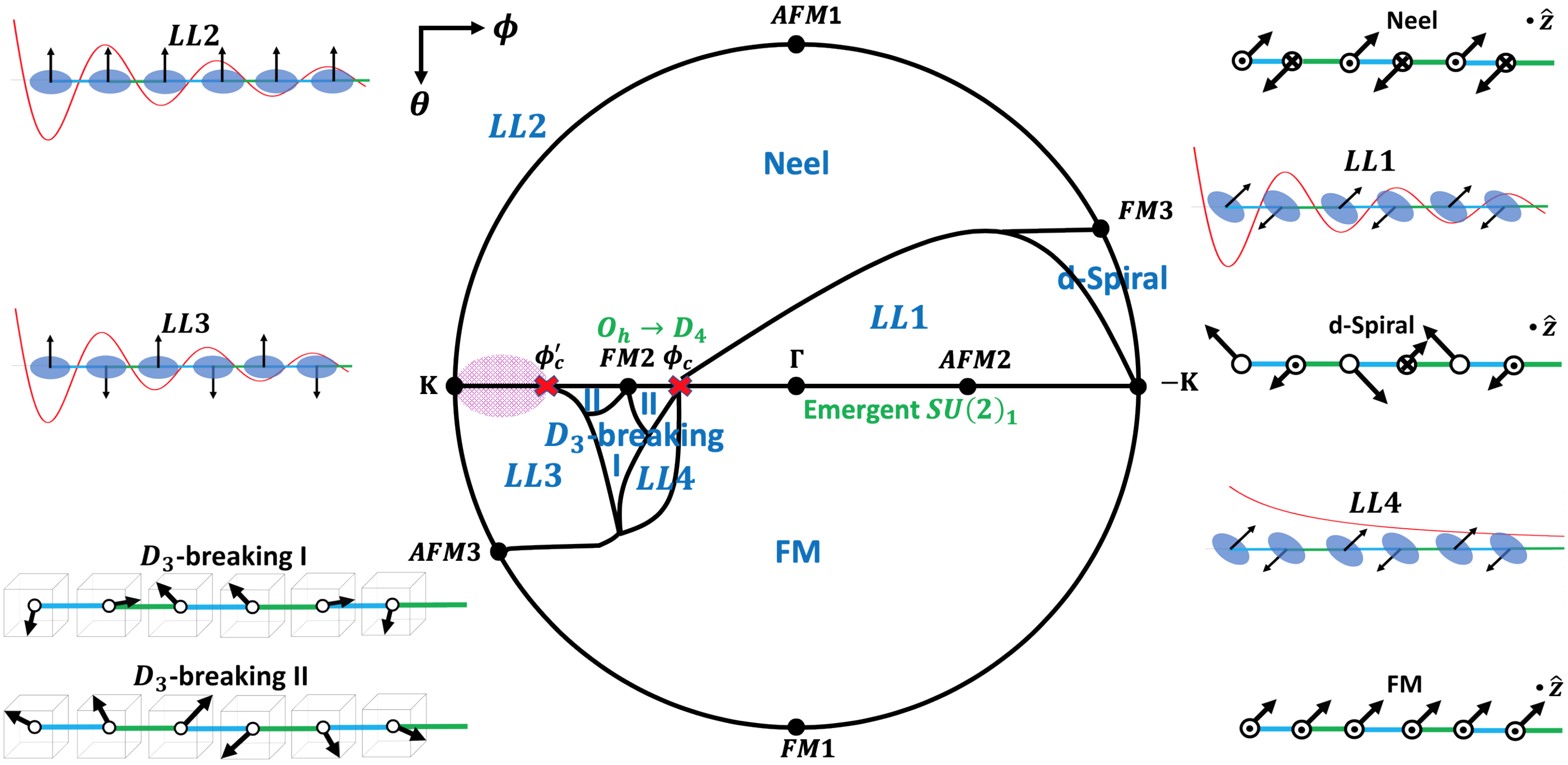}
\caption{Sketch of the phase diagram for the spin-1/2 Kitaev-Heisenberg-Gamma chain with eleven phases in total, and cartoon plots for the long-range and quasi-long-ranger orders in the corresponding phases.
The full parameter space is a two dimensional unit sphere parametrized by the polar and azimuthal angles $\theta$ and $\phi$.
Due to the equivalence $(K,J,-\Gamma)\simeq (K,J,\Gamma)$,
only the front half of the unit sphere corresponding to $\theta,\phi\in[0,\pi]$ is shown.
The $(\theta,\phi)$ coordinates of the $K$, $-K$, $\Gamma$, AFM1, FM1, AFM2, FM2, AFM3, FM3 points are $(\frac{\pi}{2},0)$, $(\frac{\pi}{2},\pi)$, $(\frac{\pi}{2},\frac{\pi}{2})$, $(0,\#)$, $(\pi,\#)$, $(\frac{\pi}{2},\frac{\pi}{4})$, $(\frac{\pi}{2},\frac{3\pi}{4})$, $(\pi-\arctan(2),0)$ and $(\arctan(2),\pi)$, respectively,
in which the symbol ``$\#$" is used when the value of $\phi$ can be arbitrarily chosen. 
We note that: 
 the ``Emergent SU(2)$_1$"  and ``$O_h\rightarrow D_4$" phases on the equator are parts of  the ``LL1" and ``$D_3$-breaking II" phases, respectively, hence are denoted by green (instead of blue) colors for clarification;
the ``LL2" phase locates on the circular boundary of the half sphere;
the nature of the phase diagram in a narrow region 
close to the $K$ point
 remains unclear, which are denoted by the hatched magenta color  in the figure;
 and all the other phases are extended in a finite area in the phase diagram.
 In the cartoon plots, the $z$-direction in spin space is chosen to be pointing vertically up, except for the  "Neel", ``d-Spiral"  and ``FM" phases where the black dot in the upper-right corner indicates that the $z$-axis is chosen to be perpendicular to the plane.
In the LL$i$ ($i=1,2,3,4$) phases, the black arrows represent a site-dependent quantization axis for the direction of longitudinal fluctuations.
In all cartoon plots, the spin directions refer to Eq. (\ref{eq:Ham}) without any sublattice rotation.
} 
\label{fig:phase}
\end{figure*}

The phase diagrams of the Kitaev-Heisenberg model in the 1D chain \cite{Agrapidis2018} and the two-leg ladder cases \cite{Agrapidis2019,Catuneanu2019} have been analyzed before. 
In particular, the two-leg ladder model already has a phase diagram quite similar to the model on the 2D honeycomb lattice \cite{Chaloupka2013}.
On the other hand, the Kitaev and Gamma couplings dominate over the Heisenberg coupling in $\alpha$-RuCl$_3$ \cite{Gordon2019}.
Thus a good approach is to first consider the Kitaev-Gamma model and then treat the Heisenberg term as a small perturbation.
Along this line of logic, the phase diagram of the 1D spin-1/2 Kitaev-Gamma chain has been investigated in Ref. [\onlinecite{Yang2020a}],
which shows an emergent SU(2)$_1$ phase and a rank-1 spin ordered phase with $O_h\rightarrow D_4$ symmetry breaking, 
where $O_h$ is the full octahedral group and  $D_n$ is the dihedral group of order $2n$.

In this work, we include a nonzero Heisenberg term, and make a comprehensive study of the phase diagram of the spin-1/2 Kitaev-Heisenberg-Gamma chain in its full parameter space, 
using a combination of extensive analytical, DMRG and  ED calculations. 
Here we note that although the Kitaev coupling in the existing materials  is FM,
it is worth also to study the region with AFM Kitaev coupling, 
since there may be $f$-electron systems realizing AFM Kitaev materials \cite{Motome2020}.
Thus, our work provides a road-map to the exotic physics in both the existing and potential Kitaev materials.
A strikingly rich phase diagram is found with a total of nine  distinct phases as shown in Fig. \ref{fig:phase},
where the Kitaev, Gamma, and Heisenberg couplings are parametrized as $K=\sin(\theta)\cos(\phi)$, $\Gamma=\sin(\theta)\sin(\phi)$, and $J=\cos(\theta)$, respectively.
We note that the ``Emergent SU(2)$_1$"  and ``$O_h\rightarrow D_4$" phases on the equator in Fig. \ref{fig:phase} are parts of  the ``LL1" and ``$D_3$-breaking II" phases, respectively,
hence they are written in green (rather than blue) color in Fig. \ref{fig:phase}.
We also note that the nature of the phase diagram in a narrow region close to the $K$ point (i.e., $(\theta=\pi/2,\phi=0)$) remain unclear, which is denoted by the hatched magenta color in Fig. \ref{fig:phase} and will be left for further study. 
The plot in Fig. \ref{fig:phase} is only schematic, and the precise phase boundaries are shown in Fig. \ref{fig:LL_K} as determined by DMRG numerics. 
Also, only half of the parameter space within the range $\theta,\phi\in[0,\pi]$ is shown due to the equivalence $(\theta,\phi)\simeq(\theta,2\pi-\phi)$ (see Eq. (\ref{eq:G_minusG})).

From an analytic point of view, there are six special points with explicit or hidden SU(2) symmetries, which provide starting points for a perturbative analysis in the regions nearby.
More precisely, the AMF1 and FM1 points located at the north and south poles are explicitly SU(2) symmetric  with AFM and FM couplings, respectively.
On the other hand, the AFM2 and FM2 points have hidden SU(2) symmetry as revealed by a six-sublattice rotation \cite{Chaloupka2015}, and the AFM3 and FM3 are SU(2) symmetric after a four-sublattice rotation \cite{Chaloupka2013,Chaloupka2015}.

For the regions close to the AFM points, we perform an RG analysis which is controllable as long as the deviations from the AFM points are not large. 
The low energy field theory is first obtained by directly projecting the perturbation Hamiltonian to the low energy SU(2)$_1$ degrees of freedom,
and then subject to an RG analysis.
We have also performed a careful symmetry analysis in each region to 
verify that the projected perturbation Hamiltonian contains all the symmetry allowed
terms up to relevant and marginal couplings  classified by the SU(2)$_1$ critical theory.
In this way, five phases can be understood, including the Luttinger liquid phases ``LL1", ``LL2" and ``LL3" (``LL" is ``Luttinger liquid" for short), and the ordered phases ``N\'eel" and ``FM" with N\'eel and FM orders, respectively,
which are all confirmed by our ED and DMRG numerical calculations.
The Luttinger parameters in each Luttinger liquid phase are determined numerically, and a color plot of the results are shown in Fig. \ref{fig:LL_K}.
We note that the above symmetry and RG analysis has a wider applicable range not limited to the spin-1/2 Kitaev-Heisenberg-Gamma model.
The conclusions also hold when another off-diagonal symmetric $\Gamma^\prime$ term \cite{Rau2014} and terms involving non-nearest-neighbors are included as long as the lattice symmetries are respected and the perturbations are small;
though of course, the coupling constants in the low energy theory will be renormalized.

There are three remaining phases which cannot be captured by the above RG analysis.
We find a narrow ``d-Spiral" phase close to the FM Kitaev point in the upper hemisphere as shown in Fig. 
\ref{fig:phase}.
When $\Gamma=0$, there is a duality transformation for the Kitaev-Heisenberg chain which maps the interval $[-K,\text{FM3}]$ to  $[\text{FM2},-K]$ on the circular boundary of Fig. \ref{fig:phase}.
This establishes $[\text{FM3},-K]$ to have a spiral spin order as a consequence of the FM order in the arc $[\text{FM2},-K]$,
and the symmetry breaking of the spiral order is determined to be $(\mathbb{Z}_2\times \mathbb{Z}_2)\ltimes D_{4d}\rightarrow (\mathbb{Z}_2\times \mathbb{Z}_2)\ltimes D_2$.
Then the symmetry breaking pattern in the ``d-Spiral" phase is inferred from the symmetry breaking of the spiral order in the arc $[\text{FM2},-K]$  based on the assumption that there is no phase transition between the ``d-Spiral" and the spiral orders.
The thus obtained symmetry breaking $D_{4d}\rightarrow D_2$ in the ``d-Spiral" phase predicts a ``distorted-spiral" pattern of the spin alignments, which is supported by our DMRG numerics. 

To understand the ``LL4" and ``$D_3$-breaking" phases, we take a perturbative Luttinger liquid approach.
The strategy is to separate the Hamiltonian in the six-sublattice rotated frame into two parts, 
such that one part is of the easy-plane XXZ type, and the other part is taken as a perturbation.
Analysis shows that the first order effect of the perturbation Hamiltonian vanishes, 
hence the Luttinger liquid behavior can be stabilized in a window of intermediate values of $J$,
which is identified as the ``LL4" phase.
DMRG numerics provide evidence for the existence of the ``LL4" phase as shown in Fig. \ref{fig:LL_K}.

When $J$ approaches zero, the Luttinger parameter diverges,
hence higher order effects eventually become important and drive the system into a strong coupling limit.
By a careful combination of strong coupling and symmetry analysis,
two different types of symmetry breaking patterns are identified for small $|J|$,
namely, $D_{3d}\rightarrow \mathbb{Z}_2^{\text{(I)}}$ and $D_{3d}\rightarrow \mathbb{Z}_2^{\text{(II)}}$, where $\mathbb{Z}_2^{\text{(I)}}$ and $\mathbb{Z}_2^{\text{(II)}}$ are two different $\mathbb{Z}_2$ groups.
Since $D_{3d}/\mathbb{Z}_2\cong D_3$, both  phases break the $D_3$ symmetry, which is the origin of the name ``$D_3$-breaking" for the phases in Fig. \ref{fig:phase}.
The type of $D_3$-breaking order is selected by the sign of the coupling constant, the determination of which requires a third order perturbation calculation.
We do not perform such a difficult high order perturbation calculation, but instead turn to a classical analysis.
Based on the classical analysis, we find that the ``$D_3$-breaking" region can be divided into two subregions corresponding to the ``$D_3$-breaking I" and the ``$D_3$-breaking II" phases as shown  in Fig. \ref{fig:phase}, which have $\mathbb{Z}_2^{\text{(I)}}$ and $\mathbb{Z}_2^{\text{(II)}}$ as the unbroken symmetry groups, respectively.
Our DMRG numerics provide evidence for the existence of these two $D_3$-breaking phases.
 
The rest of the paper is organized as follows,
where each section is made self-contained for the convenience of the readers who are interested in specific phases in the phase diagram.
In Sec. \ref{sec:Ham}, the model Hamiltonian is introduced,
and the sublattice rotations are discussed which reveal the hidden SU(2) symmetric points AFM$i$ and FM$i$ ($i=1,2,3$) shown  in Fig. \ref{fig:phase}.
In Sec. \ref{sec:LL1_phase} and Sec. \ref{sec:Neel_phase},
we combine  RG calculations, symmetry analysis, and numerics together to study the ``LL1" and the ``N\'eel" phases, respectively.
A brief description of the ``LL2" phase is also included in Sec. \ref{sec:Neel_phase}.
In Sec. \ref{sec:d_sp}, the ``d-Spiral" phase is studied. 
The symmetry breaking pattern is identified  to be $D_{4d}\rightarrow D_2$, and the spin alignments are shown to exhibit a ``distorted" spiral pattern.
In Sec. \ref{sec:LL3_phase}, the ``LL3" phase is investigated, again by a combination of RG,  symmetry, and numerical analysis. 
Sec. \ref{sec:LL_classical} is devoted to a discussion of the ``LL4" and ``$D_3$-breaking I, II" phases.
In Sec. \ref{sec:FM_phase}, the ``FM" phase is discussed.
Finally in Sec. \ref{sec:summary}, we briefly summarize the main results and open questions of the paper.

\section{Model Hamiltonian}
\label{sec:Ham}

\subsection{The Hamiltonian}

\begin{figure}[h]
\includegraphics[width=8.6cm]{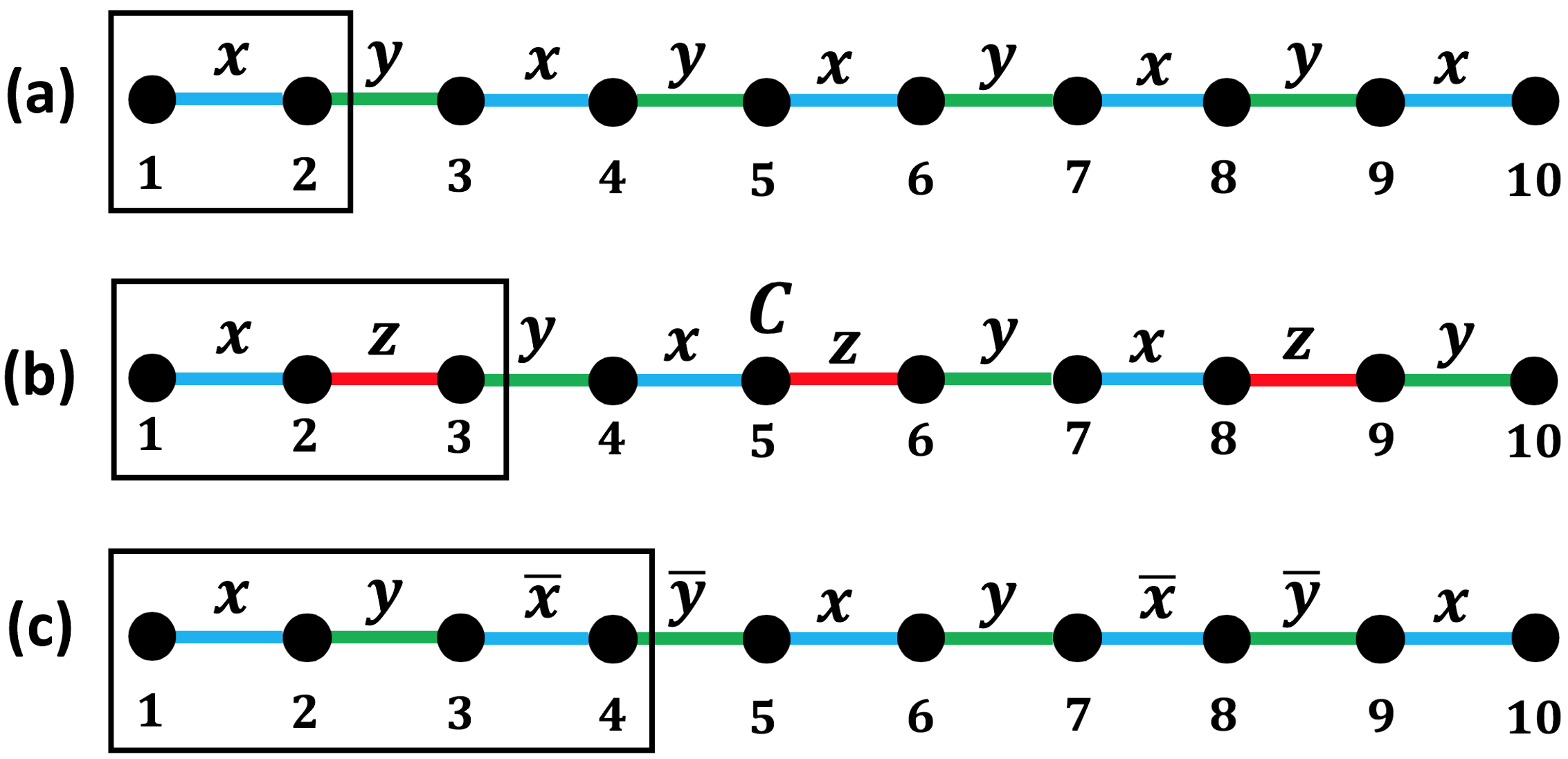}
\caption{The bond patterns of the Kitaev-Heisenberg-Gamma chain (a) without any sublattice rotation,
(b) after the six-sublattice rotation, and (c) after the four-sublattice rotation.
} \label{fig:bonds}
\end{figure}

We consider a spin-1/2 Kitaev-Heisenberg-Gamma ($KH\Gamma$) chain \cite{Rau2014} in zero magnetic field defined as
\begin{flalign}
H=\sum_{<ij>\in\gamma\,\text{bond}}\big[ KS_i^\gamma S_j^\gamma+ J\vec{S}_i\cdot \vec{S}_j+\Gamma (S_i^\alpha S_j^\beta+S_i^\beta S_j^\alpha)\big],\nn\\
\label{eq:Ham}
\end{flalign}
in which $i,j$ are two sites of nearest neighbors;
$\gamma=x,y$ is the spin direction associated with the $\gamma$ bond shown in Fig. \ref{fig:bonds} (a);
$\alpha\neq\beta$ are the two remaining spin directions other than $\gamma$;
$K$, $J$ and $\Gamma$,
are the Kitaev, Heisenberg and Gamma couplings, respectively.
The terms in $H$ are spelled out explicitly in Supplementary Materials \cite{SM}.
Throughout this work, we parametrize  
 $K,J,\Gamma$ as 
 \begin{eqnarray}
 J&=&\cos(\theta),\nn\\ 
 K&=&\sin(\theta)\cos(\phi),\nn\\
\Gamma&=&\sin(\theta)\sin(\phi),
\label{eq:parametrization}
\end{eqnarray}
in which $\theta\in[0,\pi]$ and $\phi\in[0,2\pi]$.
 
It is straightforward to observe that a global spin rotation 
$R(\hat{z},\pi):(S_i^x,S_i^y,S_i^z)\rightarrow (S_i^y,-S_i^x,S_i^z)$
leaves $K$ and $J$ invariant but changes the sign of $\Gamma$,
in which $R(\hat{n},\theta)$ represents a rotation in spin space around the $\hat{n}$-direction by an angle $\theta$.
Hence, 
\begin{eqnarray}
(K,J,-\Gamma) \simeq(K,J,\Gamma),
\label{eq:G_minusG}
\end{eqnarray} 
i.e., $(\theta,\phi)\simeq(\theta,2\pi-\phi)$.
Due to this equivalence, the phase diagram will be studied within the parameter range $\theta\in(0,\pi)$, $\phi\in(0,\pi)$.
It is apparent that the north and south poles where $K,\Gamma$ vanish
have explicit SU(2) symmetries:
$\theta=0$ corresponds to the AFM Heisenberg model, and $\theta=\pi$ is the FM Heisenberg model.
These two SU(2) symmetric points are denoted as AFM1 and FM1 in the phase diagram in Fig. \ref{fig:phase}.

\subsection{Sublattice transformations and hidden SU(2) symmetries}

In this subsection, we briefly review the six- and four-sublattice rotations \cite{Chaloupka2013,Chaloupka2015,Stavropoulos2018,Yang2020a},
and show that they unveil several points in the parameter space which have hidden SU(2) symmetries.
These hidden SU(2) symmetric points provide starting points for a perturbative study of the regions surrounding them.

\subsubsection{The six-sublattice rotation}

The six-sublattice rotation $U_6$ is defined as \cite{Stavropoulos2018,Yang2020a}
\begin{eqnarray}
\text{Sublattice $1$}: & (x,y,z) & \rightarrow (x^{\prime},y^{\prime},z^{\prime}),\nn\\ 
\text{Sublattice $2$}: & (x,y,z) & \rightarrow (-x^{\prime},-z^{\prime},-y^{\prime}),\nn\\
\text{Sublattice $3$}: & (x,y,z) & \rightarrow (y^{\prime},z^{\prime},x^{\prime}),\nn\\
\text{Sublattice $4$}: & (x,y,z) & \rightarrow (-y^{\prime},-x^{\prime},-z^{\prime}),\nn\\
\text{Sublattice $5$}: & (x,y,z) & \rightarrow (z^{\prime},x^{\prime},y^{\prime}),\nn\\
\text{Sublattice $6$}: & (x,y,z) & \rightarrow (-z^{\prime},-y^{\prime},-x^{\prime}),
\label{eq:6rotation}
\end{eqnarray}
in which "Sublattice $i$" ($1\leq i \leq 6$) represents all the sites $i+6n$ ($n\in \mathbb{Z}$) in the chain, and we have abbreviated $S^\alpha$ ($S^{\prime \alpha}$) as $\alpha$ ($\alpha^\prime$) for short ($\alpha=x,y,z$).
The Hamiltonian $H^\prime=U_6 H U_6^{-1}$ in the six-sublattice rotated frame is
\begin{eqnarray}
H^\prime&=\sum_{<ij>\in \gamma\,\text{bond}}\big[ -KS_i^\gamma S_j^\gamma-\Gamma (S_i^\alpha S_j^\alpha+S_i^\beta S_j^\beta) \nn\\
& -J(S_i^\gamma S_j^\gamma+S_i^\alpha S_j^\beta+S_i^\beta S_j^\alpha)\big],
\label{eq:6rotated}
\end{eqnarray}
in which $\gamma=x,z,y$ has a three-site periodicity shown in Fig. \ref{fig:bonds} (b),
and $\vec{S}_i^\prime$ is denoted as $\vec{S}_i$ for simplicity.
The terms in $H^\prime$ are spelled out explicitly in Supplementary Materials \cite{SM}.

It is clear from Eq. (\ref{eq:6rotated}) that while the Kitaev and Gamma terms acquire a form similar to 
the Heisenberg model (but with unequal couplings along different spin directions) in the six-sublattice rotated frame, the Heisenberg $J$ term loses its form.
Indeed, $H^\prime$ is SU(2) invariant when $K=\Gamma$, $J=0$.
Combining $U_6$ with Eq. (\ref{eq:G_minusG}), we see that $(\theta=\pi/2,\phi=\pi/4)$ and $(\theta=\pi/2,\phi=3\pi/4)$ have hidden SU(2) symmetries with FM and AFM couplings, respectively,
which are denoted as the FM2 and AFM2 points in the phase diagram shown in Fig. \ref{fig:phase}.

\subsubsection{The four-sublattice rotation}
\label{sec:4rotation}

The four-sublattice rotation $U_4$ is defined as \cite{Chaloupka2013,Chaloupka2015},
\begin{eqnarray}
\text{Sublattice $1$}: & (x,y,z) & \rightarrow (-x^{\prime},y^{\prime},-z^{\prime}),\nn\\ 
\text{Sublattice $2$}: & (x,y,z) & \rightarrow (-x^{\prime},-y^{\prime},z^{\prime}),\nn\\
\text{Sublattice $3$}: & (x,y,z) & \rightarrow (x^{\prime},-y^{\prime},-z^{\prime}),\nn\\
\text{Sublattice $4$}: & (x,y,z) & \rightarrow (x^{\prime},y^{\prime},z^{\prime}),
\label{eq:4rotation}
\end{eqnarray}
in which ``Sublattice $i$" ($1\leq i \leq 4$) represents all the sites $i+4n$ ($n\in \mathbb{Z}$) in the chain, and we have again dropped the spin symbol $S$ for simplicity.
The Hamiltonian $H^{\prime\prime}=U_4 H U_4^{-1}$ in the four-sublattice rotated frame acquires the form
\begin{eqnarray}
H^{\prime\prime}&=\sum_{<ij>\in \gamma\,\text{bond}}\big[ (K+2J)S_i^\gamma S_j^\gamma-J \vec{S}_i\cdot \vec{S}_j \nn\\
& +\epsilon(\gamma) \Gamma (S_i^\alpha S_j^\beta+S_i^\beta S_j^\alpha)\big],
\label{eq:4rotated}
\end{eqnarray}
in which the bonds $\gamma=x,y,\bar{x},\bar{y}$ has a four-site periodicity as shown in Fig. \ref{fig:bonds} (c);
the function $\epsilon(\gamma)$ is defined as $\epsilon(x)=\epsilon(y)=-\epsilon(\bar{x})=-\epsilon(\bar{y})=1$;
$S_i^{\bar{\gamma}}=S_i^{\gamma}$;
and $\vec{S}_i^{\prime\prime}$ is denoted as $\vec{S}_i$ for short.
The terms in $H^{\prime\prime}$ are spelled out explicitly in Supplementary Materials \cite{SM}.

When $\Gamma=0$, $U_4$ defines a duality transformation for the Kitaev-Heisenberg chain parametrized by the two coupling constants $(K,J)$,
and there is the equivalence \cite{Chaloupka2013}
\begin{eqnarray}
(K,J,\Gamma=0)\simeq (K+2J,-J,\Gamma=0).
\label{eq:duality_KH}
\end{eqnarray}
If further $K+2J=0$, $H^{\prime\prime}$ describes an SU(2) symmetric Heisenberg model with a coupling constant equal to $-J$.
The two hidden SU(2) symmetric points $(\theta=\pi-\arctan(2),\phi=0)$ and $(\theta=\arctan(2),\phi=\pi)$ thus revealed are denoted as AFM3 and FM3 in Fig. \ref{fig:phase}, respectively. 

It is straightforward to observe that the Kitaev points are self-dual under $U_4$. 
Setting $\Gamma=0$ and normalizing the transformed parameters according to $K^{\prime2}+J^{\prime2}+\Gamma^{\prime 2}=1$ (where $K^\prime=K+2J$, $J^\prime=-J$, $\Gamma^\prime=0$), $U_4$ establishes the equivalences: 
\begin{eqnarray}
&[\text{AFM1},K]  \simeq [K,\text{AFM3}],\nn\\
&[\text{AFM3},\text{FM1}]  \simeq  [\text{AFM1},\text{FM3}],\nn\\
&[\text{FM1},-K]  \simeq  [\text{FM3},-K],
\label{eq:arc_equiv}
\end{eqnarray}
in which $[A,B]$ represents the arc between the points $A$ and $B$ on the circular boundary of Fig. \ref{fig:phase}.

\subsection{Summary of the phase diagram}


\begin{figure*}[htbp]
\includegraphics[width=13.0cm]{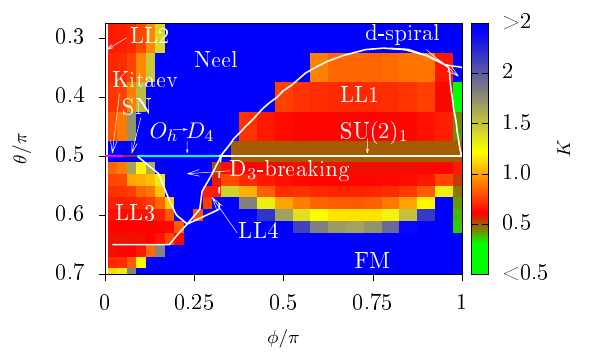}
\caption{Color plots of the numerically extracted values of the Luttinger parameter in the ranges $\theta\in[0.3\pi,0.7\pi]$ and $\phi\in[0,\pi]$.
The blue color is used to denote the situation where  no reliable value of the Luttinger parameter can be extracted.
The phase boundaries are plotted as solid white lines, 
except the line segment separating the ``LL4" and the ``FM" phases close to the equator which is represented by a dashed white line, since the precise shape of the phase boundary in that region cannot be clearly identified due to the percolation of the Luttinger liquid behaviors in the ``LL1" phase into the lower hemisphere. 
The names of the phases are written in white letters, where ``SN" is ``Spin-Nematic" for short.
DMRG numerics are performed on systems of $L=96$ sites with open boundary conditions.
} \label{fig:LL_K}
\end{figure*}

In this subsection,  we make a  quick summary of the phase diagram shown in Fig. \ref{fig:phase}.

On the equator, the ``$O_h\rightarrow D_4$" and ``Emergent SU(2)$_1$" phases
occupy the $\phi$-intervals $[\phi_c^\prime,\phi_c]$ and $[\phi_c,\pi]$, respectively, 
in which $\phi_c^\prime\simeq 0.10\pi$ and $\phi_c\simeq 0.33\pi$ as determined in Ref. \onlinecite{Yang2020a}.
When $J\neq 0$, the ``Emergent SU(2)$_1$" and ``$O_h\rightarrow D_4$" phases extend to the ``LL1" and ``$D_3$-breaking II" phases, respectively.
We note that the nature of the narrow hatched region close to the $K$ point highlighted with magenta color in Fig. \ref{fig:phase}  remains unclear, 
which will be left for further study.

For the four Luttinger liquid phases in Fig. \ref{fig:phase}, the Luttinger parameters have been calculated numerically based on the method described in Ref. \onlinecite{Laflorencie2006}.
The results are shown as color plots in Fig. \ref{fig:LL_K}, in which the blue color is used to signify the situation where  no reliable Luttinger parameter can be extracted.
As can be seen from Fig. \ref{fig:LL_K}, it is a very nice result that nearly all the phase boundaries in Fig. \ref{fig:phase} can be determined from this Luttinger parameter calculation.
We note that the phase boundaries in Fig. \ref{fig:phase} are only schematic,
and their precise shapes should be referred to Fig. \ref{fig:LL_K}.

The cartoon plots for the phases with $J\neq 0$ are collected in Fig. \ref{fig:phase}, all referring to the original frame without any sublattice rotation.
The patterns of the spin orientations in the ordered phases ``Neel", ``d-Spiral", ``$D_3$-breaking I, II" and ``FM"  are plotted,
where the $z$-direction is chosen to be vertical for the ``$D_3$-breaking I, II" phases, and perpendicular to the plane for the other ordered phases.
The quasi-long range orders in the Luttinger liquid phases ``LL$i$" ($i=1,2,3,4$) are also shown,
in which the $z$-axes are all fixed to be pointing upwards.
The black arrows represent the site-dependent directions of the quantization axes for the longitudinal fluctuations,
whereas the shaded blue ellipses represent the plane of the transverse fluctuations
which dominate over the longitudinal fluctuations in all the four Luttinger liquid phases.
We note that since the low energy Hamiltonian for the ``LL4" phase in the six-sublattice rotated frame has an FM-type quasi-long range order, there is no oscillation accompanying the power decay in the cartoon plot of the ``LL4" phase in Fig. \ref{fig:phase}.

Finally, we make a comment about the ED and DMRG numerics that we have performed in this work. 
For systems with open boundary conditions, the DMRG method was used on chains with length up to $L=144$ sites.
For some of the calculations, such as the ground state energy computations determining the boundaries of the phases, we used ED on chains up to $L=24$ sites long, while DMRG with periodic boundary conditions was used for chains of $L=36$ sites.  
In all the cases, we have checked that our DMRG results are converged using up to $m=1000$ states with a truncation error below $10^{-7}$.


\section{The ``LL1" phase}
\label{sec:LL1_phase}

\begin{figure}[h]
\includegraphics[width=8.0cm]{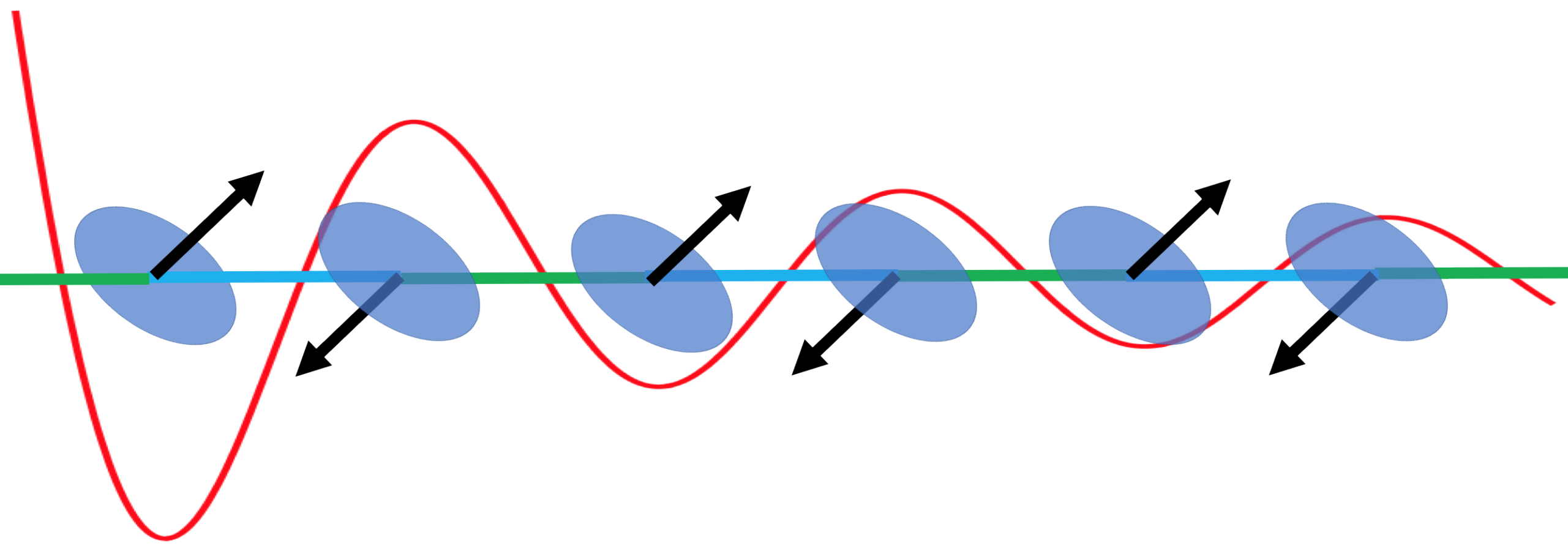}
\caption{Site-dependent quantization axes for the longitudinal fluctuations in the ``LL1" phase within the original frame. 
The black arrows denote the directions of the quantization axes, and the solid blue ellipses represent the transverse fluctuations.
The red line represents the AFM quasi-long range order for the longitudinal and transverse fluctuations defined in terms of the six-sublattice rotated frame.
The $z$-direction in spin space is chosen to be pointing upwards, and the $y$-direction is along the chain pointing to the right.
} \label{fig:LL1_quantize}
\end{figure}

In this section, we show that the region denoted by ``LL1" in Fig. \ref{fig:phase}
is described by the gapless Luttinger liquid theory.
The system exhibits a site-dependent quantization axis for the longitudinal fluctuations within the original frame as shown in Fig. \ref{fig:LL1_quantize}.
The strategy for analyzing the ``LL1" phase is to take the ``Emergent SU(2)$_1$" phase of the Kitaev-Gamma chain \cite{Yang2020a} on the equator of Fig. \ref{fig:phase} as the unperturbed Hamiltonian,
and treat the Heisenberg term  as a small perturbation using a perturbative RG analysis.
To facilitate analysis, we work in the six-sublattice rotated frame defined by Eq. (\ref{eq:6rotation}) throughout this section unless otherwise stated. 

\subsection{Brief review of the ``Emergent  SU(2)$_1$" phase}
\label{sec:review_SU2}

When $J=0$ in Eq. (\ref{eq:Ham}), the system reduces to a Kitaev-Gamma chain 
and has been studied in Ref. \onlinecite{Yang2020a}.
In this section, we briefly review the ``Emergent SU(2)$_1$" phase when $\phi_c\leq\phi\leq \pi$ and $\theta=\pi/2$,
which provides an RG perturbative starting point for analyzing the ``LL1" and ``FM" phases on the two different sides of the equator.
Due to the equivalence $\Gamma\simeq -\Gamma$, we will consider the equivalent region in the other half of the equator, i.e., $\theta=\pi/2$, $\phi\in(\pi,2\pi-\phi_c)$. 
In particular, the point $\phi=5\pi/4$ has explicit SU(2) symmetry in the rotated frame.

The low energy degrees of freedom in the emergent SU(2)$_1$ phase are
$\vec{J}_L,\vec{J}_R,g$, where $\vec{J}_L,\vec{J}_R$ are the WZW left and right currents
and $g$ is the SU(2)$_1$ primary field which is a $2\times 2$ matrix.
At low energies, the lattice spin operators $\vec{S}_i$'s can be expressed in terms of $\vec{J}_L,\vec{J}_R$ and $g$ using the following modified nonabelian bosonization formula \cite{Yang2020a},
\begin{eqnarray}
\frac{1}{a}S_j^{\alpha} = D_{[j]}^{\alpha} (J_L^\alpha+J_R^\alpha)+C_{[j]}^{\alpha} (-)^j \frac{1}{a} i\text{tr}(g\sigma^\alpha),
\label{eq:Nonabel}
\end{eqnarray}
in which $[j]$ ($1 \leq[j]\leq 3$) is defined as $j\equiv [j] \mod 3$,
$\sigma^\alpha$ ($\alpha=x,y,z$) are the three Pauli matrices,
and the WZW primary field $g$ is taken as dimensionless.
Symmetry constraints lead to the following relations among the coefficients \cite{Yang2020a}
\begin{flalign}
&\mathcal{E}_{1}^{x}=\mathcal{E}_{1}^{y}=\mathcal{E}_{2}^{x}=\mathcal{E}_{2}^{z}=\mathcal{E}_{3}^{y}=\mathcal{E}_{3}^{z}(=\mathcal{E}_1),\nn\\
&\mathcal{E}_{1}^{z}=\mathcal{E}_{2}^{y}=\mathcal{E}_{3}^{x}(=\mathcal{E}_2),
\end{flalign}
in which $\mathcal{E}=C,D$.
The low energy Hamiltonian of the Kitaev-Gamma chain is \cite{Yang2020a}
\begin{eqnarray}
H=\frac{2\pi}{3}v\int dx (\vec{J}_L\cdot \vec{J}_L+\vec{J}_R\cdot \vec{J}_R)-g_c\int dx \vec{J}_L\cdot \vec{J}_R,
\label{eq:Low_ham_KG}
\end{eqnarray}
in which $v$ is the spin velocity, and $g_c>0$ is the marginally irrelevant coupling.
The values of both $v$ and $g_c$ depend on the microscopic details,
which in principle can be obtained from DMRG numerical calculations on the Kitaev-Gamma chain.
We note that the spin-spin correlation functions can be calculated using Eq. (\ref{eq:Nonabel}) and Eq. (\ref{eq:Low_ham_KG}).

\subsection{Low energy Hamiltonian}
\label{sec:LL1_low_ham}

In this subsection, we derive the low energy perturbation Hamiltonian
by directly projecting the Heisenberg Hamiltonian
to the low energy space.
The low energy field theory is found to be the same as that of an XXZ chain with a quantization axis along the $(1,1,1)$-direction.
In practice, the spin operators within
\begin{eqnarray}
H^J=\sum_{<ij>\in\gamma\,\text{bond}} -J(S_i^\gamma S_j^\gamma+S_i^\alpha S_j^\beta+S_i^\beta S_j^\alpha)
\label{eq:Heisenberg_rot}
\end{eqnarray}
are replaced by $\vec{J}_L,\vec{J}_R$ and $g$ 
using the modified nonabelian bosonization formula in Eq. (\ref{eq:Nonabel}).
This method is essentially a first order perturbation treatment of the Heisenberg term,
and the coupling constants thus obtained are not accurate in the sense that they acquire renormalizations along the RG flow.
On the other hand, by performing a careful symmetry analysis on the low energy theory, we have justified the use of this first order perturbation method in capturing the essential physics.


To obtain the perturbation Hamiltonian, 
the following operator product expansion (OPE) formula for the N\'eel order fields is needed,
\begin{flalign}
&N^\lambda(x+a) N^\mu(x) = 
2\delta_{\lambda\mu} -4\pi a\epsilon^{\lambda\mu \nu} (J_L^\nu-J_R^\nu )\nn\\
&-(2\pi a)^2 \big[2J_L^{\{\lambda} J_R^{\mu\}} +J_L^{[\lambda} J_L^{\mu]} +J_R^{[\lambda} J_R^{\mu]}\nn\\
& +\delta_{\lambda\mu} (-2\vec{J}_L\cdot \vec{J}_R +\frac{1}{3}\vec{J}_L\cdot\vec{J}_L+\frac{1}{3}\vec{J}_R\cdot\vec{J}_R) \big]+...,
\label{eq:OPE_N}
\end{flalign}
in which $x$ is a spatial coordinate,
$a$ is the lattice constant, 
$\lambda,\mu=x,y,z$ are spin directions,
$\{...\}$ and $[...]$  in the superscripts denote symmetrization and antisymmetrization of the 
indices, respectively,
and only terms up to quadratic order in the WZW currents are kept.
 Eq. (\ref{eq:OPE_N}) can be derived from the affine symmetry of the SU(2)$_1$ WZW model as discussed in detail in Supplementary Materials \cite{SM}.

Plugging Eq. (\ref{eq:Nonabel}) into Eq. (\ref{eq:Heisenberg_rot})
and using  Eq. (\ref{eq:OPE_N}), we obtain
\begin{flalign}
&H^F\rightarrow-\frac{1}{3} Ja\int dx\big[
u_1(J_L^0-J_R^0) 
+u_2J^0_LJ^0_R
+u_3 \vec{J}_L\cdot \vec{J}_R\nn\\
&~~~~+u_4(J_L^0J_L^0+J_R^0J_R^0)
+u_5(\vec{J}_L\cdot \vec{J}_L+\vec{J}_R\cdot \vec{J}_R)
\label{eq:low_HF}
\big],
\end{flalign}
in which only the relevant and marginal terms are kept;
 the arrow ``$\rightarrow$"  indicates that it is not an exact equality but only a projection;
$\vec{J}_s\cdot \vec{J}_{s^\prime}=\sum_{i=0,1,2}J_s^iJ_{s^\prime}^i$ ($s,s^\prime=L,R$)
where
\begin{eqnarray}
J^0_s&=& \frac{1}{\sqrt{3}} (J^x_s+J^y_s+J^z_s),\nn\\
J^1_s&=& \frac{1}{\sqrt{6}} (2J^x_s-J^y_s-J^z_s),\nn\\
J^2_s&=& \frac{1}{\sqrt{2}} (J^y_s-J^z_s);
\end{eqnarray}
and the coefficient $u_2$ is 
\begin{eqnarray}
u_2&=&\frac{3}{2} [(D_1)^2+(D_2)^2]+12\pi^2[(C_1)^2+(C_2)^2].
\label{eq:u2}
\end{eqnarray}
The values of the other coefficients $u_i$'s ($i=1,3,4,5$) are not important for our purpose and can be found in Supplementary Materials \cite{SM},
in which a detailed derivation of Eq. (\ref{eq:low_HF}) is also included.
Furthermore, we have performed
a careful symmetry analysis in the low energy Hamiltonian showing that Eq. (\ref{eq:low_HF}) contains all the relevant and marginal terms allowed by symmetry
within the SU(2)$_1$ WZW model (for details, see Supplementary Materials \cite{SM}).
Therefore, the low energy Hamiltonian in Eq. (\ref{eq:low_HF}) is enough and complete to capture the physics as long as $|J|$ is small.
 
In the SU(2)$_1$ WZW theory,
$J_\lambda^0J_\lambda^0$ ($\lambda=L,R$)
is equal to $\frac{1}{3}\vec{J}_\lambda\cdot \vec{J}_\lambda$ (see Supplementary Materials \cite{SM}),
hence it does not give an independent contribution.
In addition, the inversion breaking term $J_L^0-J_R^0$
can be eliminated by a chiral rotation \cite{Garate2010,Gangadharaiah2008,Schnyder2008}.
As a result,
the only nontrivial SU(2) breaking term in the
low energy theory of the KHG chain is $J^0_LJ^0_R$.
This shows that the low energy physics is the same as that of an XXZ chain with a quantization axis along the $(1,1,1)$-direction.
Whether the system remains gapless or develops an order 
is determined by the sign of the coupling $-\frac{1}{3}Ju_2$.

To gain a simple understanding as to why the $(1,1,1)$-direction is special,
here we give a brief description of the symmetry group in the six-sublattice rotated frame.
With a nonzero Heisenberg term, the symmetry transformations of the Hamiltonian $H^\prime$ in Eq. (\ref{eq:6rotated}) are
\begin{eqnarray}
1.&T &:  (S_i^x,S_i^y,S_i^z)\rightarrow (-S_{i}^x,-S_{i}^y,-S_{i}^z)\nn\\
2.&R_I I&: (S_i^x,S_i^y,S_i^z)\rightarrow (-S_{10-i}^z,-S_{10-i}^y,-S_{10-i}^x)\nn\\
3.& R_aT_a&:  (S_i^x,S_i^y,S_i^z)\rightarrow (S_{i+1}^z,S_{i+1}^x,S_{i+1}^y),
\label{eq:sym_Jneq0}
\end{eqnarray}
in which $T$ is time reversal; 
$T_a$ is translation by one lattice site;
$I$ is the spatial inversion around the point $C$ in Fig. \ref{fig:bonds} (b);
and $R_a=R(\hat{n}_a,-2\pi/3)$, $R_I=R(\hat{n}_I,\pi)$ where 
\begin{eqnarray}
\hat{n}_a=\frac{1}{\sqrt{3}}(1,1,1)^T, ~ \hat{n}_I=\frac{1}{\sqrt{2}}(1,0,-1)^T.
\label{eq:na_nI}
\end{eqnarray}
In Eq. (\ref{eq:sym_Jneq0}), the choice of the inversion center in the definition of $I$ is well defined modulo three.
If another inversion center $5+3n$ is chosen, then $S_{10-i}^\alpha$ ($\alpha=x,y,z$) in the second line in Eq. (\ref{eq:sym_Jneq0}) should be correspondingly replaced by $S_{10+6n-i}^\alpha$.
According to Eq. (\ref{eq:sym_Jneq0}), the symmetry group $G_1$ of $H^\prime$ is
\begin{eqnarray}
G_1=\mathopen{<}T, R_aT_a, R_I I\mathclose{>}.
\label{eq:group_G1}
\end{eqnarray} 
In the continuum limit, the lattice is coarse-grained and the effect of $T_a$ is smeared.
Hence, it is not a surprise that the symmetry operation $R_aT_a$ picks out $\hat{n}_a$ (i.e., the $(1,1,1)$-direction)  to be the quantization axis in the low energy theory (which is also justified by the symmetry analysis on the low energy Hamiltonian as discussed in Supplementary Materials \cite{SM}).

\begin{figure}[h]
\includegraphics[width=7.0cm]{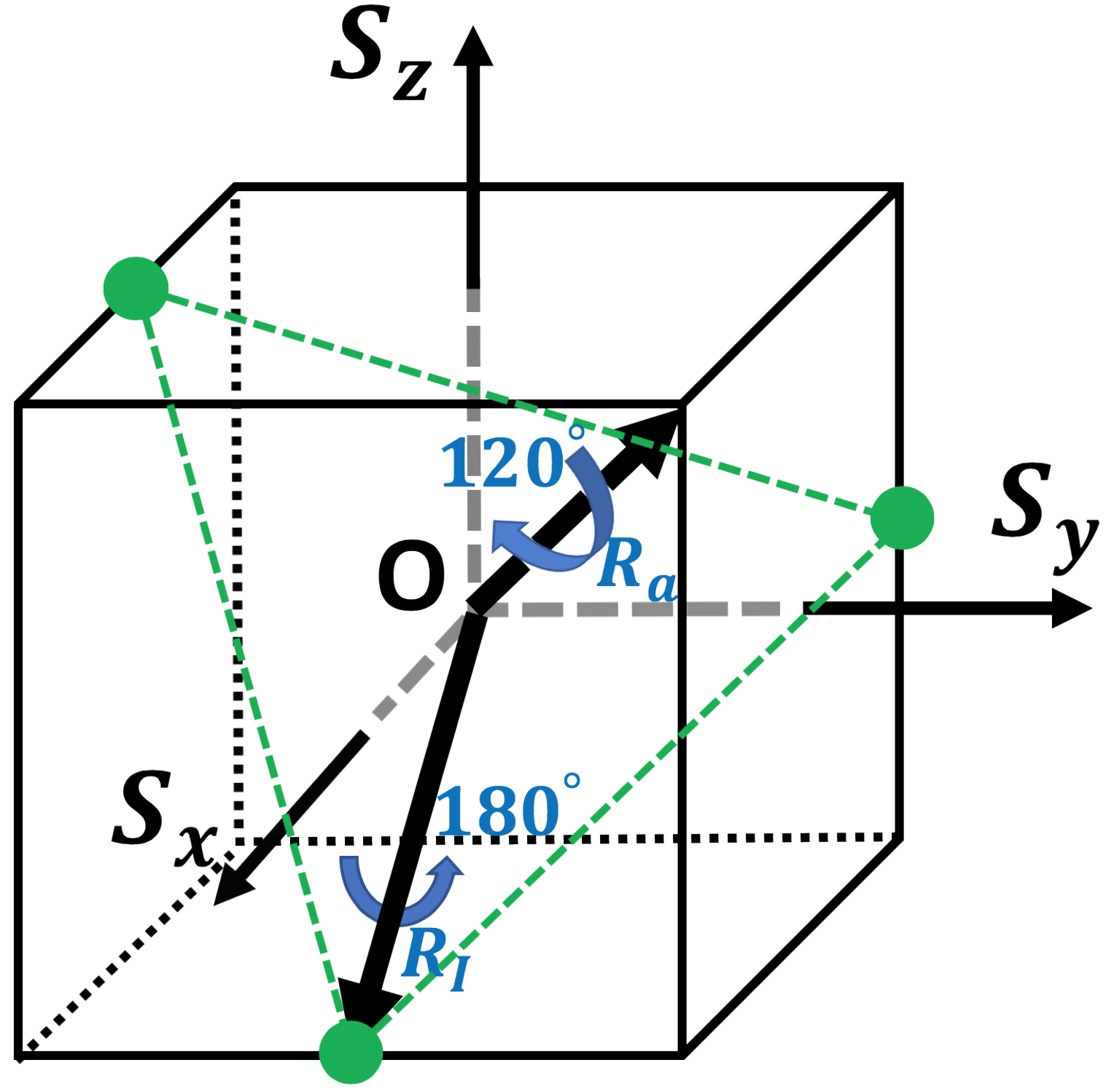}
\caption{$R_a$ and $R_I$ as symmetry transformations of a regular triangle in the spin space.
The vertices of the regular triangle are represented by the  three solid green circles.
The axes are labeled as $S_\alpha$ ($\alpha=x,y,z$) to emphasize that we are considering the spin space.
} \label{fig:D3d_geometry}
\end{figure}

Finally, we make a comment on the group structure of $G_1$ which will be used in Sec. \ref{sec:LL_classical}.
Since $T_{3a}=(R_aT_a)^3$ belongs to $G_1$, it is legitimate to consider the quotient group $G_1/\mathopen{<}T_{3a}\mathclose{>}$.
In our case, time reversal plays the role of ``inversion" in the spin space, since $T$ changes the signs of the spin operators.
Neglecting the spatial components in $R_aT_a$  and $R_II$, the spin operations $R_a$ and $R_I$ generate the symmetry group of a regular triangle as shown in 
Fig. \ref{fig:D3d_geometry}.
Since $D_{3d}=D_3\times\mathbb{Z}_2$ in which $D_3$ is the symmetry group of a regular triangle and $\mathbb{Z}_2$ is generated by the inversion operation \cite{Coxeter1965},
we see that $\mathopen{<}T,R_a,R_I\mathclose{>}\cong D_{3d}$.
As proved in Supplementary Materials \cite{SM},
$\mathopen{<}T,R_aT_a,R_II
\mathclose{>}/\mathopen{<}T_{3a}\mathclose{>}$ is isomorphic to $D_{3d}$ even when the spatial components in the operations are included.
As a result, 
$G_1/\mathopen{<}T_{3a}\mathclose{>}\simeq D_{3d}$.   
This shows that the group structure of $G_1$ is
\begin{eqnarray}
G_1\simeq D_{3d}\ltimes 3\mathbb{Z},
\label{eq:D3Z}
\end{eqnarray}
in which  $3\mathbb{Z}=\mathopen{<}T_{3a}\mathclose{>}$.

\subsection{Phase diagram}
\label{sec:RG_KG_H}

\begin{figure}[h]
\includegraphics[width=7.0cm]{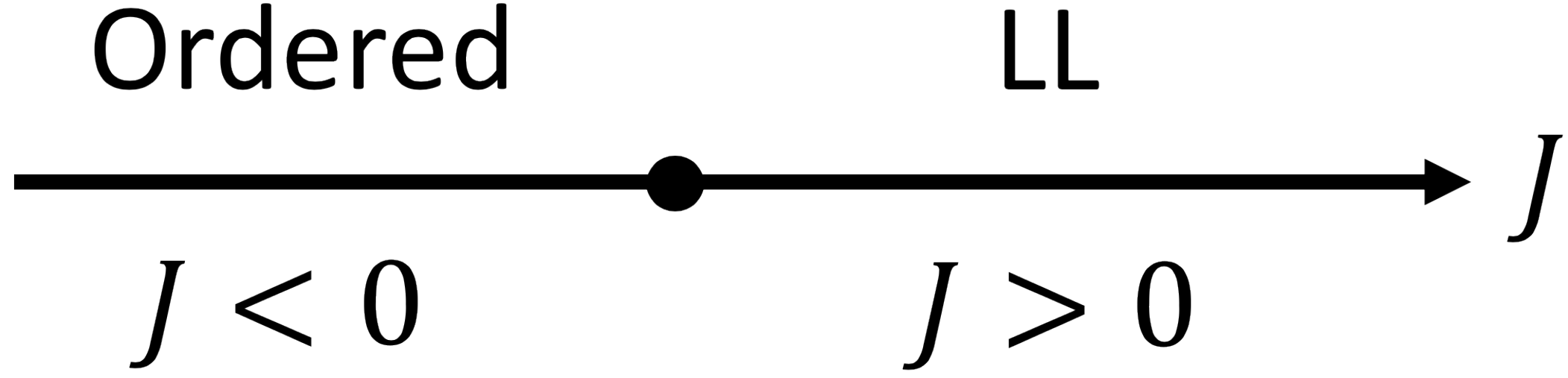}
\caption{Phase diagram of the KHG chain by tuning $J$,
where $\phi\in(\pi+\phi_c,2\pi)$,
in which ``LL" is ``Luttinger liquid" for short.
} \label{fig:phase_J}
\end{figure}

In a perturbative treatment of the Heisenberg term, 
we will assume that $|J|\ll va,g_ca$.
With a nonzero Heisenberg term, the low energy Hamiltonian is modified to 
\begin{eqnarray}
H&=&\frac{2\pi}{3}v^\prime\int dx (\vec{J}_L\cdot \vec{J}_L+\vec{J}_R\cdot \vec{J}_R)\nn\\
&&-2\pi v^\prime \int dx [ y_\perp (J_L^1J_R^1+J_L^2J_R^2)+y_\parallel J_L^0J_R^0],
\label{eq:low_Ham_KHG}
\end{eqnarray}
in which according to Eq. (\ref{eq:low_HF}),
\begin{eqnarray}
v^\prime&=&v-\frac{1}{6\pi}Jau_5,\nn\\ 
y_\perp& =& \frac{1}{2\pi v^\prime} (g_c+\frac{1}{3} Jau_3),\nn\\
y_\parallel &=& \frac{1}{2\pi v^\prime} [g_c+\frac{1}{3} Ja(u_3+u_2)].
\end{eqnarray}
In particular, since $u_2$ is always positive (as can be seen from Eq. (\ref{eq:u2})),
we have $y_\parallel>y_\perp$  when $J>0$, and $y_\parallel<y_\perp$ when $J<0$.
We note that in Eq. (\ref{eq:low_Ham_KHG}),
the chiral term $J_L^0-J_R^0$ term is dropped 
according to the discussions in Sec. \ref{sec:LL1_low_ham}.
The RG flow equations of $y_\perp$ and $y_\parallel$ are of the Kosterlitz-Thouless (KT) type \cite{Giamarchi2003}:
\begin{eqnarray}
\frac{dy_\perp}{dl} &=&-y_\perp y_\parallel,\nn\\
\frac{dy_\parallel}{dl} &=&-y_\perp^2,
\label{eq:RG_LL1_KT}
\end{eqnarray}
which are obtained by integrating out the modes with wavelengths between $e^la$ and $e^{l+dl}a$.
It is well known that the system flows to a strong coupling limit when $y_\parallel<y_\perp$,
and remains gapless when $y_\parallel>y_\perp$ \cite{Giamarchi2003}.

The phase diagram by tuning $J$ is shown in Fig. \ref{fig:phase_J}.
Notice that this phase diagram can be easily understood from intuitive physical arguments.
The $J>0$ case introduces an FM-like term into the Hamiltonian since the coupling is $-J$ in the rotated frame.
Thus the AFM fluctuations in the longitudinal direction are suppressed and the system has an easy-plane anisotropy, corresponding to the ``LL" phase in Fig. \ref{fig:phase_J},
where ``LL" is ``Luttinger liquid" for short.
Similarly, the $J<0$ case has an easy-axis anisotropy which corresponds to the ``Ordered" phase in Fig. \ref{fig:phase_J}.
We note that Fig. \ref{fig:phase_J} is for a fixed background Kitaev-Gamma chain,
i.e., the value of $\phi$ in Fig. \ref{fig:phase} is fixed.
By varying $\phi$, the gapless phase in Fig. \ref{fig:phase_J}
extends to the ``LL1" phase in Fig. \ref{fig:phase},
whereas the ``Ordered" phase in Fig. \ref{fig:phase_J} is below the equator,
which will be shortly shown to be the ``FM" phase in Fig. \ref{fig:phase}.
We also note that the $0$-direction (i.e., the longitudinal direction) in Eq. (\ref{eq:low_Ham_KHG}) is  the $(111)$-direction,
which becomes staggered in the original frame according to Eq. (\ref{eq:6rotation}).
A sketch of the site-dependent  quantization axis is shown in Fig. \ref{fig:LL1_quantize}.
However,  the quantization axes in Fig. \ref{fig:LL1_quantize} are not precise since they can be distorted due to the renormalization effects of the spin operators along the RG flow similar as the origins of the  $C_1,C_2$ coefficients discussed in Eq. (\ref{eq:Nonabel}).

\begin{figure*}[htbp]
\includegraphics[width=18.0cm]{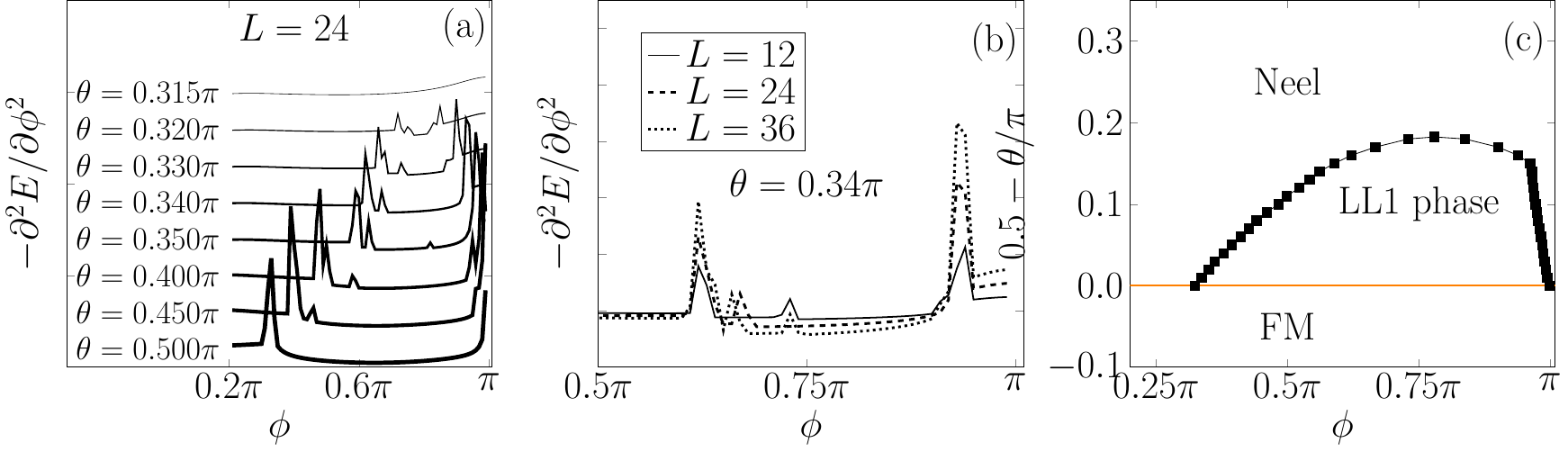}
\caption{(a) Horizontal scan of $\partial^2 E/\partial\phi^2$ as a function of $\phi$ for fixed values of $\theta$, 
(b) finite size scaling of $\partial^2 E/\partial\phi^2$,
(c) phase boundary of the ``LL1" phase.
In (a), the curves for different angles are shifted vertically. 
In (c), the phase boundary is determined from the positions of the largest two peaks for fixed $\theta$'s in (a).
ED calculations are performed on periodic systems of $L=24$ sites in (a), 
$L=12, 24$ sites in (b). In (b), DMRG simulations with periodic boundary conditions were performed on $L=36$ sites chain.
} \label{fig:LL1}
\end{figure*}

When $J$ is negative, we have seen that the system develops a N\'eel long range order along the $(111)$-direction. 
However, the alignments of the spins are distorted due to the difference in $C_1$ and $C_2$.
Using Eq. (\ref{eq:Nonabel}) and assuming a nonzero expectation value for $N^0=\frac{1}{\sqrt{3}} (N^x+N^y+N^z)$, 
the spin polarizations have a six-site periodicity,
\begin{eqnarray}
\left<\vec{S}_{1+6n}\right> &=&\frac{1}{\sqrt{3}}N (C_1,C_1,C_2)^T,\nn\\
\left<\vec{S}_{2+6n} \right>&=&\frac{1}{\sqrt{3}}N (-C_1,-C_2,-C_1)^T,\nn\\
\left<\vec{S}_{3+6n} \right>&=&\frac{1}{\sqrt{4}}N (C_2,C_1,C_1)^T,\nn\\
\left<\vec{S}_{4+6n} \right>&=&\frac{1}{\sqrt{4}}N (-C_1,-C_1,-C_2)^T,\nn\\
\left<\vec{S}_{5+6n} \right>&=&\frac{1}{\sqrt{3}}N (C_1,C_2,C_1)^T,\nn\\
\left<\vec{S}_{6+6n} \right>&=&\frac{1}{\sqrt{3}}N (-C_2,-C_1,-C_1)^T,
\label{eq:LL1_FM_6sub}
\end{eqnarray}
in which $n\in\mathbb{Z}$ and  $N=\left<N^0\right>$.

Recall that the results in Eq. (\ref{eq:LL1_FM_6sub}) are within the six-sublattice rotated frame.
To obtain the spin orientations in the original frame, 
it is enough to perform the inverse of the transformation defined in Eq. (\ref{eq:6rotation}).
It is straightforward to verify that 
\begin{eqnarray}
\left<\vec{S}_i^{(0)}\right> \equiv\frac{1}{\sqrt{3}} N(C_1,C_1,C_2)^T,
\label{eq:FM_LL1}
\end{eqnarray}
where the superscript ``$(0)$" is used to denote the spin operators in the original frame.
Clearly, Eq. (\ref{eq:FM_LL1}) represents an FM order along the $(C_1C_1C_2)$-direction.
This provides an understanding of the FM phase in Fig. \ref{fig:phase} in the region close to the equator.

\subsection{Numerical results}
\label{sec:LL1_K_value}

To determine the range of the ``LL1" phase, we study the ground state energy $E$ as a function of $\theta,\phi$.
ED calculations are performed for periodic systems of $L=24$ sites.
In Fig. \ref{fig:LL1} (a), $\partial^2 E/\partial \phi^2$ is scanned horizontally for several fixed values of $\theta$,
in which different curves are shifted vertically.
The peaks correspond to divergent positions of $\partial^2 E/\partial \phi^2$, indicating first order phase transitions.
We emphasize that only the two largest peaks in Fig. \ref{fig:LL1} (a) for each $\theta$ represent first order phase transitions, which determine the left and right boundary points of the black dome in Fig. \ref{fig:LL1} (c) at the corresponding cut of $\theta$.
The middle smaller peaks in Fig. \ref{fig:LL1} (a) do not scale with systems sizes as shown in Fig. \ref{fig:LL1} (b),
therefore there is no phase transition at these positions in the thermodynamic limit.
The absence of phase transition in the middle region is also confirmed by our numerical calculations of the Luttinger parameter which will be discussed shortly later.
Fig. \ref{fig:LL1} (c) shows the phase boundary of the ``LL1" phase determined from Fig. \ref{fig:LL1} (a), as represented by the curve connecting the small solid black squares.
Starting from the ``LL1" phase, the system transits into the ``N\'eel" phase and the ``d-spiral" phases by going upwards and rightwards, respectively, as shown in Fig. \ref{fig:LL1} (c).
The ``N\'eel" and ``d-spiral" phases will be discussed in Sec. \ref{sec:Neel_phase} and Sec. \ref{sec:d_sp}.

Next, we numerically determine the Luttinger parameter $K$ in the ``LL1" phase
following the method described in Ref. \onlinecite{Laflorencie2006}.
The low energy field theory is given by the following Luttinger liquid Hamiltonian 
\begin{eqnarray}
H=\frac{1}{2} \int dx (\frac{u}{\mathcal{K}} (\nabla \varphi)^2 +u\mathcal{K} (\nabla \theta)^2),
\label{eq:LL_liquid}
\end{eqnarray}
in which $u$ is the velocity and $\mathcal{K}$ is the Luttinger liquid parameter.
While $\mathcal{K}=1/2$ corresponds to the SU(2) symmetric case, the symmetry is reduced to U(1) when $\mathcal{K}\neq 1/2$.
Define the Hamiltonian density $h(x)$ as 
\begin{flalign}
h(x)=KS_x^\gamma S_{x+1}^\gamma+ J\vec{S}_x\cdot \vec{S}_{x+1}+\Gamma (S_x^\alpha S_{x+1}^\beta+S_x^\beta S_{x+1}^\alpha).
\end{flalign}
For a finite size system with an open boundary condition, the energy density $\langle h(x)\rangle$ contains a uniform part $E_U(x)$ and a staggered part $E_A(x)$,
where \cite{Laflorencie2006}
\begin{eqnarray}
E_A(x)\propto \frac{1}{[\frac{L}{\pi}\sin(\frac{\pi x}{L})]^\mathcal{K}},
\label{eq:LL_EA}
\end{eqnarray}
in which $x=j a$ ($j\gg 1$) is the distance measured from the boundary of the system.
Consider a general function $f(j)$ ($j\in \mathbb{Z}$) which contains a uniform part $u(j)$ and a stagger part $s(j)$, i.e., 
\begin{eqnarray}
f(j)=u(j)+(-)^j s(j).
\end{eqnarray}
A three-point formula can be used to extract $u(j)$ and $s(j)$, as
\begin{eqnarray}
u(j)&=& \frac{1}{4}f(j-1)+\frac{1}{2}f(j)+\frac{1}{4}f(j+1),\nn\\
s(j)&=& (-)^j\big[-\frac{1}{4}f(j-1)+\frac{1}{2}f(j)-\frac{1}{4}f(j+1)\big].
\label{eq:three_point}
\end{eqnarray}
Hence $E_A$ can be obtained from the numerical result of $h(x)$ by applying Eq. (\ref{eq:three_point}).

We have used this method to study the Luttinger parameters in the whole ``LL1" phase, and the results are displayed in Fig. \ref{fig:LL_K} where the magnitudes of $\mathcal{K}$ are represented by different colors.
This provides direct numerical evidence for the entire region enclosed by the phase boundary determined in Fig. \ref{fig:LL1} to be a Luttinger liquid phase.
On the other hand, as can be seen from Fig. \ref{fig:LL_K}, the Luttinger liquids percolate into the ``FM" phase.
However, we note that this is a finite size artifact. 
As discussed in Sec. \ref{sec:RG_KG_H}, the phase transition between the ``LL1" and ``FM" phases is second order.
Since a gap $E_g\sim e^{-\text{const}/\sqrt{|J|}}$ opens exponentially slowly in the ``FM" phase close to the transition line \cite{Affleck1988}, 
the crossover system size $L_c$ (only above which an order can be observed) grows exponentially at small $J$, i.e., $L_c\sim e^{\text{const}/\sqrt{|J|}}$.
Thus the Luttinger liquid behavior can still be observed in an extended region in the ``FM" phase in a finite size system.

\section{The ``N\'eel" phase}
\label{sec:Neel_phase}

\begin{figure}[htbp]
\includegraphics[width=7.5cm]{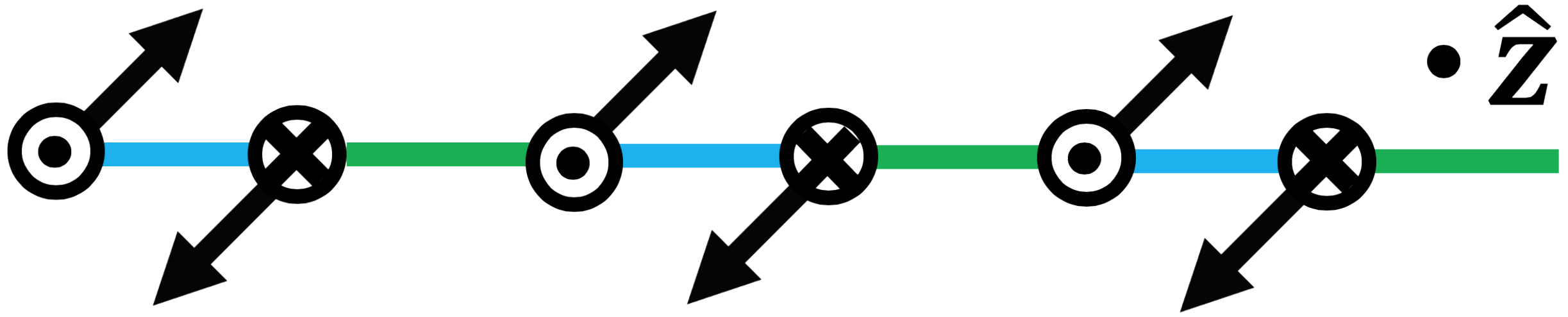}
\caption{Spin alignments in the N\'eel phase within the original frame.
The $z$-direction in spin space is chosen to be perpendicular to the plane,
and the $x$-direction is along the chain to the right.
} 
\label{fig:picture_Neel}
\end{figure}

In this section, we study the ``N\'eel" phase shown in Fig. \ref{fig:phase}.
The spin alignments within the original frame are plotted in Fig. \ref{fig:picture_Neel}.
The strategy that we take to analyze the ``N\'eel" phase is to perform a perturbative RG analysis in the neighborhood of the AFM1 point located at the north pole in Fig. \ref{fig:phase}.
Numerics provide evidence for the  N\'eel ordering to hold in the entire region marked as ``N\'eel"  in Fig. \ref{fig:phase}.
Throughout the section, we work in the original frame,
and the discussions will be based on the Hamiltonian given in Eq. (\ref{eq:Ham}).

\begin{figure}
\includegraphics[width=7.5cm]{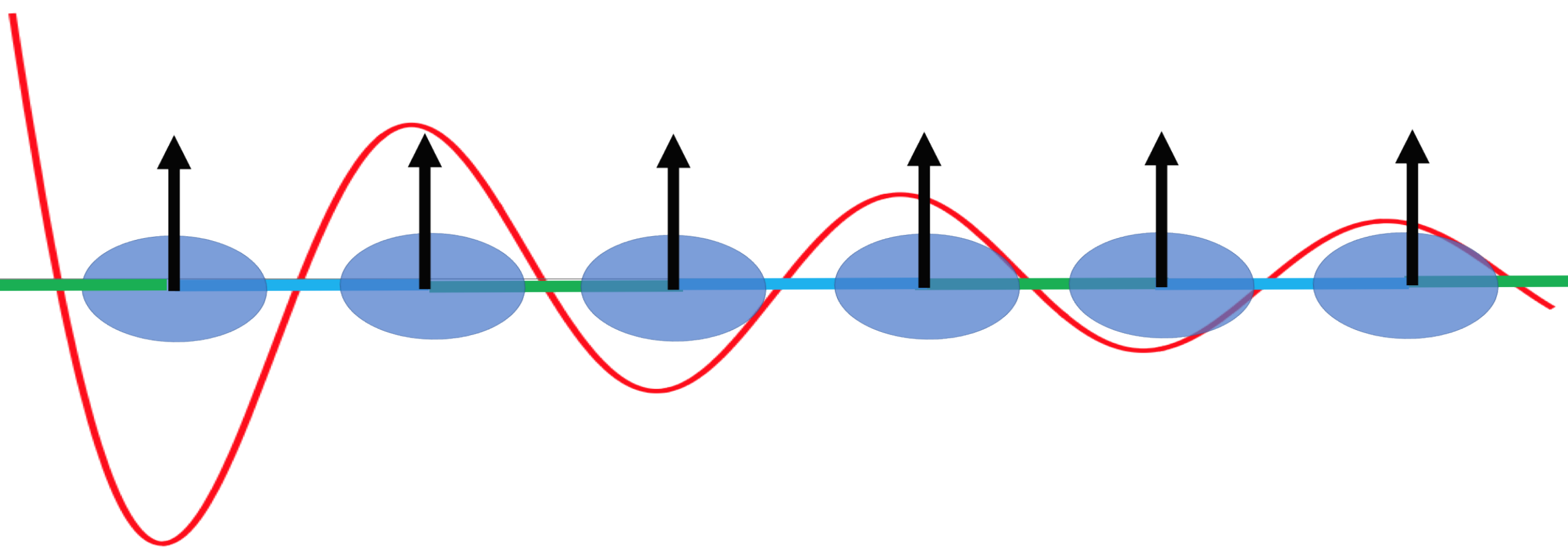}
\caption{Site-dependent quantization axes for the longitudinal fluctuations in the ``LL2" phase within the original frame. 
The black arrows denote the directions of the quantization axes, and the solid blue ellipses represent the transverse fluctuations.
The red line represents the AFM quasi-long range order for the longitudinal and transverse fluctuations.
The $z$-direction in spin space is chosen to be pointing upwards, and the $y$-direction is along the chain.
} 
\label{fig:LL2_quantization}
\end{figure}

\subsection{RG analysis}
\label{eq:RG_Neel}

We can perform a first order perturbation treatment of the Kitaev and Gamma terms,
by projecting $H_K$ and $H_\Gamma$ to the low energy degrees of freedom using the nonabelian bosonization formula \cite{Affleck1988},
\begin{flalign}
\frac{1}{a} S_i^\alpha=J_L^\alpha+J_R^\alpha
+(-)^{i+1} \frac{c}{2\pi a} i\text{tr} (g\sigma^\alpha).
\label{eq:nonabelian_Heisenberg}
\end{flalign}
in which $D_1=D_2=1$, $C_1=C_2=\frac{c}{2\pi}$
 where $c$ is a constant.
Comparing with Eq. (\ref{eq:Nonabel}),  the SU(2) symmetry is not broken in Eq. (\ref{eq:nonabelian_Heisenberg}). 
Explicit calculations show that (for details, see Supplementary Materials \cite{SM}),
\begin{flalign}
&H^K\rightarrow Ka \int dx \big[
-\frac{c^2}{2\pi^2 a^2} +\frac{3+2c^2}{6} (\vec{J}_L\cdot\vec{J}_L+\vec{J}_R\cdot\vec{J}_R)\nn\\
&-\frac{1}{2} (J_L^zJ_L^z+J_R^zJ_R^z)+\vec{J}_L\cdot \vec{J}_R-(1+2c^2)J_L^zJ_R^z
\big],
\label{eq:first_order_KHm}
\end{flalign}
and
\begin{eqnarray}
H^\Gamma\rightarrow 2c^2\Gamma a\int dx \big[(J_L^x+J_L^y)J_R^z+J_L^z(J_R^x+J_R^y)\big].
\label{eq:Neel_HGamma}
\end{eqnarray}
We have performed a careful symmetry analysis of the low energy field theories 
for both the $\Gamma=0$ and $\Gamma\neq 0$ cases as discussed in details in 
Supplementary Materials \cite{SM},
where it is shown that the first order perturbation Hamiltonian is already enough and complete to capture the low energy physics.
Notice that  the low energy Hamiltonian is of an XXZ type in the absence of $\Gamma$.
Since the coefficient of the $J_L^zJ_R^z$ term is $-Ka(1+2c^2)$ when $\Gamma=0$,
the system has easy-plane and easy-axis anisotropies for $K>0$ and $K<0$, respectively.

The Luttinger liquid phase when $K>0$ (and $\Gamma=0$) revealed by this analysis is the ``LL2" phase shown in Fig. \ref{fig:phase}.
The ``LL2" phase has already been discussed in Ref. \cite{Agrapidis2018}, and our DMRG numerics have confirmed the existence of the ``LL2" phase as shown in Fig. \ref{fig:LL_K}.
We note that the percolation of the Luttinger liquid behavior in the ``LL2" phase to nonzero $\phi$'s is a finite size artifact. 
The quantization axis for the longitudinal fluctuation in the ``LL2" phase is along the $z$-axis as shown in Fig. \ref{fig:LL2_quantization}.
However, similar to the case of the ``LL1" phase, the quantization axes in Fig. \ref{fig:LL2_quantization} are not precise since they can be distorted due to the renormalization effects of the spin operators along the RG flow.

Next we perform an RG analysis of the low energy Hamiltonian in Eq. (\ref{eq:Neel_HGamma}) for nonzero $\Gamma$.
We will see that it belongs to the XYZ class as long as $\Gamma \neq 0$, and an N\'eel order develops.
The low energy perturbation Hamiltonian can be written as
\begin{eqnarray}
(J_L^xJ_L^yJ_L^z) \left(\begin{array}{ccc}
0 & 0 &2c^2\Gamma a\\
0 & 0 &2c^2\Gamma a\\
2c^2\Gamma a & 2c^2\Gamma a&-(1+2c^2)Ka\\
\end{array}\right)
\left(\begin{array}{c}
J_R^x\\
J_R^y\\
J_R^z
\end{array}
\right).
\label{eq:low_Neel}
\end{eqnarray}
The matrix kernel in Eq. (\ref{eq:low_Neel}) can be straightforwardly diagonalized.
The eigenvalues are 
\begin{eqnarray}
E_1&=&0,\nn\\ 
E_{\pm}&=&c^2\Gamma a (k\pm\sqrt{8+k^2}),
\end{eqnarray}
with the corresponding unnormalized eigenvectors
\begin{eqnarray}
\Psi_0&=&(-1,1,0)^T,\nn\\ 
\Psi_\pm&=&(\frac{1}{4} (k\mp\sqrt{8+k^2}),\frac{1}{2} (k\mp\sqrt{8+k^2}),1)^T,
\end{eqnarray}
 in which $k=\frac{2+4c^2}{3c^2}\frac{K}{\Gamma}$.
Notice that as long as $\Gamma> 0$,
we always have $E_+>0$ and $E_-<0$
regardless of  the value of $K$.

Next, let's perform a rotation of the coordinate system,
such that
\begin{eqnarray}
 \hat{x}^\prime=\Psi_-/|\Psi_-|,
\hat{y}^\prime=\Psi_+/|\Psi_+|,
\hat{z}^\prime=\Psi_0/|\Psi_0|.
\label{eq:prime_directions}
\end{eqnarray}
Then in terms of the new coordinates, the marginal terms become
\begin{eqnarray}
-2\pi v \int dx (\lambda_x J_L^{\prime x} J_R^{\prime x}+\lambda_y J_L^{\prime y} J_R^{\prime y} +\lambda_z J_L^{\prime z} J_R^{\prime z}),
\end{eqnarray}
in which $\lambda_x=(g_c+E_-)/(2\pi v)$,
$\lambda_y=(g_c+E_+)/(2\pi v)$, and
$\lambda_z=g_c/(2\pi v)$. 
In particular, when $K=0$, we have
\begin{eqnarray}
\lambda_x^{(0)}&=&\frac{1}{2\pi v} (g_c+\sqrt{2}c^2\Gamma a),\nn\\
\lambda_y^{(0)}&=&\frac{1}{2\pi v} (g_c-\sqrt{2}c^2\Gamma a),\nn\\
\lambda_z^{(0)}&=&\frac{1}{2\pi v} g_c,
\label{eq:initial_RG}
\end{eqnarray}
in which $\Gamma$ is assumed to be small,
and only terms up to $O(\Gamma)$ are kept.
The superscript ``$(0)$" in Eq. (\ref{eq:initial_RG}) is used to indicate that these are the initial values of the 
couplings before the RG flow.
In what follows, we consider the case $\Gamma>0$, $K=0$ for simplicity.
The case of a nonzero $K$ can be  discussed similarly.

The RG flow equations of $\lambda_x,\lambda_y,\lambda_z$ are
\cite{Affleck1988}
\begin{eqnarray}
\frac{d\lambda_x}{dl} &=&-\lambda_y \lambda_z,\nn\\
\frac{d\lambda_y}{dl} &=&-\lambda_z\lambda_x,\nn\\
\frac{d\lambda_z}{dl} &=&-\lambda_x\lambda_y,
\label{eq:RG_LL1}
\end{eqnarray}
which can be obtained from the OPE formula for the WZW current operators \cite{Affleck1988}.
There are three constants of motion for Eq. (\ref{eq:RG_LL1})
which take the following forms when $K=0$:
\begin{eqnarray}
\lambda_x^2(l)-\lambda_y^2(l)&=& \frac{c^2g_c\Gamma a}{\sqrt{2}\pi^2 v^2}  \nn\\
\lambda_y^2(l)-\lambda_z^2(l)&=& -\frac{c^2g_c\Gamma a}{2\sqrt{2}\pi^2 v^2}  \nn\\
\lambda_z^2(l)-\lambda_x^2(l)&=&  -\frac{c^2g_c\Gamma a}{2\sqrt{2}\pi^2 v^2}. 
\label{eq:3hyper}
\end{eqnarray}
Solving Eq. (\ref{eq:RG_LL1})  with initial conditions given in Eq. (\ref{eq:initial_RG}) (see Supplementary Materials \cite{SM} for derivation),
we obtain
\begin{eqnarray}
\lambda_x,\lambda_z\sim \Lambda, ~ \lambda_y\sim -\Lambda,
\end{eqnarray}
where $\Lambda\rightarrow \infty$
as $l\rightarrow \infty$.
It is clear that the system flows to strong couplings at low energies.


To identify the phase corresponding to the strong coupling limit,
we perform a chiral rotation $R_L(\hat{y},\pi)$, which maps  
$(J_L^{\prime x},J_L^{\prime y},J_L^{\prime z})$ to $(-J_L^{\prime x},J_L^{\prime y},-J_L^{\prime z})$,
but leaves $J_R^{\prime\alpha}$ ($\alpha=x,y,z$) unchanged.
Then it is clear that the three couplings after the chiral rotation become $\tilde{\lambda}_x,\tilde{\lambda}_y,\tilde{\lambda}_z\simeq-\Lambda\rightarrow-\infty$.
On the other hand, as shown in Ref. \onlinecite{Affleck1988},
this strong coupling limit corresponds to a dimer order, i.e., 
\begin{eqnarray}
\text{tr}\tilde{g}\neq 0,
\label{eq:tilde_g}
\end{eqnarray}
where $\tilde{g}=U_Lg$ is the SU(2) WZW field after the chiral rotation,
in which $U_L=e^{i\frac{1}{2}\sigma^y \pi}=i\sigma^y$ is the chiral rotation matrix.
Therefore, Eq. (\ref{eq:tilde_g}) implies that 
\begin{eqnarray}
i\text{tr}(g\sigma^y)\neq 0,
\end{eqnarray}
i.e., there is a N\'eel order along the $y^\prime$-direction.
In terms of the original coordinates, the N\'eel ordering is along
$
\hat{y}^\prime=(\frac{1}{2},\frac{1}{2},\frac{1}{\sqrt{2}})^T
$,
as can be seen from Eq. (\ref{eq:prime_directions}).

\subsection{Symmetry breaking}

We make a discussion on the symmetry breaking pattern in the N\'eel phase.
The symmetry transformations in the original frame for a general Kitaev-Heisenberg-Gamma chain are 
\begin{eqnarray}
1.&T &:  (S_i^x,S_i^y,S_i^z)\rightarrow (-S_{i}^x,-S_{i}^y,-S_{i}^z)\nn\\
2.& T_{2a}&:  (S_i^x,S_i^y,S_i^z)\rightarrow (S_{i+2}^x,S_{i+2}^y,S_{i+2}^z)\nn\\
3.&T_a I&: (S_i^x,S_i^y,S_i^z)\rightarrow (S_{-i+1}^x,S_{-i+1}^y,S_{-i+1}^z)\nn\\
4.&R(\hat{n}_N,\pi)T_a&: (S_i^x,S_i^y,S_i^z)\rightarrow (-S_{i+1}^y,-S_{i+1}^x,-S_{i+1}^z),\nn\\
\label{eq:Sym_Neel_KHG}
\end{eqnarray}
in which $\hat{n}_N=\frac{1}{\sqrt{2}}(1,-1,0)^T$.
Since $T_{2a}=[R(\hat{n}_N,\pi)T_a]^2$,
we conclude that the symmetry group $G_N$ is 
\begin{eqnarray}
G_N=\mathopen{<}T,T_aI,R(\hat{n}_N,\pi)T_a
\mathclose{>}.
\label{eq:GN}
\end{eqnarray}

The only broken symmetry in the ``N\'eel" phase is the time reversal symmetry.
The unbroken symmetry group $H_N$ can be determined as 
\begin{eqnarray}
H_N=\mathopen{<}R(\hat{n}_N,\pi)T_a,T T_aI
\mathclose{>}.
\label{eq:HN}
\end{eqnarray}
It is straightforward to verify that the most general  pattern of the spin orderings which is invariant under $H_N$ is given by
\begin{eqnarray}
\vec{S}_j=(-)^j(a,a,b)^T.
\label{eq:spinorder_Neel}
\end{eqnarray}
The values of the parameters $a,b$ in Eq. (\ref{eq:spinorder_Neel}) cannot be determined from pure symmetry analysis, and in general depend on $K$ and $\Gamma$.
A plot of the spin alignments in Eq. (\ref{eq:spinorder_Neel}) is shown in Fig. \ref{fig:picture_Neel}.

\subsection{Numerical results}

\begin{figure*}[htbp]
\includegraphics[width=18.0cm]{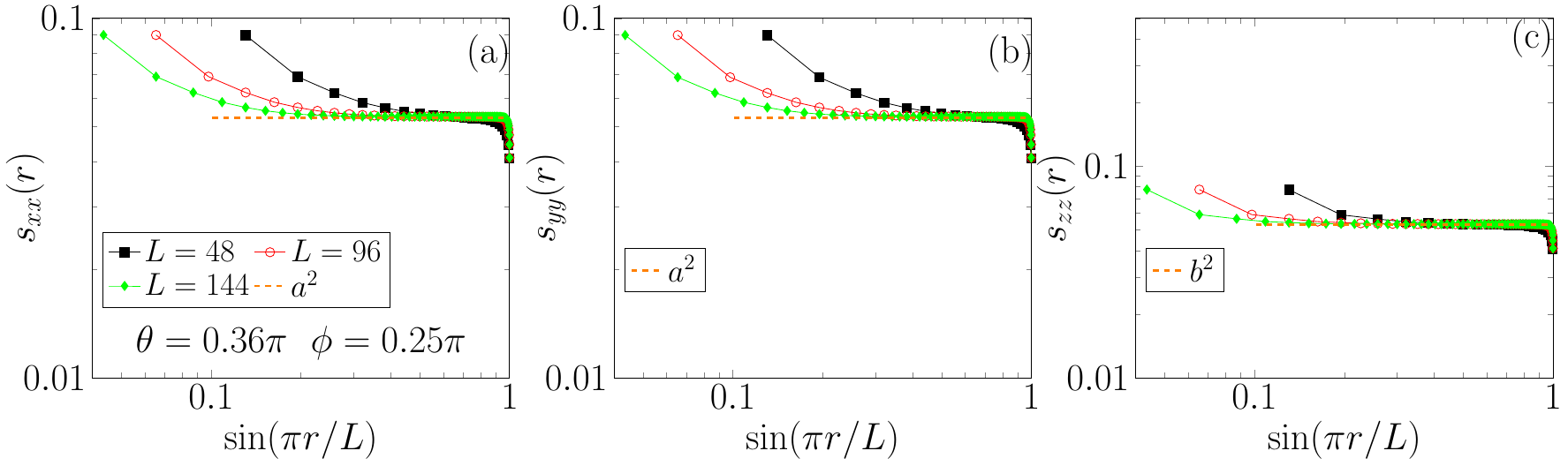}
\caption{Staggered components (a) $s_{xx}(r)$, (b) $s_{yy}(r)$, (c) $s_{zz}(r)$ of the corresponding correlation functions $\langle S_1^\alpha S_{1+r}^\alpha\rangle$ ($\alpha=x,y,z$) as functions of $\sin(\pi r/L)$.
DMRG numerics are performed on three system sizes $L=48,96,144$  with open boundary conditions at a representative point $(\theta=0.36\pi,\phi=0.25\pi)$ in the ``N\'eel" phase.
} 
\label{fig:Neel}
\end{figure*}

Numerics provide evidence for the ``N\'eel" phase to occupy the entire region above the equator in Fig. \ref{fig:phase} excluding the ``LL1" and the ``d-Spiral" phases.
Fig. \ref{fig:Neel} shows  $ s_{\alpha\alpha}(r)$ ($\alpha=x,y,z$) vs. $\sin(\pi r/L)$ extracted from the three-point formula in Eq. (\ref{eq:three_point})
at a representative point $(\theta=0.36\pi,\phi=0.25\pi)$ in the N\'eel phase,
where $(-)^rs_{\alpha\alpha}(r)$ is the staggered component of the spin-spin correlation function $\langle S_1^\alpha S_{1+r}^\alpha\rangle$.
The DMRG numerical results for three different system sizes $L=48,96,144$ are displayed with black, red and green curves, respectively.
As can be seen clearly from Fig. \ref{fig:Neel}, all the three $ s_{\alpha\alpha}(r)$ ($\alpha=x,y,z$)
approach constant values when $r\gg1$, indicating a N\'eel long range order.
In addition, the  patterns of the spin orderings are fully consistent with Eq. (\ref{eq:spinorder_Neel}), where $a^2,b^2$ can be determined from the asymptotic values of $s_{\alpha\alpha}(r)$ at large $r$.

\section{The ``$\text{d-Spiral}$" phase}
\label{sec:d_sp}

\begin{figure}[h]
\includegraphics[width=8.5cm]{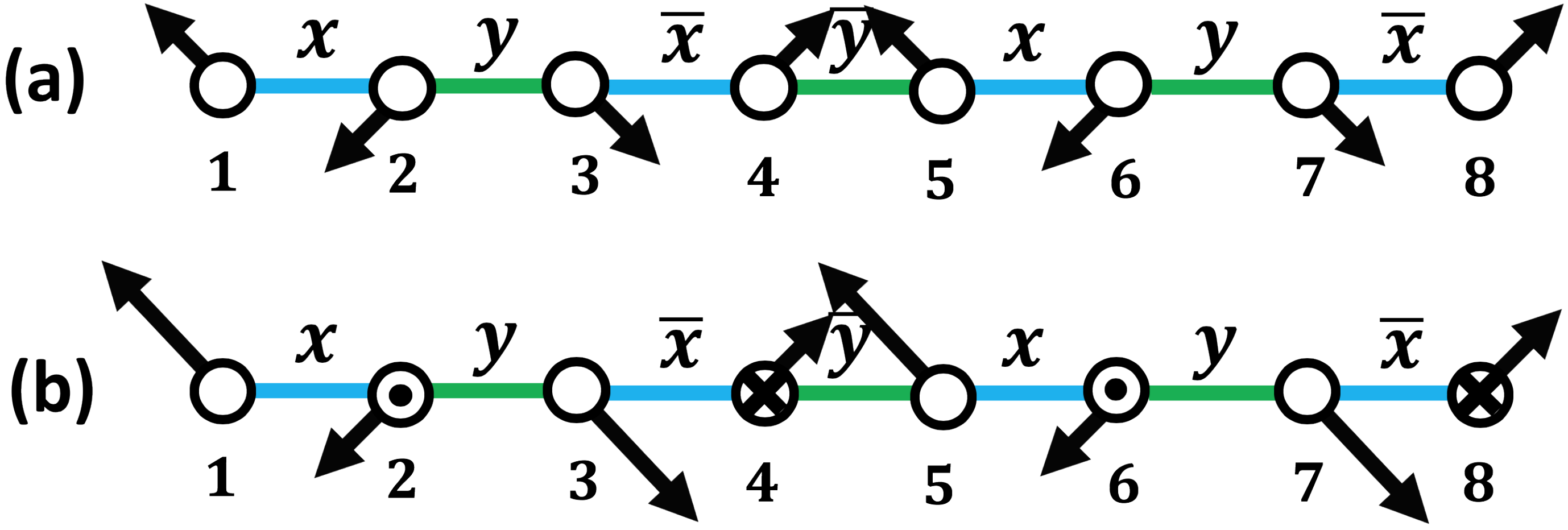}
\caption{Spin orientations within the original frame in the (a) spiral phase and
(b) ``d-Spiral" phase.
In both (a) and (b), the $z$-direction in spin space is chosen to be perpendicular to the plane,
and the $x$-direction is along the chain.
The hollow circles are used to denote an absence of a $z$-component of the spin orientations. 
In (b), the dot and cross are used to indicate that there are components along the $\hat{z}$ and $-\hat{z}$ directions, respectively
} 
\label{fig:d_spiral}
\end{figure}

In this section, we study the ``d-Spiral" phase in Fig. \ref{fig:phase}.
The spin orientations within the original frame are shown in Fig. \ref{fig:d_spiral}.
The symmetry breaking pattern in the ``d-Spiral" phase is inferred from the symmetry breaking in the spiral phase \cite {Agrapidis2018} of the Kitaev-Heisenberg chain  based on the assumption that there is no phase transition between the ``d-Spiral" and the spiral phases.
The spin orientations in the ``d-Spiral" phase are predicted to exhibit a ``distorted-spiral" pattern, which is supported by our DMRG numerics. 
Throughout this section, we work in the four-sublattice rotated frame defined in Eq. (\ref{eq:4rotation}) unless otherwise stated.

\subsection{Symmetry breaking for $\Gamma=0$}
\label{sec:sym_group_orig}

\subsubsection{Symmetry group for $\Gamma=0$}
\label{sec:sym_group_gamma_0}

When $\Gamma=0$, the system reduces to a Kitaev-Heisenberg chain even in the four-sublattice rotated frame by virtue of the duality property as mentioned in Eq. (\ref{eq:duality_KH}).
Hence, in this case, there is no difference between the original and four-lattice rotated frames if the symmetry group structure is concerned.

The symmetry transformations of the Kitaev-Heisenberg chain are
\begin{eqnarray}
1.&T &:  (S_i^x,S_i^y,S_i^z)\rightarrow (-S_{i}^x,-S_{i}^y,-S_{i}^z)\nn\\
2.& T_{2a}&:  (S_i^x,S_i^y,S_i^z)\rightarrow (S_{i+2}^x,S_{i+2}^y,S_{i+2}^z)\nn\\
3.&T_a I&: (S_i^x,S_i^y,S_i^z)\rightarrow (S_{-i+1}^x,S_{-i+1}^y,S_{-i+1}^z)\nn\\
4.&R(\hat{y},\pi)&: (S_i^x,S_i^y,S_i^z)\rightarrow (-S_{i}^x,S_{i}^y,-S_{i}^z)\nn\\
5.&R(\hat{z},-\frac{\pi}{2})T_a&: (S_i^x,S_i^y,S_i^z)\rightarrow (-S_{i+1}^y,S_{i+1}^x,S_{i+1}^z),\nn\\
\label{eq:Sym_Neel}
\end{eqnarray}
in which  the inversion center for $I$ is taken to be the site $0$.
We note that all the other symmetry transformations can be generated by the operations in Eq. (\ref{eq:Sym_Neel}) as discussed in 
Supplementary Materials \cite{SM}.
Thus, the symmetry group of the Kitaev-Heisenberg chain is
\begin{flalign}
G_0&=\mathopen{<}
T,T_{2a},T_aI,R(\hat{y},\pi),R(\hat{z},-\frac{\pi}{2})T_a
\mathclose{>}.
\label{eq:group_G0}
\end{flalign}

Next we briefly describe the group structure of $G_0$ (for a detailed proof, see Supplementary Materials \cite{SM}).
Since $T_{4a}=[R(\hat{z},-\frac{\pi}{2})T_a]^4$ belongs to $G_0$,
it is legitimate to consider the quotient group $G_0/\mathopen{<}T_{4a}\mathclose{>}$,
whose group structure is (see Supplementary Materials \cite{SM})
\begin{eqnarray}
G_0/\mathopen{<}T_{4a}\mathclose{>}=[(\mathbb{Z}_2\times \mathbb{Z}_2) \ltimes D_{4d}],
\label{eq:G0_d_sp}
\end{eqnarray}
in which from left to right, $\mathbb{Z}_2=\mathopen{<}T_aI\mathclose{>}$,
$\mathbb{Z}_2=\mathopen{<} T_{2a} \mathclose{>} \mod T_{4a} $, and 
\begin{eqnarray}
D_{4d}=\mathopen{<}T,R(\hat{z},-\frac{\pi}{2})T_a,R(\hat{y},\pi)T_a I
\mathclose{>} /\mathopen{<} T_{4a}\mathclose{>}.
\label{eq:D4d_expression}
\end{eqnarray}
This shows that 
\begin{eqnarray}
G_0=[(\mathbb{Z}_2\times \mathbb{Z}_2) \ltimes D_{4d}]\ltimes 4\mathbb{Z},
\label{eq:G0_d_sp}
\end{eqnarray}
in which $4\mathbb{Z}=\mathopen{<}T_{4a}\mathclose{>}$.

\begin{figure}[h]
\includegraphics[width=6.5cm]{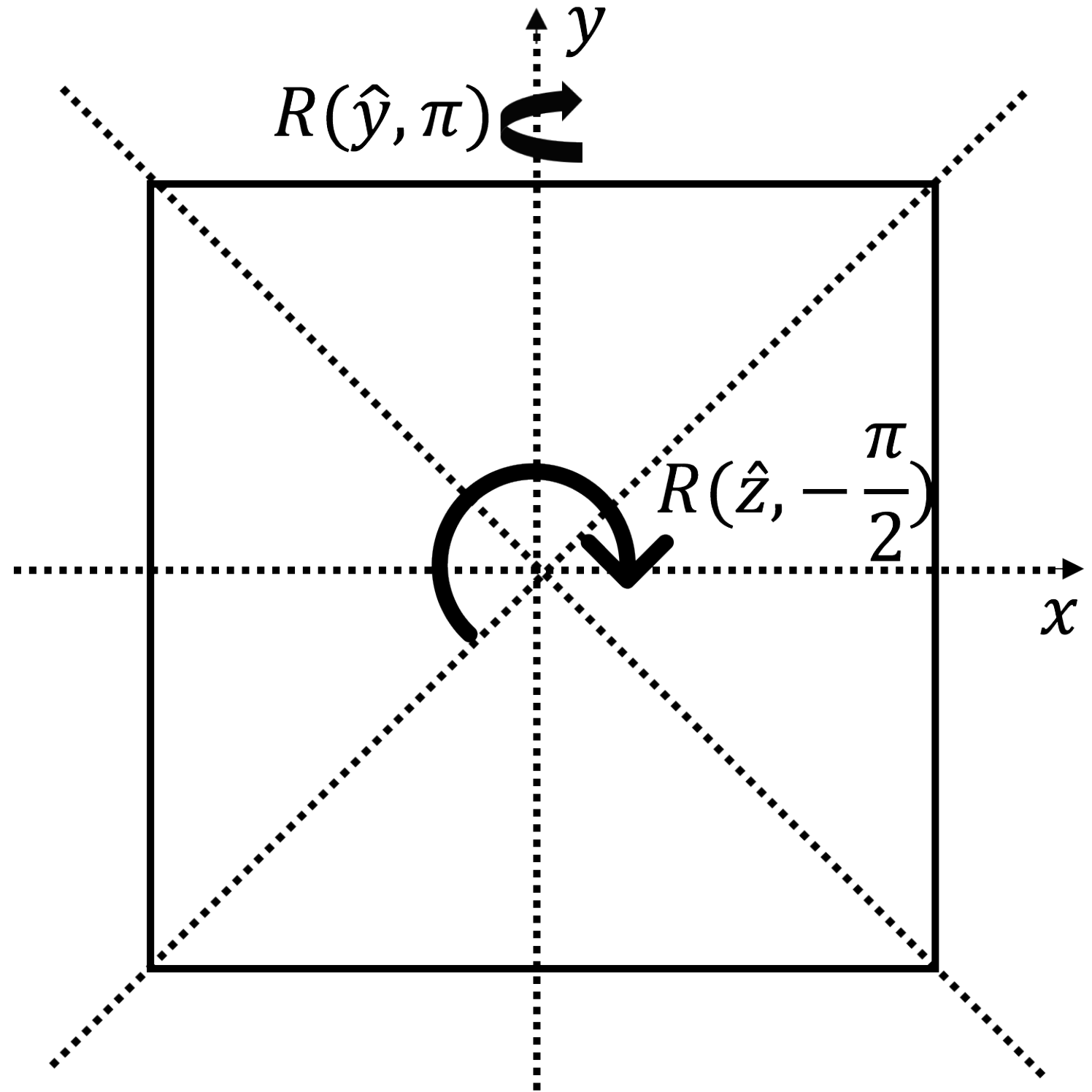}
\caption{$R(\hat{y},\pi)$ and $R(\hat{z},-\pi/2)$ as symmetry operations of a square.
Within the $xy$-plane, the effect of the  rotation $R(\hat{y},\pi)$ on the square is the same as that of a reflection with respect to the $x=0$ plane.
In the above figure, the $z$-direction is perpendicular to the plane.
} 
\label{fig:D4d}
\end{figure}

We make some comments on the geometrical meaning for the origin of $D_{4d}$.
Notice that $D_{4d}=D_4 \times \mathbb{Z}_2$ in which $D_4$ is the symmetry group of a square containing four reflections and four rotations, and $\mathbb{Z}_2=\mathopen{<}i\mathclose{>}$ where $i$ is the inversion operation.
In our case, time reversal acts as ``inversion" in spin space, since it flips the signs of the spin operators. 
Neglecting the spatial components in $R(\hat{z},-\frac{\pi}{2})T_a$  and $R(\hat{y},\pi)T_a I$, the spin operations $R(\hat{z},-\frac{\pi}{2})$ and $R(\hat{y},\pi)$ generate the symmetry group of a square as shown in 
Fig. \ref{fig:D4d}.
As proved in Supplementary Materials \cite{SM},
$\mathopen{<}T,R(\hat{z},-\frac{\pi}{2})T_a,R(\hat{y},\pi)T_a I
\mathclose{>}/\mathopen{<}T_{4a}\mathclose{>}$ is isomorphic to $D_{4d}$ even when the spatial components in the operations are included.

\subsubsection{Symmetry breaking pattern for $\Gamma=0$}

In this subsection, we  discuss  the spin ordering and the symmetry breaking pattern for the $\Gamma=0$ case, i.e., the Kitaev-Heisenberg chain,
and later extend the analysis to $\Gamma \neq 0$ in Sec. \ref{sec:Gamma_neq_0}.
The spin ordering for $\theta\in [-K,FM3]$ located on the circular boundary in Fig. \ref{fig:phase}
has been worked out in Ref. \onlinecite{Agrapidis2018} which exhibits a spiral pattern with a four-site periodicity.
On the other hand, this region of parameter can be mapped to the FM-xy phase in Fig. \ref{fig:phase} by the four-sublattice rotation.
Therefore, the system is FM aligned along $\pm(110)$-direction in the four-sublattice rotated frame. 
In what follows, we determine the symmetry breaking pattern corresponding to this FM spin order within the four-sublattice rotated frame.

Let's consider the little group of the FM-xy order,
in which the spin alignments are $\vec{S}_i \equiv (f,f,0)^T$.
Apparently, the spin alignments are invariant under both $T_aI$ and $T_{2a}$.
Furthermore, although $R(\hat{z},-\frac{\pi}{2})T_a$ and $R(\hat{y},\pi)T_aI$ do not leave the FM-xy order invariant, the spin orientations  are invariant under the combinations
\begin{eqnarray}
c&=&(R(\hat{z},-\frac{\pi}{2})T_a)^2\cdot T,\nn\\
d&=&R(\hat{z},-\frac{\pi}{2})T_a\cdot R(\hat{y},\pi)T_aI.
\end{eqnarray}
It is easy to verify that  $c$ and $d$ satisfy $c^2=d^2=(cd)^2=e$ modulo $T_{4a}$.
Comparing with the generator-relation representation for the group $D_n$ \cite{Coxeter1965}
\begin{eqnarray}
D_n=\mathopen{<} \alpha,\beta| \alpha^n=\beta^2=(\alpha\beta)^2=e \mathclose{>},
\label{eq:generator_Dn}
\end{eqnarray}
we see that relations in Eq. (\ref{eq:generator_Dn}) are are satisfied for $c,d$ with $n=2$,
hence the group generated by $c,d$ (modulo $T_{4a}$) is isomorphic to $D_2$.
Therefore, the little group of the FM-xy order is
\begin{eqnarray}
[(\mathbb{Z}_2\times \mathbb{Z}_2)\ltimes D_2]\ltimes 4\mathbb{Z}.
\end{eqnarray}
Combining with the expression for $G_0$ in Eq. (\ref{eq:G0_d_sp}), we conclude that the symmetry breaking pattern is
\begin{eqnarray}
(\mathbb{Z}_2\times \mathbb{Z}_2)\ltimes D_{4d}\rightarrow (\mathbb{Z}_2\times \mathbb{Z}_2)\ltimes D_2,
\label{eq:sym_breaking_d_sp_0}
\end{eqnarray}
in which $T_{4a}$ has been dropped for simplicity. 
In particular, since $|D_{4d}/D_2|=4$,
the ground states are four-fold degenerate in which the spins align  ferromagnetically along $(\pm 1,\pm 1,0)$-directions.
On the other hand, by transforming back to the original frame, it can be seen that 
 the spin orientations  show a spiral pattern as plotted in Fig. \ref{fig:d_spiral} (a).

\subsection{Symmetry breaking for $\Gamma\neq0$}
\label{sec:sym_4rotation}

\subsubsection{Symmetry group for $\Gamma\neq 0$}

When $\Gamma$ is nonzero, the symmetry operations of $H^{\prime\prime}$ in the four-sublattice rotated frame in Eq. (\ref{eq:4rotated}) can be verified to be the following,
\begin{eqnarray}
1.&T &:  (S_i^x,S_i^y,S_i^z)\rightarrow (-S_{i}^x,-S_{i}^y,-S_{i}^z)\nn\\
2.& R(\hat{y},\pi)T_aI&:  (S_i^x,S_i^y,S_i^z)\rightarrow (-S_{5-i}^x,S_{5-i}^y,-S_{5-i}^z)\nn\\
3.&R(\hat{z},-\frac{\pi}{2})T_a&: (S_i^x,S_i^y,S_i^z)\rightarrow (-S_{i+1}^y,S_{i+1}^x,S_{i+1}^z),\nn\\
\label{eq:sym_LL3}
\end{eqnarray}
 in which the inversion center is taken to be site $2$ in Fig. \ref{fig:bonds} (c).
Therefore, the symmetry group is 
\begin{eqnarray}
G_3=\mathopen{<} T,R(\hat{y},\pi)T_a I,R(\hat{z},-\frac{\pi}{2})T_a \mathclose{>}.
\label{eq:generate_G3}
\end{eqnarray}
As discussed in Sec. \ref{sec:sym_group_gamma_0}, the group structure of $G_3$ is 
\begin{eqnarray}
G_3\simeq D_{4d}\ltimes 4\mathbb{Z},
\label{eq:G3_structure}
\end{eqnarray}
in which $D_{4d}$ is given by Eq. (\ref{eq:D4d_expression}), and $4\mathbb{Z}=\mathopen{<} T_{4a} \mathclose{>}$.
Comparing Eq. (\ref{eq:G3_structure}) with Eq. (\ref{eq:G0_d_sp}), 
$G_3$ is explicitly a subgroup of $G_0$,
and there are two more generators in $G_0$ than in $G_3$, i.e., $T_a I$ and $T_{2a}$.
 
\begin{figure*}[htbp]
\includegraphics[width=18cm]{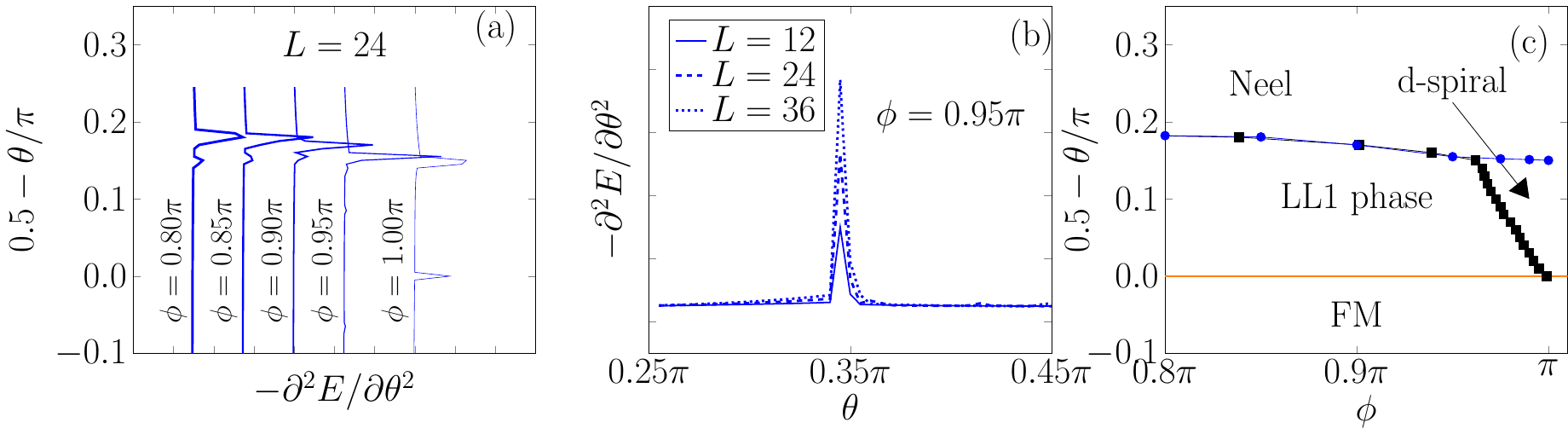}
\caption{(a) Vertical scans of $\partial^2 E/\partial\theta^2$ as a function of $\theta$ for fixed values of $\phi$, 
(b) finite size scaling of $\partial^2 E/\partial\theta^2$ as a function of $\theta$ for $\phi=0.95\pi$,
(c) phase boundary of the ``LL1" phase.
In (a), the curves for different angles are shifted horizontally.  
ED calculations are performed on periodic systems of $L=24$ sites in (a) and $L=12,24$ sites in (b).
DMRG numerics are performed for $L=36$ sites in (b).
} 
\label{fig:d_spiral_energy}
\end{figure*}
 
\begin{figure*}[htbp]
\includegraphics[width=18cm]{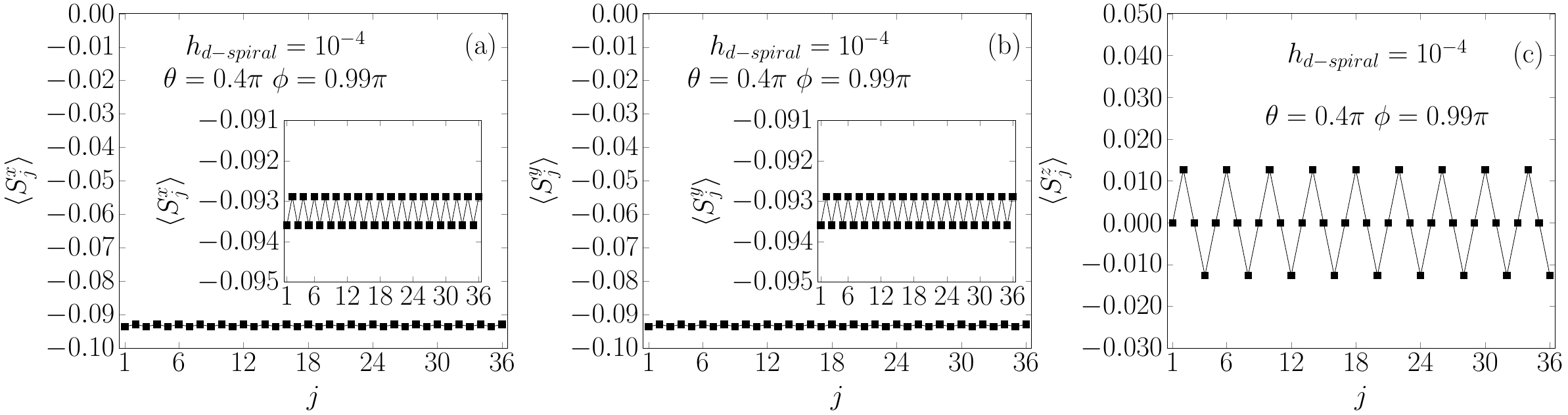}
\caption{Measured expectation values of (a) $\langle S_j^x\rangle$, (b) $\langle S_j^y\rangle$ and (c) $\langle S_j^z\rangle$ as functions of $j$ under a small field $h=10^{-4}$ along the $(110)$-direction.
The inset figures in (a) and (b) are zoom-in of the spin expectation values.
DMRG numerics are performed on a system of $L=36$ sites with periodic boundary conditions.
} 
\label{fig:d_spiral_num}
\end{figure*}

\subsubsection{Symmetry breaking pattern for $\Gamma\neq 0$}
\label{sec:Gamma_neq_0}

In this subsection, we discuss the symmetry breaking and the spin ordering for the $\Gamma\neq 0$ case.
Let's assume that $T_{4a}$ is not broken, and we'll consider the quotient group $G_3^\prime=G_3/\mathopen{<}T_{4a}\mathclose{>}$ in what follows.
If there is no phase transition when $\Gamma$ is tuned from nonzero to zero, 
then the symmetry breaking for a nonzero $\Gamma$ has to be
\begin{eqnarray}
 D_{4d}\rightarrow D_2,
\end{eqnarray}
as a natural extension of the pattern in Eq. (\ref{eq:sym_breaking_d_sp_0}) for the $\Gamma=0$ case.

Assuming the symmetry breaking to be $D_{4d}\rightarrow D_2$, next we solve the most general spin ordering which is invariant under $D_2$.
Requiring the invariance of the spin ordering under the two generators
$(R(\hat{z},-\frac{\pi}{2})T_a)^2\cdot T$ and $R(\hat{z},-\frac{\pi}{2})T_a\cdot R(\hat{y},\pi)T_aI$,
the most general $D_2$-invariant spin configuration can be determined as
\begin{eqnarray}
\vec{S}_1&=&(f,f,0)^T,\nn\\
\vec{S}_2&=&(k,k,h)^T,\nn\\
\vec{S}_3&=&(f,f,0)^T,\nn\\
\vec{S}_4&=&(k,k,-h)^T.
\label{eq:Spin_align_d_sp_4rot}
\end{eqnarray}
Rotating back to the original frame, the spin ordering is
\begin{eqnarray}
\vec{S}_1^{(0)}&=&(-f,f,0)^T,\nn\\
\vec{S}_2^{(0)}&=&(-k,-k,h)^T,\nn\\
\vec{S}_3^{(0)}&=&(f,-f,0)^T,\nn\\
\vec{S}_4^{(0)}&=&(k,k,-h)^T.
\label{eq:spin_align_d_sp_orig}
\end{eqnarray}
Eq. (\ref{eq:spin_align_d_sp_orig}) exhibits a ``distorted-spiral' pattern,
which is the origin of the name ``d-Spiral" for the phase 
where ``d" is ``distorted" for short.
A schematic plot of the spin alignments in Eq. (\ref{eq:spin_align_d_sp_orig}) 
is shown in Fig. \ref{fig:d_spiral} (b). 
Taking $f=k$, $h=0$, the spin orientations in Eq. (\ref{eq:spin_align_d_sp_orig}) reduce back to the $\Gamma=0$ case in Fig. \ref{fig:d_spiral} (a). 
We note that the other three degenerate ground states can be obtained by applying the 
operations in the equivalent classes in $D_{4d}/D_2$.

\subsection{Numerical results}
\label{sec:numerics_dspiral}

We first determine the range of the ``d-Spiral" phase by studying the ground state energy $E(\theta,\phi)$ as a function of $\theta,\phi$.
Fig. \ref{fig:d_spiral_energy} (a) shows the second order derivative $\partial^2 E(\theta,\phi)/\partial \theta^2$ in vertical scans by varying $\theta$ for several fixed values of $\phi$,
where ED numerics are performed for a periodic system of $L=24$ sites.
The sequence of the uppermost divergent peaks in Fig. \ref{fig:d_spiral_energy} (a) correspond to first order phase transitions which determine the upper boundary of the ``d-Spiral" phase represented by the solid blue circles connected with the blue line in Fig. \ref{fig:d_spiral_energy} (c).
To further confirm the nature of the peaks as first order phase transitions, we have studied the finite size scaling behaviors of $\partial^2 E(\theta,\phi)/\partial \theta^2$.
Fig. \ref{fig:d_spiral_energy} (b) displays the numerical results for $\partial^2 E(\theta,\phi)/\partial \theta^2$ at $\phi=0.95\pi$ with three different system sizes $L=12,24,36$.
As is clear from Fig. \ref{fig:d_spiral_energy} (b), $\partial^2 E(\theta,\phi)/\partial \theta^2$ scales linearly with $L$, indicating a first order phase transition.
On the other hand, we find no divergence in $\partial^2 E(\theta,\phi)/\partial \theta^2$ when the lower boundary of the ``d-Spiral" phase is traversed.
Recall that there is a divergence in $\partial^2 E(\theta,\phi)/\partial \phi^2$ as a function of $\phi$ as discussed in Fig. \ref{fig:LL1} (c), which determines the lower boundary between the ``d-Spiral" phase and the ``LL1" phase
shown by the solid black squares in Fig. \ref{fig:d_spiral_energy} (c) within the region $\phi\in[0.96\pi,\pi]$.

Next we discuss the numerical evidence for the spin ordering in Eq. (\ref{eq:Spin_align_d_sp_4rot}) within the four-sublattice rotated frame.
Before going to that, 
we mention a subtlety in numerical calculations.
The four symmetry breaking ground states only become exactly degenerate in the thermodynamic limit. 
The ground state of a finite size system can be some arbitrary linear combination of the four states,
and the coefficients depend on the system size and numerical details.
Because of this, random cancellations occur if the expectation values of the spin operators $\langle\vec{S}_i\rangle$ are directly computed.
To circumvent this problem, we measure $\langle\vec{S}_i\rangle$ in the presence of a small field along $(110)$-direction, which is able to polarize the system to reside in one of the four nearly degenerate states so that the spins orient according to the pattern in Eq. (\ref{eq:Spin_align_d_sp_4rot}).
The value of this field should satisfy $\Delta E\ll  |h|L \ll E_g$, where $\Delta E$ is the finite size energy splitting among the four states, $E_g$ is the gap between the ground state multiplet and the excitations, and $|h|$ is the magnitude of the applied field.
Such choice of field  leads to a degenerate perturbation within the ground state quartet, but no mixing is induced between the ground state subspace and the excitations.

Fig. \ref{fig:d_spiral_num} shows the measured values of $S_j^\alpha$ ($\alpha=x,y,z$) at a representative point $(\theta=0.4\pi,\phi=0.99\pi)$ in the ``d-Spiral" phase.
The data are obtained by performing DMRG numerics on a periodic system of $L=36$ sites under an $h=10^{-4}$ field along the $(110)$-direction.
As can be seen from Fig. \ref{fig:d_spiral_num}, the pattern of the spin orientations are fully consistent with Eq. (\ref{eq:Spin_align_d_sp_4rot}), with extracted values $f,k,h$ as $f\simeq -0.0936$, $k\simeq -0.0928879$, and $h\simeq 0.0127$.
This provides evidence for the existence of the ``d-Spiral" phase.
We note that numerics have also been done for several other points in the ``d-Spiral" phase which are all consistent with Eq. (\ref{eq:Spin_align_d_sp_4rot}). 

\section{The ``LL3" phase}
\label{sec:LL3_phase}

\begin{figure}[h]
\includegraphics[width=7.0cm]{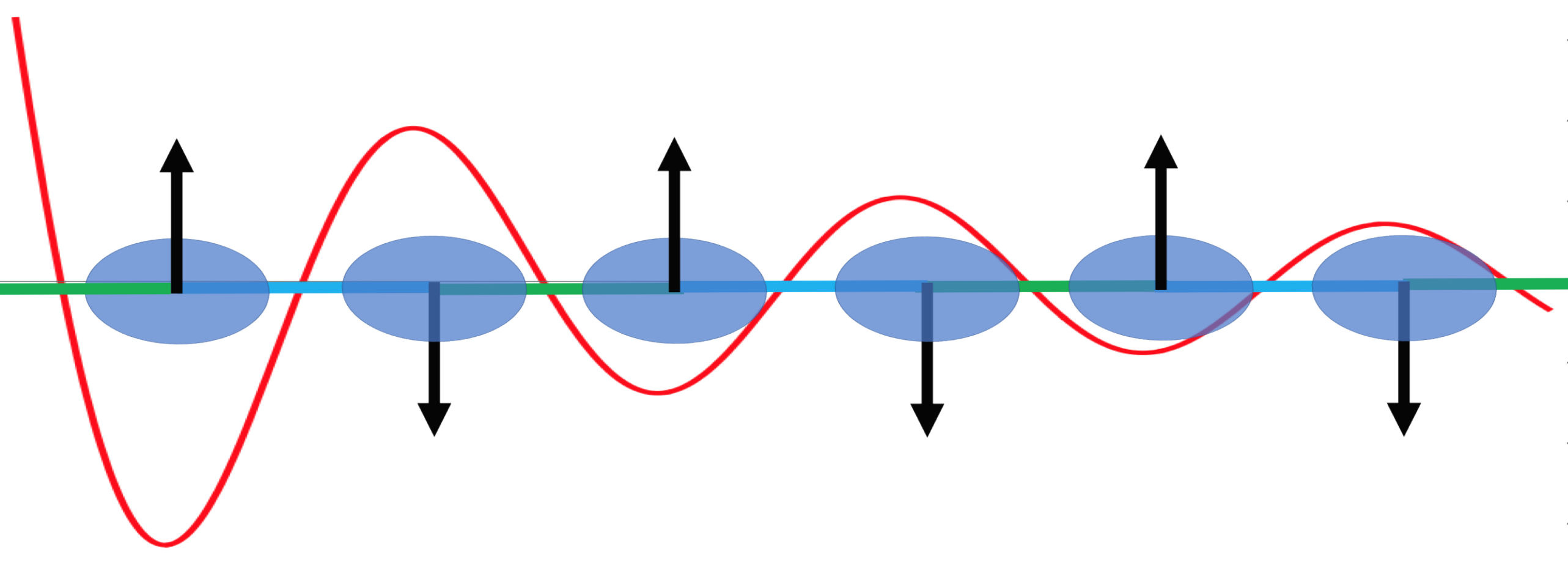}
\caption{Site-dependent quantization axes for the longitudinal fluctuations in the ``LL3" phase within the original frame. 
The black arrows denote the directions of the quantization axes, and the blue ellipses represent the transverse fluctuations.
The red line represents the AFM quasi-long range order for the longitudinal and transverse fluctuations defined in terms of the four-sublattice rotated frame.
The $z$-direction in spin space is chosen to be pointing upwards, and the $y$-direction is along the chain to the right.
} \label{fig:LL3_quantization}
\end{figure}

In this section, we study the ``LL3" phase in Fig. \ref{fig:phase}, and demonstrate that the low energy physics in this phase is described by the Luttinger liquid theory.
The system exhibits a site-dependent quantization axis for the longitudinal fluctuations within the original frame as shown in Fig. \ref{fig:LL3_quantization}.
Recall that as discussed in Sec. \ref{sec:4rotation},
the four-sublattice rotation reveals a hidden SU(2) symmetric AFM point, i.e.,  the AFM3 point in Fig. \ref{fig:phase}.
We will study the region close to the AFM3 point
using again a combination of RG and symmetry analysis.
Numerics provide evidence for Luttinger liquid behaviors in the entire region of the ``LL3" phase as shown in Fig. \ref{fig:LL_K}.
To facilitate analysis, we work in the four-sublattice rotated frame defined by Eq. (\ref{eq:4rotation}) throughout this section unless otherwise stated.

\subsection{Perturbative analysis}
\label{sec:sym_LL3}

We first remark that in the absence of $\Gamma$,
there is a duality transformation for the Kitaev-Heisenberg chain
as discussed in Sec. \ref{sec:4rotation}.
This establishes the equivalence between the arcs $[K,\text{AFM3}]$
and $[\text{AFM1},K]$ on the circular boundary of the half sphere in Fig. \ref{fig:phase}.
Since the latter (i.e., the ``LL2" phase) has been shown to be described by Luttinger liquid theory
in Sec. \ref{eq:RG_Neel},
we conclude that $[K,\text{AFM3}]$ is also in a Luttinger liquid phase \cite{Agrapidis2018}.

\begin{figure}[h]
\includegraphics[width=7.0cm]{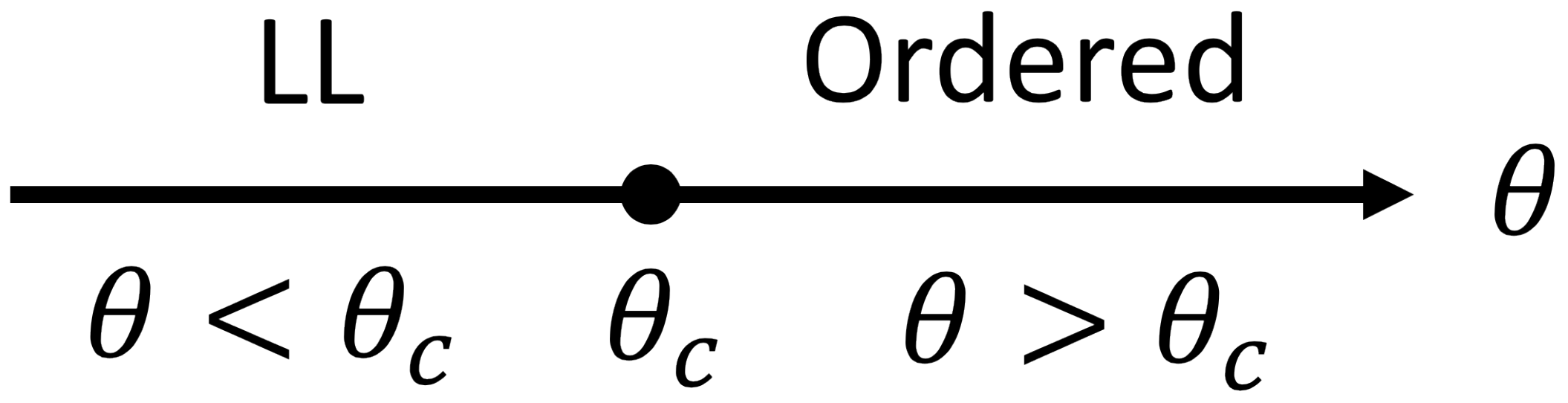}
\caption{Phase diagram in the vicinity of the AFM3 point for fixed values of $\phi$ but changing $\theta$. 
The critical value $\theta_c$ depends on $\phi$.
} \label{fig:phase_RG_LL3}
\end{figure}

Next we include a nonzero $\Gamma$.
The Kitaev and Gamma terms will be treated as perturbations to the 
SU(2)$_1$ WZW theory which describes the low energy physics of the AFM3 point in the four-sublattice rotated frame.
We perform a first order projection of the perturbing Hamiltonian to obtain the low energy field theory of the system.
 According to Eq. (\ref{eq:4rotated}),
 the perturbing Hamiltonian $\Delta H^{\prime\prime}$ is 
 \begin{flalign}
\Delta H^{\prime\prime}&=\sum_{<ij>\in \gamma\,\text{bond}}\big[ (K+2J)S_i^\gamma S_j^\gamma\nn\\
&~~~+\epsilon(\gamma) \Gamma (S_i^\alpha S_j^\beta+S_i^\beta S_j^\alpha)\big].
\label{eq:4rotated_perturb}
\end{flalign} 
Notice that all of the operators $\vec{J}_L,\vec{J}_R,g$ are invariant under translation by two sites,
whereas the $\Gamma$ term in Eq. (\ref{eq:4rotated_perturb}) is staggered every two sites.
Therefore, the $\Gamma$ term vanishes after  projecting to the the low energy degrees of freedom and thus has no effect.
This means that at least up to first order projection,
the low energy Hamiltonian of the  $KH\Gamma$ chain is of an XXZ type, having the same coupling constants as an Kitaev-Heisenberg chain.
Based on this analysis, 
the phase diagram for fixed values of $\phi$ but changing $\theta$ in the vicinity of the AFM3 point can be derived as shown in Fig. \ref{fig:phase_RG_LL3}.
When $\theta>\theta_c$, the system flows to a strong coupling limit where an order develops,
whereas when $\theta<\theta_c$, the system remains gapless.
Up to first order approximation, $\theta_c$ is equal to $\pi-\arctan(2) $ independent of $\phi$. 

However, a concern at this point is whether higher order effects will spoil the XXZ-type low energy Hamiltonian,
and in particular, whether the gapless Luttinger liquid phase in Fig. \ref{fig:phase_RG_LL3} is stable with respect to high order perturbations.
To resolve this  question, we have performed a careful symmetry analysis.
The symmetry group of the system in the four-sublattice rotated frame has been demonstrated to be $G_3\cong D_{4d} \ltimes 4\mathbb{Z}$ in Sec. \ref{sec:sym_4rotation}.
We are able to show that up to relevant and marginal couplings, the symmetry allowed terms other than the XXZ-type low energy Hamiltonian only include the chiral term $J_L^z-J_R^z$.
Since this chiral term can be eliminated by a chiral rotation \cite{Garate2010,Gangadharaiah2008,Schnyder2008},
the low energy physics remains to be of the XXZ type.
The detailed symmetry analysis is included in Supplementary Materials \cite{SM}.
This justifies the validity of the first order perturbation analysis and the phase diagram in Fig. \ref{fig:phase_RG_LL3}.
However, we emphasize that the value of $\theta_c$ can be shifted due to high order effects,
and the actual value of $\theta_c$ is a function of $\phi$.

In the ``LL3" phase, based on the above analysis, the quantization axis for the longitudinal fluctuations in the four-sublattice rotated frame is along the $z$-direction.
Notice that by performing the inverse of the four-sublattice rotation defined in  Eq. (\ref{eq:4rotation}), the $z$-direction becomes staggered in the original frame, hence the quantization axes in the ``LL3" phase  are drawn in a staggered manner in Fig. \ref{fig:LL3_quantization}.
On the other hand, we emphasize that  the quantization axes in Fig. \ref{fig:LL3_quantization} are not precise, since they can acquire site-dependent distortions due to the renormalization effects of the spin operators along the RG flow similar to the situation in Sec. \ref{sec:RG_KG_H}.

We also note that  the ordered phase in Fig. \ref{fig:phase_RG_LL3} is an N\'eel order along $z$-direction,
which is in the four-sublattice rotation  frame.
As can be checked by performing the  inverse of the four-sublattice rotation,
the Neel-$z$ order becomes an FM-$z$ order in the original frame.
But this is not accurate due to the possible distortions of the quantization axes.
To identify the precise direction of the ordering, we first figure out the unbroken symmetries of the N\'eel-$z$ order in the four-sublattice rotated frame, and then solve the most general form of the spin orientations which are invariant under the unbroken symmetry group.

The full symmetry group of the model in the four-sublattice rotated frame is given by $G_3$ in Eq. (\ref{eq:generate_G3}).
As can be easily checked, the unbroken symmetry group $H_3$ corresponding to the Neel-$z$ order is 
\begin{eqnarray}
H_3=\mathopen{<} R(\hat{y},\pi)T_aI,TR(\hat{z},-\frac{\pi}{2})T_a \mathclose{>}.
\label{eq:H3_generators}
\end{eqnarray}
Since $T_{4a}=(TR(\hat{z},-\frac{\pi}{2})T_a)^4$ is an unbroken symmetry operation, we will consider spins in a unit cell in what follows where all site indices should be understood as modulo four. 
The most general spin alignments which are invariant under the group $H_3$ in Eq. (\ref{eq:H3_generators}) can then be solved as
\begin{eqnarray}
\vec{S}_1&=& (-a,a,-b)^T,\nn\\
\vec{S}_2&=& (-a,-a,b)^T,\nn\\
\vec{S}_3&=& (a,-a,-b)^T,\nn\\
\vec{S}_4&=& (a,a,b)^T,
\end{eqnarray}
in which $a,b$ are two parameters depending on the values of $K,\Gamma,J$.
Performing the inverse of the four-sublattice rotation defined in Eq. (\ref{eq:4rotation}), the spin ordering in the original frame can be determined as
\begin{eqnarray}
\vec{S}_i^{(0)}\equiv (a,a,b)^T,
\end{eqnarray}
which is an FM order along the $(a,a,b)$-direction, providing another understanding to the ``FM" phase in Fig. \ref{fig:phase}.

\subsection{Numerical results}

The Luttinger parameter $\mathcal{K}$ can be extracted numerically using the same method as Sec. \ref{sec:LL1_K_value}.
DMRG numerics are performed to calculate the energy density on an open system of $L=96$ sites.
We have studied the Luttinger parameter in the entire ``LL3" phase and the results are shown in Fig. \ref{fig:LL_K}.
On the right hand side of the phase boundary between the ``LL3" and ``$D_3$-breaking" phases, no Luttinger parameter can be extracted, and in fact, the transition line in Fig. \ref{fig:phase} between these two phases is determined in this way. 
On the other hand, as discussed in Sec. \ref{sec:sym_LL3}, the phase transition between the ``LL3" and ``FM" phases is second order.
Since a gap $E_g\sim e^{-\text{const}/\sqrt{|\theta-\theta_c|}}$ opens exponentially slowly in the ``FM" phase close to the transition line \cite{Affleck1988}, 
the Luttinger liquid behaviors percolate into the ``FM" phase in a finite size system,
which smears the phase transition as can be seen from Fig. \ref{fig:LL_K}.


\section{The ``LL4" and ``$D_3$-breaking I, II" phases}
\label{sec:LL_classical}

\begin{figure}[h]
\includegraphics[width=8.0cm]{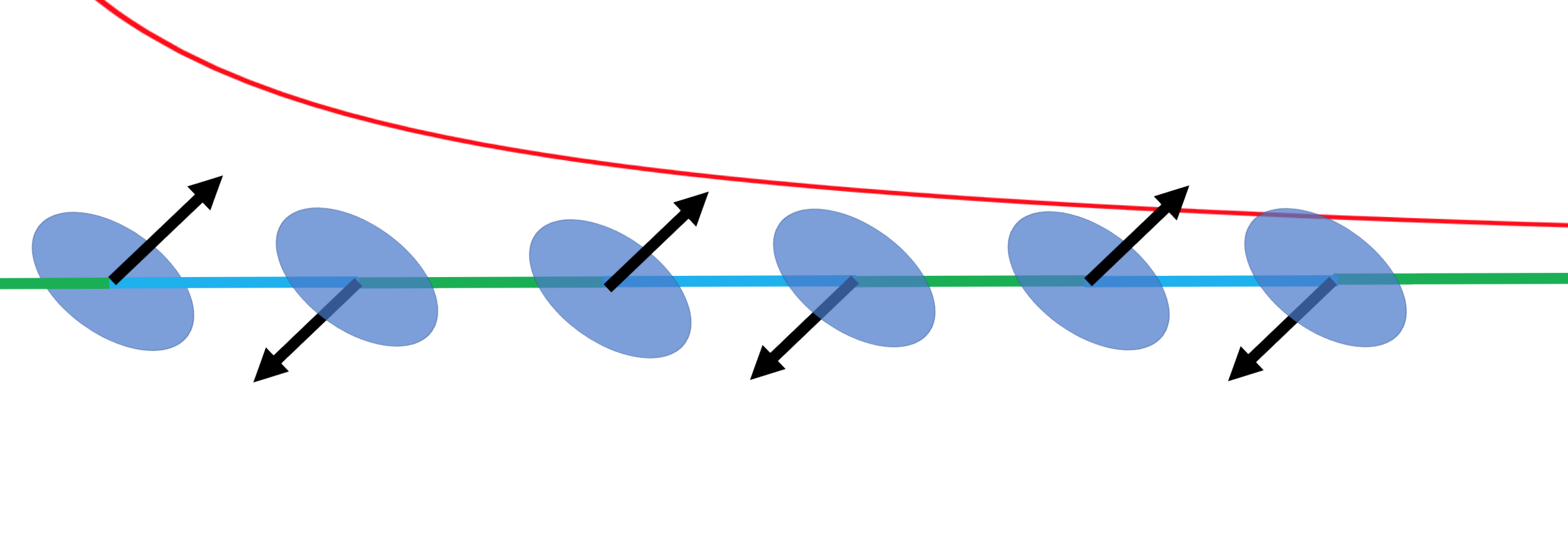}
\caption{Site-dependent quantization axes for the longitudinal fluctuations in the ``LL4" phase within the original frame. 
The black arrows denote the directions of the quantization axes, and the blue ellipses represent the transverse fluctuations.
The red line represents the FM quasi-long range order for the longitudinal and transverse fluctuations defined in terms of the six-sublattice rotated frame.
The $z$-direction in spin space is chosen to be pointing upwards, and the $y$-direction is along the chain to the right.
} 
\label{fig:LL4_quantization}
\end{figure}

\begin{figure}[h]
\includegraphics[width=8.5cm]{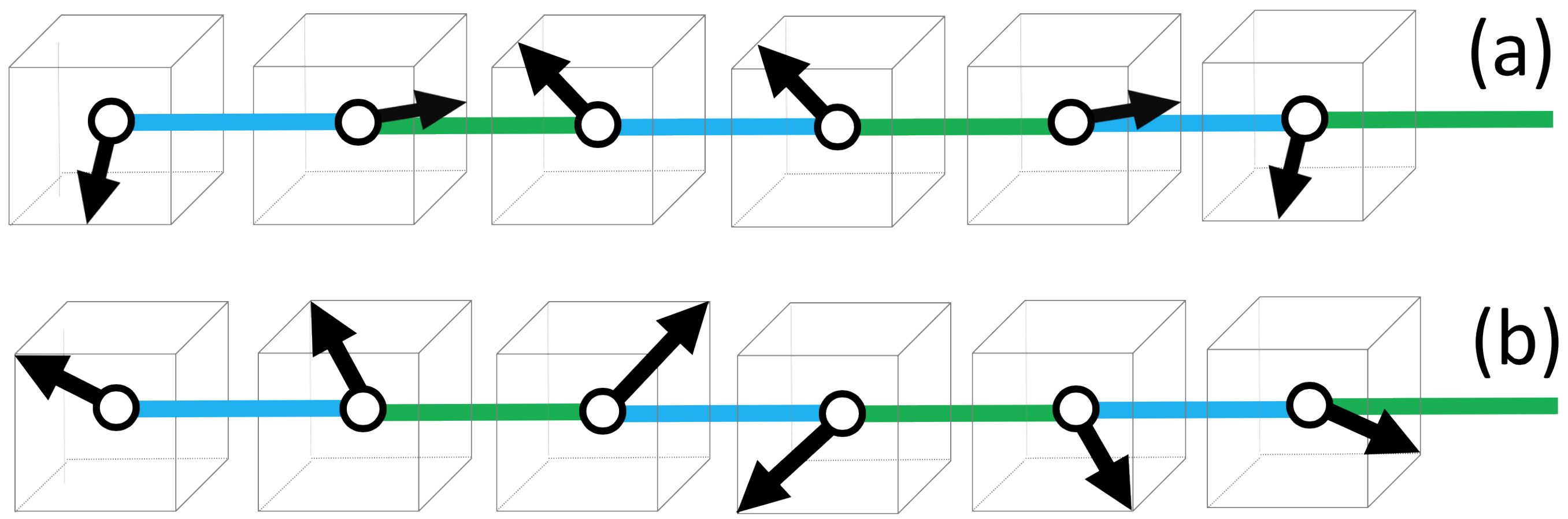}
\caption{Spin alignments in (a) ``$D_3$ breaking I" and (b) ``$D_3$ breaking II" phases within the original frame.
The $z$-direction in spin space is chosen to be pointing upwards, and the $y$-direction is along the chain to the right.
} 
\label{fig:spins_D3}
\end{figure}

In this section, we study the ``LL4" and the ``$D_3$-breaking I, II" phases in Fig. \ref{fig:phase}. 
In the ``LL4" phase, the system exhibits a site-dependent quantization axis for the longitudinal fluctuations within the rotated frame as shown in Fig. \ref{fig:LL4_quantization}.
In the ```$D_3$-breaking I, II" phases, the spin orientations have six-site periodicities within the original frame as shown in Fig. \ref{fig:spins_D3}.
To facilitate analysis, we work in the six-sublattice rotated frame throughout this section unless otherwise stated. 

The strategy is to separate the Hamiltonian into two parts, where one part is of the easy-plane XXZ type, and the other part is taken as a perturbation.
We demonstrate that the first order projection of the perturbation term to the low energy degrees of freedom remains to be of the XXZ type, 
indicating that there exists a region of  Luttinger liquid which is stable against the perturbation.

However, the Luttinger parameter diverges as $J\rightarrow 0$, and higher order terms eventually become relevant driving the system into an ordered phase.
As discussed in Sec. \ref{sec:LL1_low_ham}, 
the symmetry group in the six-sublattice rotated frame is isomorphic to $D_{3d}\ltimes 3\mathbb{Z}$.
We demonstrate that there are two types of orders both having six-fold degenerate ground states.
The symmetry breaking patterns are $D_{3d}\rightarrow \mathbb{Z}_2^{\text{(I)}}$ and $D_{3d}\rightarrow \mathbb{Z}_2^{\text{(II)}}$ in the two ordered phases, respectively, 
where $\mathbb{Z}_2^{\text{(I)}}$ and $\mathbb{Z}_2^{\text{(II)}}$ are two different $\mathbb{Z}_2$ groups.
Since $D_{3d}/\mathbb{Z}_2\cong D_3$, both ordered phases break $D_3$ symmetry (albeit different $D_3$ groups), and the two phases are named as ``$D_3$-breaking I" and ``$D_3$-breaking II".
The regions occupied by these two different $D_3$-breaking orders are determined by a classical analysis discussed in Sec. \ref{sec:D3_classical}.
On the other hand, when $|J|$ becomes large, the system is driven into a N\'eel ordered phase along the (111)-direction in the six-sublattice rotated frame, which corresponds to the ``FM" phase in the original frame in Fig. \ref{fig:phase}.

Finally in Sec. \ref{sec:LL4_numerics}, we present and discuss DMRG numerical results,
which provide evidence for the existence of  the ``LL4" and the two $D_3$-breaking phases.

\subsection{Perturbative Luttinger liquid analysis and the ``LL4" phase}
\label{sec:LL4_LL_analysis}

We will perform a perturbative analysis in a neighborhood of the FM2 point in the lower hemisphere.
Thus $\Gamma, K>0$ and $J<0$.
Performing the following two-sublattice rotation to the six-sublattice rotated spin operators $S_j^\alpha$'s,
\begin{eqnarray}
S^{\prime x}_j &=&(-)^j( -\frac{1}{\sqrt{6}} S^x_j+\sqrt{\frac{2}{3}} S_j^y-\frac{1}{\sqrt{6}} S^z_j )\nn\\
S^{\prime y}_j &=&(-)^j( -\frac{1}{\sqrt{2}} S^x_j + \frac{1}{\sqrt{2}} S_j^z) \nn\\
S^{\prime z}_j &=& \frac{1}{\sqrt{3}} S^x_j+\frac{1}{\sqrt{3}} S_j^y+\frac{1}{\sqrt{3}} S_j^z,
\label{eq:rotate_6rot}
\end{eqnarray}
the Hamiltonian in Eq. (\ref{eq:6rotated}) becomes
\begin{eqnarray}
\mathcal{H}^{\prime}=H_{XXZ}+\Delta H+\Delta H^{(2)},
\label{eq:Hprime_tilde}
\end{eqnarray}
in which 
\begin{flalign}
&H_{\text{XXZ}}=\sum_j \big[\Gamma (S_j^{\prime x}S_{j+1}^{\prime x}+S_j^{\prime y}S_{j+1}^{\prime y}) -(\Gamma+J) S_j^{\prime z}S_{j+1}^{\prime z}\big],\nn\\
&\Delta H=\sum_{i=1,2,3}\sum_n \Delta H_{i+3n,i+3n},\nn\\
&\Delta H^{(2)}=\sum_{i=1,2,3 }\sum_n \Delta H_{i+3n,i+3n}^{(2)},
\label{eq:Hxxz}
\end{flalign}
where
\begin{flalign}
&\Delta H_{1+3n,2+3n}=-\frac{1}{2}J(S_{1+3n}^{\prime x}S_{2+3n}^{\prime x}-S_{1+3n}^{\prime y}S_{2+3n}^{\prime y})\nn\\
&~~~+\frac{\sqrt{3}}{2}J(S_{1+3n}^{\prime x}S_{2+3n}^{\prime y}+S_{1+3n}^{\prime y}S_{2+3n}^{\prime x}),\nn\\
&\Delta H_{2+3n,3+3n}=-\frac{1}{2}J(S_{2+3n}^{\prime x}S_{3+3n}^{\prime x}-S_{2+3n}^{\prime y}S_{3+3n}^{\prime y})\nn\\
&~~~ -\frac{\sqrt{3}}{2}J(S_{2+3n}^{\prime x}S_{3+3n}^{\prime y}+S_{2+3n}^{\prime y}S_{3+3n}^{\prime x}),\nn\\
&\Delta H_{3+3n,4+3n}=J(S_{3+3n}^{\prime x}S_{4+3n}^{\prime x}-S_{3+3n}^{\prime y}S_{4+3n}^{\prime y}),
\label{eq:DeltaHwave}
\end{flalign}
and 
\begin{eqnarray}
\Delta H^{(2)}_{1+3n,2+3n}&=&(\Gamma-K) S_{1+3n}^xS_{2+3n}^x\nn\\
\Delta H^{(2)}_{2+3n,3+3n}&=&(\Gamma-K) S_{2+3n}^zS_{3+3n}^z\nn\\
\Delta H^{(2)}_{3+3n,4+3n}&=&(\Gamma-K) S_{3+3n}^yS_{4+3n}^y.
\end{eqnarray}
Notice that the $S_j^\alpha$'s in $\Delta H^{(2)}$ can be written in terms of $S_j^{\prime \alpha}$'s via Eq. (\ref{eq:rotate_6rot}).
We note that the purpose of this additional two-sublattice rotation 
is to make the transverse directions to have an AFM coupling in $H_{\text{XXZ}}$. 

Clearly, $H_{\text{XXZ}}$ corresponds to an easy-plane XXZ chain when $J<0$, $\Gamma>0$.
The low energy theory is described by the Luttinger liquid Hamiltonian in Eq. (\ref{eq:LL_liquid}), and the Luttinger parameter $\mathcal{K}$ is known to be \cite{Giamarchi2003}
\begin{eqnarray}
\mathcal{K}=\frac{\pi}{2\arccos(\frac{J_{\parallel}}{J_{\perp}})},
\label{eq:K_approx}
\end{eqnarray}
in which $J_{\parallel}=\Gamma-|J|$ and $J_{\perp}=\Gamma$.

Next, we add $\Delta H$ and $\Delta H^{(2)}$ as perturbations.
Let's first consider $\Delta H^{(2)}$.
In the Luttinger liquid phase, the continuum limit can be taken at low energies.
To avoid the staggered sign, we enlarge the unit cell to six sites, and consider the following sum
\begin{eqnarray}
\Delta H^{(2)}=(\Gamma-K) \sum_n(\Delta H^{(2),a}_n+\Delta H^{(2),b}_n),
\end{eqnarray}
in which 
\begin{eqnarray}
\Delta H^{(2),a}_n&=&S_{1+3n}^xS_{2+3n}^x+S_{3+3n}^yS_{4+3n}^y+S_{5+3n}^zS_{6+3n}^z,\nn\\
\Delta H^{(2),b}_n&=&S_{2+3n}^zS_{3+3n}^z+S_{4+3n}^xS_{5+3n}^x+S_{6+3n}^yS_{7+3n}^y.\nn\\
\end{eqnarray}
The first order effect of $\Delta H^{(2)}$ can be obtained by projecting  $\Delta H^{(2)}$ to the low energy degrees of freedom  using the following abelian bosonization formula for the spin operators \cite{Affleck1988},
\begin{flalign}
S^{\prime z}(x)&= -\frac{1}{\sqrt{\pi}} \nabla \phi(x) + \text{const.} \frac{1}{a}(-)^n \cos(2\sqrt{\pi} \phi(x)),\nn\\
S^{\prime+}(x)&=\text{const.} \frac{1}{\sqrt{a}}e^{-i\sqrt{\pi}\theta(x)} \big[(-)^n+\cos(2\sqrt{\pi}\phi(x))\big],
\label{eq:abelian_bosonize2}
\end{flalign}
in which $a$ is the lattice constant, $n$ is the lattice site number, and  $x=na$. 
We note that the first order perturbation projection can be figured out without even carrying out any calculation.
Since different sites are smeared out in the continuum limit, 
both $\Delta H^{(2),a}_n$ and $\Delta H^{(2),b}_n$ become SU(2) symmetric $\sim \vec{S}(x)\cdot \vec{S}(x+a)$.
As a result, the effect of $\Delta H^{(2)}$ is to shift the values of $J_{\parallel}$ and $J_\perp$, as
\begin{eqnarray}
J_{\parallel}&=&\Gamma -|J|-\frac{1}{3}(\Gamma-K),\nn\\
J_{\perp}&=&\Gamma -\frac{1}{3}(\Gamma-K).
\end{eqnarray}
Since we still have $|J_{\parallel}|<|J_{\perp}|$ when $\Delta H^{(2)}$ is included,
the system remains to be gapless at least for $|\Delta|\ll1$ where $\Delta=(K-\Gamma)/\Gamma$.
Of course, the Luttinger parameter is renormalized due to $\Delta H^{(2)}$.

Next we consider the effects of $\Delta H$.
Again, the first order effect of $\Delta H$ can be obtained by projecting $\Delta H$ to the low energy degrees of freedom using Eq. (\ref{eq:abelian_bosonize2}).
In fact, it can be observed that the projection vanishes without even carrying out any calculation.
From Eq. (\ref{eq:DeltaHwave}), we obtain 
\begin{flalign}
&\Delta H_{1+6n,2+6n}+\Delta H_{3+6n,4+6n}+\Delta H_{5+6n,6+6n}=\nn\\
&-\frac{1}{2}J (\mathcal{F}_{1+6n}-2\mathcal{F}_{3+6n}+\mathcal{F}_{5+6n} )+\frac{\sqrt{3}}{2}J (\mathcal{G}_{1+6n}-\mathcal{G}_{5+6n}),
\label{eq:LL4_FG}
\end{flalign}
in which
\begin{eqnarray}
\mathcal{F}_i&=&S_{i}^{\prime x}S_{i+1}^{\prime x}-S_{i}^{\prime y}S_{i+1}^{\prime y},\nn\\
\mathcal{G}_i&=&S_{i}^{\prime x}S_{i+1}^{\prime y}+S_{i}^{\prime y}S_{i+1}^{\prime x}.
\end{eqnarray}
In the continuum limit, Eq. (\ref{eq:LL4_FG}) consists of total derivatives as can be seen by Taylor expanding $\mathcal{F}$ and $\mathcal{G}$,
which vanish in the Hamiltonian after the integration $\int dx$.
Similar analysis can be performed on the other three terms $\Delta H_{2+6n,3+6n}+\Delta H_{4+6n,5+6n}+\Delta H_{6+6n,7+6n}$.
Hence we conclude that the projection of $\Delta H$ onto the Luttinger liquid degrees of freedom vanishes, and the system should remain in the Luttinger liquid phase up to first order in $J$.

However, the above analysis of first order projection cannot be trusted if higher order effects are included.
Since $\mathcal{K}$ diverges as $J\rightarrow 0$, higher order terms may become relevant for small enough $J$.
Denote $J^{\prime\alpha}$ and $N^{\prime \alpha}$ as the uniform and staggered parts of the spin operators $S_i^{\prime \alpha}$, respectively, which, according to Eq. (\ref{eq:abelian_bosonize2}), are defined as
\begin{eqnarray}
J^{\prime x}&=&\frac{1}{a}\cos(\sqrt{4\pi} \varphi) \cos(\sqrt{\pi} \theta),\nn\\
J^{\prime y}&=&\frac{1}{a}\cos(\sqrt{4\pi} \varphi) \sin(\sqrt{\pi} \theta),\nn\\
J^{\prime z}&=&-\frac{1}{\sqrt{\pi}}\nabla \varphi,
\label{eq:bosonize_J}
\end{eqnarray}
and 
\begin{eqnarray}
N^{\prime x}&=&\frac{1}{a}\cos(\sqrt{\pi}\theta),\nn\\
N^{\prime y}&=&\frac{1}{a}\sin(\sqrt{\pi}\theta),\nn\\
N^{\prime z}&=&\frac{1}{a}\cos(\sqrt{4\pi}\varphi).
\label{eq:bosonize_N}
\end{eqnarray}
Since the scaling dimension of $N^{\prime\pm}$ ($=N^{\prime x}\pm i N^{\prime y}$) is $1/(4\mathcal{K})$, the operators involving  powers of $N^{\prime\pm}$  become relevant for sufficiently small $J$. 
On the other hand, since $N^{\prime z}$ has scaling dimension equal to $\mathcal{K}$, 
the operators involving $N^{\prime z}$ become relevant when $\mathcal{K}$ is small enough. 
Hence, in what follows, we will consider terms only involving $N^{\prime\alpha}$ ($\alpha=x,y,z$),
which should be able to capture the spin orderings for both small and large values of $J$.

Next we perform a symmetry analysis to figure out what terms are allowed in the low energy theory.
In the rotated basis defined via Eq. (\ref{eq:rotate_6rot}), 
the generators of the symmetry group $G_1$ in Eq. (\ref{eq:group_G1}) become 
\begin{eqnarray}
R_a^\prime T_a&=&U_2 R_aT_a U_2^{-1},\nn\\
R_I^\prime I&=&U_2 R_II U_2^{-1},
\label{eq:sym_prime_6rot}
\end{eqnarray}
where $U_2$ is equal to $\Pi_n R_{2n}(\hat{z}^\prime,\pi)R^\prime$ which gives the transformation in Eq. (\ref{eq:rotate_6rot}).
It is straightforward to verify that $R^\prime_a=R(\hat{z}^\prime,\frac{2\pi}{3})$ and $R^\prime_I=R(\hat{y}^\prime,\pi)$.
Under the symmetries in Eq. (\ref{eq:sym_prime_6rot}), the transformation properties of $N^{\prime\alpha}$ are
\begin{eqnarray}
T(N^{\prime+},N^{\prime -})T^{-1}&=&(N^{\prime +},N^{\prime -}) 
(-\tau_1),\nn\\
R^\prime_aT_a (N^{\prime+},N^{\prime-}) (R^\prime_aT_a)^{-1}&=&(N^{\prime+},N^{\prime-})
(\frac{1}{2}\tau_0-\frac{\sqrt{3}i}{2}\tau_3),\nn\\
R^\prime_II (N^{\prime+},N^{\prime-})(R^\prime_II)^{-1} &=&(N^{\prime+},N^{\prime-}) 
(-\tau_1),
\end{eqnarray}
and
\begin{eqnarray}
TN^{\prime z}T^{-1}&=&-N^{\prime z},\nn\\
(R^\prime_aT_a)N^{\prime z} (R^\prime_aT_a)^{-1}&=&-N^{\prime z},\nn\\
(R^\prime_II)N^{\prime z}(R^\prime_II)^{-1}&=&-N^{\prime z},
\end{eqnarray}
in which  $\tau_\alpha$ ($\alpha=1,2,3$) are the three Pauli matrices, and $\tau_0$ is the $2\times 2$ identity matrix.
Therefore, we see that the following terms are allowed in the low energy Hamiltonian,
\begin{eqnarray}
(N^{\prime +})^{6n}+(N^{\prime -})^{6n}, ~ (N^{\prime z})^{2n},
\end{eqnarray}
where $n>0$, $n\in \mathbb{Z}$.

At large $\mathcal{K}$, the symmetry allowed operator with the smallest scaling dimension ($=9/\mathcal{K}$) is 
\begin{eqnarray}
\sim J^3[(N^{\prime +})^6+(N^{\prime -})^6],
\end{eqnarray}
which is $\sim J^3\cos(6\sqrt{\pi}\theta)$ after bosonization.
Notice that this term can only appear at the level of third order perturbations, hence the coupling should be proportional to $J^3$.
When $\mathcal{K}> 4.5$, $\cos(6\sqrt{\pi}\theta)$ becomes relevant.
According to Eq. (\ref{eq:K_approx}), 
the value of $J$ which has the critical $\mathcal{K}$ value ($=4.5$) is
\begin{eqnarray}
J_{c1}(\Delta)=-\Gamma(1+\frac{\Delta}{3})(1-\cos(\frac{\pi}{9})).
\label{eq:Jc1}
\end{eqnarray}
When $\Delta=0$ (i.e.,  $\phi=0.25\pi$), the corresponding $\theta_{c1}$ determined from Eq. (\ref{eq:Jc1}) is $0.514\pi$.
Of course, the value of $J_{c1}(\Delta)$ in Eq. (\ref{eq:Jc1}) is not accurate due to high order effects.
Based on the above analysis, we see that the Luttinger liquid is stable only when $|J|>|J_{c1}(\Delta)|$ for a fixed value of $\Delta$.

On the other hand, the above analysis applies only to the region of a small $|J|$. 
By increasing $|J|$, we note that the term $(N^{\prime z})^2\sim \cos(4\sqrt{\pi}\phi)$ eventually has a smaller dimension than $\cos(6\sqrt{\pi}\theta)$.
When the Luttinger parameter $\mathcal{K}$ is smaller than $1/2$, $(N^{\prime z})^2$ becomes relevant since its scaling dimension is $4\mathcal{K}$.
Hence there exists another critical value $|J_{c2}(\Delta)|$ above which the Luttinger liquid become unstable again.
In conclusion, a Luttinger liquid can be stablized in an intermediate range $|J_{c1}(\Delta)|<|J|<|J_{c2}(\Delta)|$, i.e., $J_{c2}(\Delta)<J<J_{c1}(\Delta)$.

\subsection{The ``$D_3$-breaking" phase}

\subsubsection{The strong coupling limit for $|J|<|J_{c1}|$}

We will figure out the spin orders corresponding to the above mentioned two strong coupling limits when $|J|<|J_{c1}|$ and $|J|>|J_{c2}|$.
Let's first consider $|J|<|J_{c1}(\Delta)|$.
The term $\lambda \cos(6\sqrt{\pi}\theta)$ is minimized at the following values of the angle $\theta$,
\begin{flalign}
&\theta_n^{(\text{I})}=\frac{(2n+1)\sqrt{\pi}}{6},~\lambda>0,\nn\\
&\theta_n^{(\text{II})}=\frac{n\sqrt{\pi}}{3},~\lambda<0,
\end{flalign}
in which $\lambda$ is the coupling constant in the term $\lambda\cos(6\sqrt{\pi}\theta)$,
and $0 \leq n\leq 5$, $n\in\mathbb{Z}$.
Notice that the system is six-fold degenerate regardless of the sign of $\lambda$.
Naively, according to Eq. (\ref{eq:abelian_bosonize2}), the spins exhibit a N\'eel ordering in the $x^\prime y^\prime$-plane given by
\begin{eqnarray}
\vec{S}_i^\prime=(-)^i N_\perp (\cos(\sqrt{\pi}\theta_n),\sin(\sqrt{\pi}\theta_n),0),
\label{eq:spin_order_LL4}
\end{eqnarray}
in which $N_\perp$ is the magnitude of the ordering.

However, we note that the spin orders in Eq. (\ref{eq:spin_order_LL4}) are not precise.
This is because there can be U(1) breaking coefficients in the abelian bosonization formula Eq. (\ref{eq:abelian_bosonize2}) due to the renormalization effects in the high energy region along the RG flow, which is similar to the case encountered in Eq. (\ref{eq:Nonabel}) as discussed in details in Ref. \onlinecite{Yang2020a}.
As a result, the abelian bosonization formula Eq. (\ref{eq:abelian_bosonize2}) should only respect the discrete lattice symmetry, not the emergent U(1) symmetry at low energies.
Taking this into account, the true spin orientations will be distorted with respect to those in  Eq. (\ref{eq:spin_order_LL4}) due to the U(1) breaking coefficients.
To figure out the correct pattern of the spin orientations, 
we first determine
the symmetry breaking pattern corresponding to the spin ordering in Eq. (\ref{eq:spin_order_LL4}),
and then solve the most general form of the spin orderings which is consistent with the identified symmetry breaking. 
This approach is similar as the one used in Sec. \ref{sec:sym_LL3} to discuss the ``FM" phase based on the RG analysis in the vicinity of the AFM3 point. 

When $\lambda>0$, we choose a representative spin configuration corresponding to $n=1$ in $\theta_n^{\text{(I)}}$.
As can be seen from Eq. (\ref{eq:spin_order_LL4}), this is a N\'eel order along $y^\prime$-direction.
As can be easily checked, the N\'eel-$y^\prime$ configuration is invariant under $TT_{3a}$ and $R_I^\prime I$.
Hence the little group of the N\'eel-$y^\prime$ order is 
\begin{eqnarray}
H^{\text{(I)}}=\mathopen{<}TT_{3a},R_I^\prime I\mathclose{>},
\end{eqnarray}
which can be rewritten as
\begin{eqnarray}
H^{\text{(I)}}=\mathbb{Z}_2^{(I)}\ltimes 3\mathbb{Z},
\end{eqnarray}
where $\mathbb{Z}_2^{\text{(I)}}=\mathopen{<}R_I^\prime I\mathclose{>}$,
 and $3\mathbb{Z}=\mathopen{<} TT_{3a} \mathclose{>}$.
Recall that the symmetry group of the Hamiltonian $H^\prime$ in Eq. (\ref{eq:Hprime_tilde}) is
$G_1^\prime=\mathopen{<}T,R_a^\prime T_a,R_I^\prime I\mathclose{>}$,
which can be rewritten as 
\begin{eqnarray}
G_1^\prime=D_{3d}\times 3\mathbb{Z},
\end{eqnarray}
where $D_{3d}=\mathopen{<}T,R_a^\prime T_a,R_I^\prime I\mathclose{>} \mod TT_{3a}$, 
and $3\mathbb{Z}=\mathopen{<}TT_{3a}\mathclose{>}$.
Therefore, by taking the quotient of $\mathopen{<}TT_{3a}\mathclose{>}$, we conclude that the symmetry breaking pattern is 
\begin{eqnarray}
D_{3d}\rightarrow \mathbb{Z}_2^{\text{(I)}}.
\end{eqnarray}
Notice that since $|D_{3d}/\mathbb{Z}_2^{\text{(I)}}|=6$, the ground states are six-fold degenerate.
This order is named as ``$D_3$-breaking I" since the broken symmetry group $D_{3d}/\mathbb{Z}_2^{(I)}$ is isomorphic to $D_3$.

Next, we work out the most general form of the spin ordering which is invariant under $H^{\text{(I)}}$.
The invariance under $TT_{3a}$ clearly requires 
\begin{eqnarray}
\vec{S}_{i+3}^\prime=-\vec{S}_{i}^\prime.
\end{eqnarray}
As can be checked, further requiring the invariance under $R_I^\prime I$, the spin alignments are constrained to be
\begin{eqnarray}
\vec{S}_{1+3n}^{\prime\text{(I)}}&=& (-)^n N_\perp^{\text{(I)}} (x^\prime,y^\prime,z^\prime)^T,\nn\\
\vec{S}_{2+3n}^{\prime\text{(I)}}&=& (-)^n N_\perp^{\text{(I)}} (0,1,0)^T,\nn\\
\vec{S}_{3+3n}^{\prime\text{(I)}}&=& (-)^n N_\perp^{\text{(I)}} (-x^\prime,y^\prime,-z^\prime)^T,
\label{eq:spin_order_6rot_prime}
\end{eqnarray}
in which $x^{\prime 2}+y^{\prime 2}+z^{\prime 2}=1$.
The other five degenerate spin configurations can be obtained by performing the operations in the equivalent classes in $D_{3d}/\mathbb{Z}_2^{\text{(I)}}$ to the spin ordering in Eq. (\ref{eq:spin_order_6rot_prime}).

\begin{figure}
\includegraphics[width=7cm]{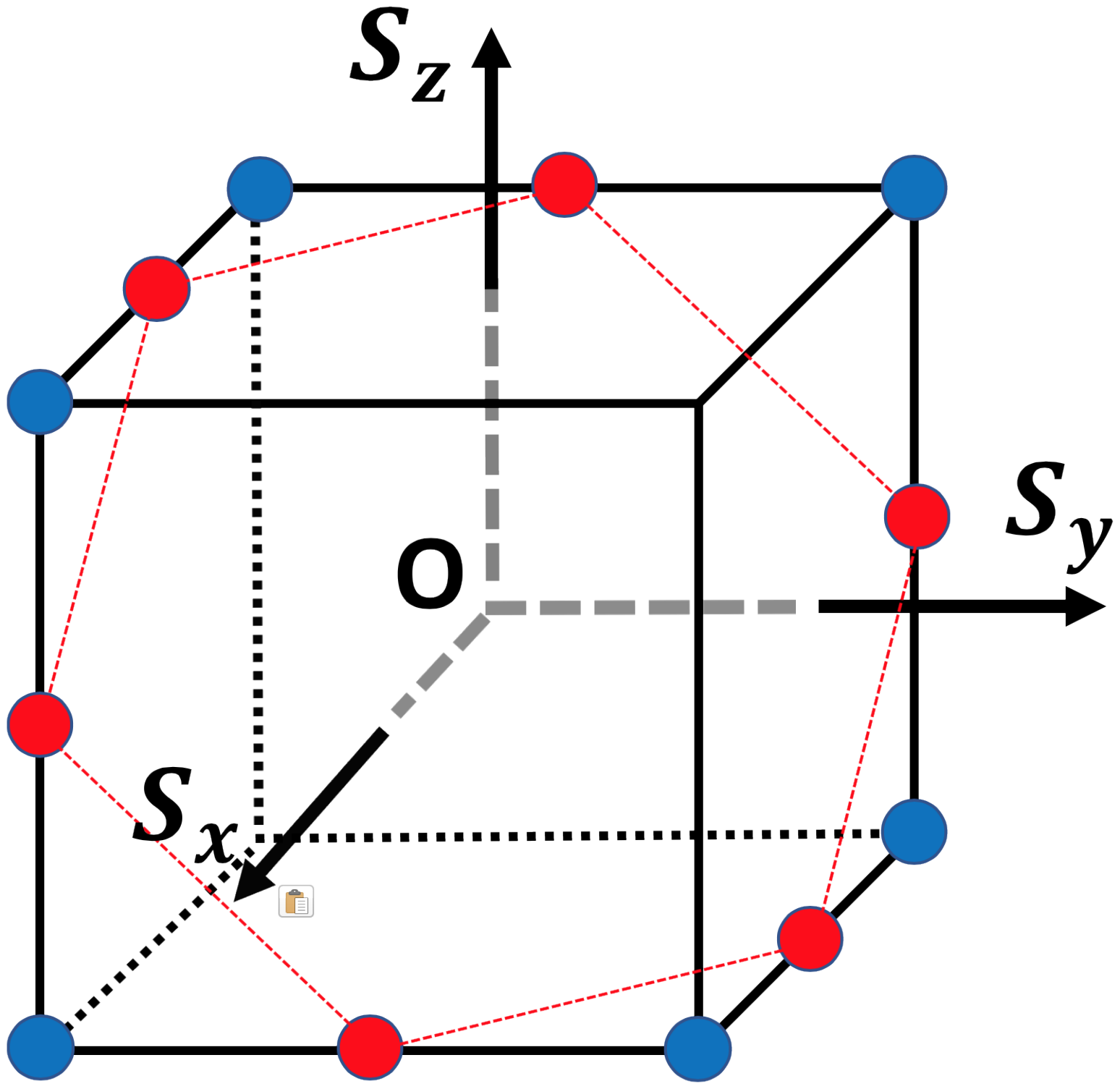}
\caption{``Center of mass" directions of the three spins within a unit cell for the six degenerate ground states 
as represented by the six red (blue) solid circles 
 in the ``$D_3$-breaking I (II)" phases in the six-sublattice rotated frame.
 In the ``$D_3$-breaking I" phase, the red circles are connected with thin dashed red lines to indicate that the hexagon is coplanar. 
 In the ``$D_3$-breaking II" phase, the plots are for $J\rightarrow 0$ according to the classical analysis.
} 
\label{fig:Spins_6rot_D3I}
\end{figure}

It is interesting to work out the spin alignments in the six-sublattice rotated frame by performing the inverse of the transformation defined in Eq. (\ref{eq:rotate_6rot}).
The result is
\begin{eqnarray}
\vec{S}_{1+3n}^{\text{(I)}}&=&  N_\perp^{\text{(I)}} (x,y,z)^T,\nn\\
\vec{S}_{2+3n}^{\text{(I)}}&=&  N_\perp^{\text{(I)}} (-\frac{1}{\sqrt{2}},0,\frac{1}{\sqrt{2}})^T,\nn\\
\vec{S}_{3+3n}^{\text{(I)}}&=&  N_\perp^{\text{(I)}} (-z,-y,-x)^T,
\label{eq:D3_order_6rot}
\end{eqnarray}
in which $x,y,z$ can be expressed through $x^\prime,y^\prime,z^\prime$ using Eq. (\ref{eq:rotate_6rot}).
Define $\vec{S}^{(\text{I})}_c=\frac{1}{\sqrt{3}N^{\text{(I)}}_{\perp}} (\vec{S}^{(\text{I})}_{1+3n}+\vec{S}^{(\text{I})}_{2+3n}+\vec{S}^{(\text{I})}_{3+3n})$
to be the  ``center of mass" direction for the three spins in a unit cell in the six-sublattice rotated frame.
Then according to Eq. (\ref{eq:D3_order_6rot}), it is clear that $\vec{S}_c^{(I)}$ is along the $(-1,0,1)$-direction. 
The ``center of mass" directions in the six degenerate ground states are represented as the six red circles as shown in Fig. \ref{fig:Spins_6rot_D3I}.
Furthermore, by performing the inverse of the six-sublattice rotation defined in Eq. (\ref{eq:6rotation}), the spin ordering in the original frame can be worked out as
\begin{eqnarray}
\vec{S}_{1+6n}^{(0),\text{(I)}}&=&N^{\text{(I)}}_\perp (x,y,z)^T,\nn\\
\vec{S}_{2+6n}^{(0),\text{(I)}}&=&N^{\text{(I)}}_\perp (\frac{1}{\sqrt{2}},-\frac{1}{\sqrt{2}},0)^T,\nn\\
\vec{S}_{3+6n}^{(0),\text{(I)}}&=&N^{\text{(I)}}_\perp (-y,-x,-z)^T,\nn\\
\vec{S}_{3+6n}^{(0),\text{(I)}}&=&N^{\text{(I)}}_\perp (-y,-x,-z)^T,\nn\\
\vec{S}_{5+6n}^{(0),\text{(I)}}&=&N^{\text{(I)}}_\perp (\frac{1}{\sqrt{2}},-\frac{1}{\sqrt{2}},0)^T,\nn\\
\vec{S}_{6+6n}^{(0),\text{(I)}}&=&N^{\text{(I)}}_\perp (x,y,z)^T,
\label{eq:D3I_spin_orient}
\end{eqnarray}
in which the superscript $(0)$ is used to denote the  original frame.
A plot of the spin orientations in Eq. (\ref{eq:D3I_spin_orient}) is shown in Fig. \ref{fig:spins_D3} (a)  where the unnormalized parameters are chosen as $x=-1$, $y=0$, $z=1$.

A similar analysis can be performed to the case $\lambda<0$.
Choosing  $n=0$ in $\theta_n^{\text{(I)}}$,
we see that the system has a N\'eel ordering along the $x^\prime$-direction.
It can be shown that the little group of the N\'eel-$x^\prime$ configuration is
\begin{eqnarray}
H^{\text{(II)}}=\mathopen{<}TT_{3a},TR_I^\prime I\mathclose{>}.
\end{eqnarray}
This time, the symmetry breaking pattern is 
\begin{eqnarray}
D_{3d}\rightarrow \mathbb{Z}_2^{\text{(II)}},
\end{eqnarray}
in which $\mathbb{Z}_2^{\text{(II)}}=\mathopen{<}TR_I^\prime I\mathclose{>}$.
The most general spin alignments invariant under $H^{\text{(II)}}$ are
\begin{eqnarray}
\vec{S}_{1+3n}^{\prime\text{(II)}}&=& (-)^n N^{\text{(II)}}_\perp (a^\prime,b^\prime,c^\prime)^T,\nn\\
\vec{S}_{2+3n}^{\prime\text{(II)}}&=& (-)^n N^{\text{(II)}}_\perp (m^\prime,0,n^\prime)^T,\nn\\
\vec{S}_{3+3n}^{\prime\text{(II)}}&=& (-)^n N^{\text{(II)}}_\perp (a^\prime,-b^\prime,c^\prime)^T,
\label{eq:spin_order_6rot_prime_II}
\end{eqnarray}
in which $a^{\prime2}+b^{\prime2}+c^{\prime2}=1$, $m^{\prime2}+n^{\prime2}=1$.
We note that the ground states are again six-fold degenerate, and the spin orientations in the other five degenerate states can be obtained by performing the operations in the equivalent classes in $D_{3d}/\mathbb{Z}_2^{\text{(II)}}$ to the above spin configuration.
This order is named as ``$D_3$-breaking II" since the broken symmetry group $D_{3d}/\mathbb{Z}_2^{\text{(II)}}$  is isomorphic to $D_3$.
We emphasize that although the broken symmetry is again isomorphic to $D_3$, it is a different $D_3$ group compared with the ``$D_3$-breaking I" case.

Similar with the ``$D_3$-breaking I" case, the spin alignments in the ``$D_3$-breaking II" phase in the six-sublattice rotated frame can be determined as
\begin{eqnarray}
\vec{S}_{1+3n}^{\text{(II)}}&=&  N_\perp^{\text{(II)}} (a,b,c)^T,\nn\\
\vec{S}_{2+3n}^{\text{(II)}}&=&  N_\perp^{\text{(II)}} (m,n,m)^T,\nn\\
\vec{S}_{3+3n}^{\text{(II)}}&=&  N_\perp^{\text{(II)}} (c,b,a)^T,
\label{eq:D3_order2_6rot}
\end{eqnarray}
in which $a,b,c,m,n$ can be expressed through $a^\prime,b^\prime,c^\prime,m^\prime,n^\prime$ using Eq. (\ref{eq:rotate_6rot}).
By performing the classical analysis discussed in Sec.\ref{sec:D3_classical},
we find that $a,b,c,m,n$ approach $1,-1,1,-1,1$, respectively, in the limit $J\rightarrow 0$ for fixed $\Delta\neq 0$.
Therefore, when $J\rightarrow 0$,
the ``center of mass" direction of the three spins in a unit cell defined as 
$\vec{S}^{\text{(II)}}_c=\frac{1}{\sqrt{3}N^{\text{(II)}}_{\perp}} (\vec{S}^{\text{(II)}}_{1+3n}+\vec{S}^{\text{(II)}}_{2+3n}+\vec{S}^{\text{(II)}}_{3+3n})$ is along the $(1,-1,1)$-direction.
The $\vec{S}^{\text{(II)}}_c$'s in the six degenerate ground states when $J\rightarrow 0$ are represented as the six blue circles as shown in Fig. \ref{fig:Spins_6rot_D3I}.
We also note that the corresponding spin orientations in the original frame are 
\begin{eqnarray}
\vec{S}_{1+6n}^{(0),\text{(II)}}&=&N_\perp^{\text{(II)}} (a,b,c)^T,\nn\\
\vec{S}_{2+6n}^{(0),\text{(II)}}&=&N_\perp^{\text{(II)}} (-m,-m,-n)^T,\nn\\
\vec{S}_{3+6n}^{(0),\text{(II)}}&=&N_\perp^{\text{(II)}} (b,a,c)^T,\nn\\
\vec{S}_{3+6n}^{(0),\text{(II)}}&=&N_\perp^{\text{(II)}} (-b,-a,-c)^T,\nn\\
\vec{S}_{5+6n}^{(0),\text{(II)}}&=&N_\perp^{\text{(II)}} (m,m,n)^T,\nn\\
\vec{S}_{6+6n}^{(0),\text{(II)}}&=&N_\perp^{\text{(II)}} (-a,-b,-c)^T.
\label{eq:D3II_spin_orient}
\end{eqnarray}
A plot of the spin orientations in Eq. (\ref{eq:D3II_spin_orient}) is shown in Fig. \ref{fig:spins_D3} (b)  where the unnormalized parameters are chosen as $a=-b=c=m=-n=1$.

Bases on the above analysis, we see that when $\theta<\theta_{c1}(\phi)$, $D_3$-breaking orders are developed, where there are two types of possible orders denoted as ``$D_3$-breaking I" and ``$D_3$-breaking II" depending on the sign of the coupling constants.
The determination of the sign of the coupling constant requires a third order perturbation.
We will not perform such a difficult calculation of third order perturbation,
but turn to a classical analysis in the $D_3$-breaking phase which will be discussed in Sec. \ref{sec:D3_classical},
where it is  verified that the region for $\theta<\theta_{c1}(\phi)$ can be divided into two subregions numbered by ``I" and ``II",
which have ``$D_3$-breaking I" and ``$D_3$-breaking II" orders correspondingly.
The solid lines separating the two $D_3$-breaking phases in Fig. \ref{fig:phase} are determined from such classical analysis.

We also note that the coupling constant $\sim J^3$ is very small when  $\theta<\theta_{c1}(\phi)$.
Hence, from an RG point of view, the system has to flow a very long time to develop such order. 
This means that a very large system size is required to observe the $D_3$-breaking orders,
making the detection of the orders very difficult numerically.
In numerics, we are not able to find a clear evidence for the six-fold ground degeneracy in either of the two ``$D_3$-breaking" phases.
Since calculating the energies of several excited states can only be performed for rather small system sizes, the above mentioned strong finite size fact is possibly the reason for the failure of identifying the degeneracy.

\subsubsection{``$O_h\rightarrow D_4$" as part of the ``$D_3$-breaking II" phase}

In this subsection, we remark that the ``$O_h\rightarrow D_4$" phase on the equator in Fig. \ref{fig:phase} is actually part of the ``$D_3$-breaking II" phase.

To see this, first notice that the representative elements of different equivalent classes in $O_h/D_4$ can be chosen such that they form the group $D_3$,
 which means that the broken symmetries of the ``$O_h\rightarrow D_4$" phase can be chosen as the $D_3$ group. 
 Secondly, notice that the unbroken symmetry group $\mathbb{Z}_2^{\text{(II)}}=\mathopen{<}TR_I I\mathclose{>}$ in the ``$D_3$-breaking II" phase is a subgroup of $D_4$. 
As discussed in Ref. \onlinecite{Yang2020a}, the spins align along $\pm \hat{\alpha}$-directions ($\alpha=x,y,z$) in the six degenerate ground states in the ``$O_h\rightarrow D_4$" phase within the six-sublattice rotated frame. 
In particular, the explicit form of the unbroken $D_4$ group for the ground state in which the spins align along the $y$-direction is $\mathopen{<} (R_aT_a)^{-1} R(\hat{z},\pi)R_I I R(\hat{z},\pi) (R_aT_a)^{-1}, TR_I I \mathclose{>}/\mathopen{<}T_{3a}\mathclose{>}$ (see Ref. \onlinecite{Yang2020a}), which contains the group $\mathbb{Z}_2^{\text{(II)}}$.
Therefore, by combining these two observations, 
we see that when a nonzero $J$ is turned on,
the full  symmetry group of the Hamiltonian reduces from $O_h$ to $D_{3d}$,
and concomitantly, the unbroken symmetry group reduces from $D_4$ to $\mathbb{Z}_2^{\text{(II)}}$.
The reductions are coordinated in such a manner that the broken symmetries remain the same, hence the system should be viewed as  residing in the same phase. 

To further validate the above analysis, notice that the spin alignments in Eq. (\ref{eq:D3_order2_6rot}) transform into the following pattern under the operation $(R_aT_a)^{-1} R(\hat{z},\pi)R_I I R(\hat{z},\pi) (R_aT_a)^{-1}$:
\begin{flalign}
&\vec{S}_1 \rightarrow (-a,b,c)^T\nn\\
&\vec{S}_2 \rightarrow (-m,n,m)^T\nn\\
&\vec{S}_3 \rightarrow (-c,b,a)^T.
\label{eq:D3_order2_6rot_y}
\end{flalign}
Hence, requiring Eq. (\ref{eq:D3_order2_6rot}) to be invariant under $(R_aT_a)^{-1} R(\hat{z},\pi)R_I I R(\hat{z},\pi) (R_aT_a)^{-1}$ in addition to $TR_II$,
we obtain $a=c=m=0$.
Thus the spins in  Eq. (\ref{eq:D3_order2_6rot_y}) are exactly along the $y$-direction,
which is the spin alignments in the ``$O_h\rightarrow D_4$" phase.
This shows that the transition of the spin alignments from the ``$D_3$-breaking II" phase to the ``$O_h\rightarrow D_4$" phase is smooth,
hence they should be viewed as  the same phase.

\subsubsection{Classical analysis}
\label{sec:D3_classical}

\begin{figure}[h]
\includegraphics[width=7.0cm]{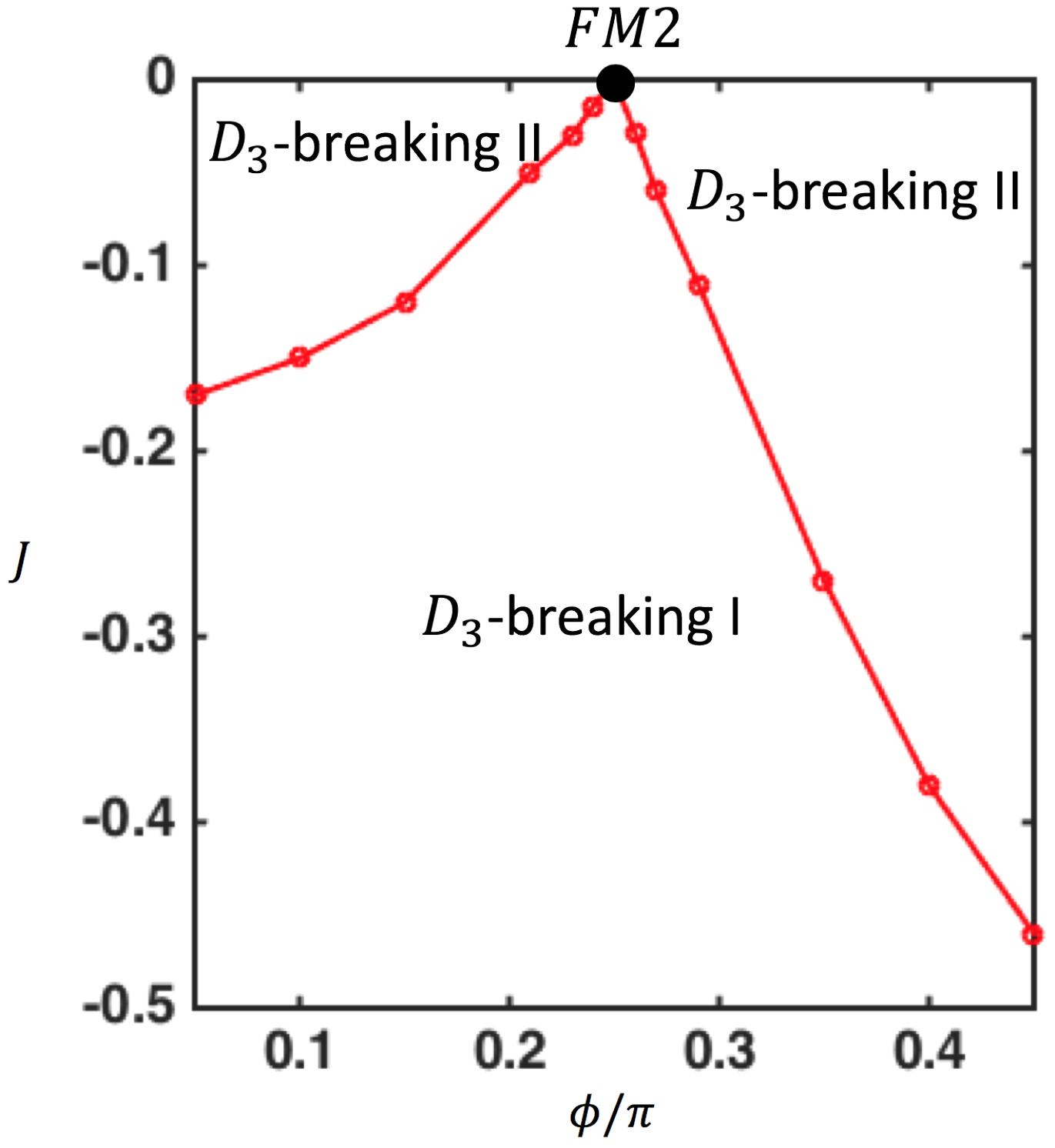}
\caption{Classical phase diagram in the vicinity of the FM2 point for negative $J$.
} 
\label{fig:classical_phase_LL4}
\end{figure}

The classical analysis is the saddle point approximation in the spin path integral formalism which is valid in the large-$S$ limit, where $S$ is the spin value.
In what follows, we neglect the quantum fluctuations of the spins and approximate them as classical three-vectors,
\begin{eqnarray}
\vec{S}_i = S\hat{n}_i,
\end{eqnarray}
in which $S$ is the spin magnitude, $\hat{n}_i=(x_i,y_i,z_i)^T$ is a unit vector.

Absorbing $S^2$ into a redefinition of $\Gamma,K,J$ by defining 
\begin{eqnarray}
\Gamma^\prime=\Gamma S^2,~K^\prime=K S^2,~J^\prime=JS^2,
\end{eqnarray}
and introducing the Lagrange multipliers $\{\lambda_i\}_{1\leq i \leq3}$ to impose the constraints $x_i^2+y_i^2+z_i^2=1$, 
the energy per unit cell  in the six-sublattice rotated frame becomes
\begin{flalign}
&F= -(K^\prime+J^\prime) x_1x_2-\Gamma^\prime(y_1y_2+z_1z_2)-J^\prime(y_1z_2+y_1z_2)\nn\\
&-(K^\prime+J^\prime) z_1z_2-\Gamma^\prime(x_1x_2+y_1y_2)-J^\prime(x_1y_2+y_1x_2)\nn\\
&-(K^\prime+J^\prime) y_1y_2-\Gamma^\prime(z_1z_2+x_1x_2)-J^\prime(z_1x_2+x_1z_2)\nn\\
&-\sum_{i=1}^3 \frac{1}{2}\lambda_i (x_i^2+y_i^2+z_i^2-1),
\label{eq:energy_KHG}
\end{flalign}
in which the free energy corresponding to a general spin-$S$ $KH\Gamma$ chain is considered.
We have numerically studied the minimization of the classical free energy,
and verified that the solution of Eq. (\ref{eq:spin_order_6rot_prime}) (Eq. (\ref{eq:spin_order_6rot_prime_II})) corresponds to the global minimum of the free energy in the range denoted by the ``$D_3$-breaking I" (``$D_3$-breaking II") phase in Fig. \ref{fig:classical_phase_LL4}.

\begin{figure*}[htbp]
\includegraphics[width=18.0cm]{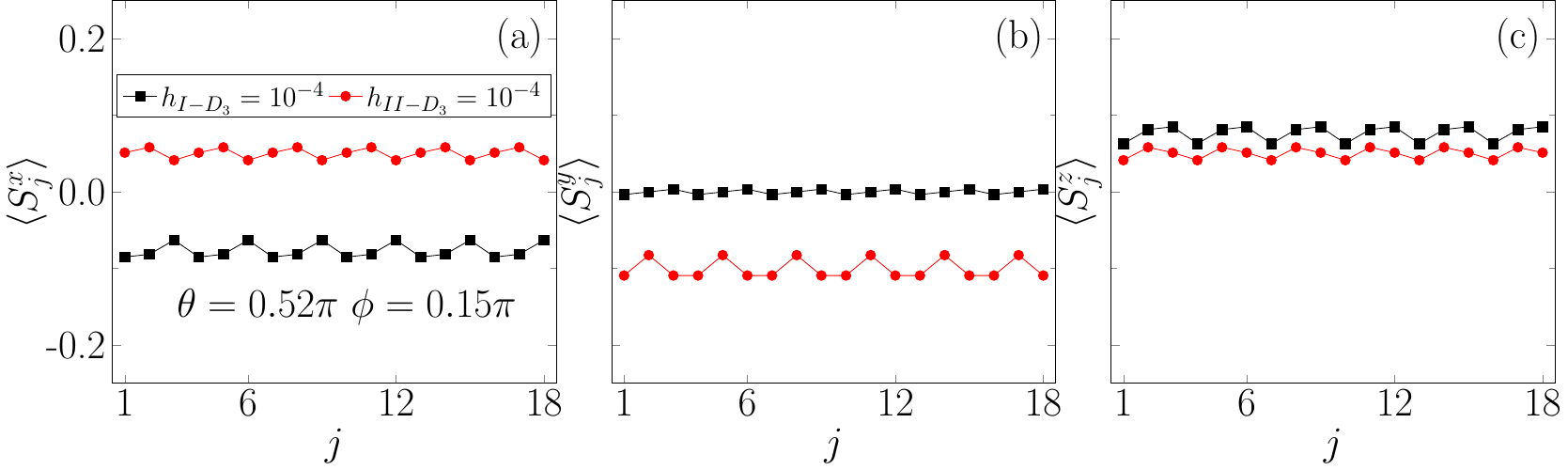}
\caption{(a) $\langle S_j^x\rangle$, (b) $\langle S_j^y\rangle$, (c) $\langle S_j^z\rangle$ vs $j$ under $h_{\text{I}}$ (black squares) and $h_{\text{II}}$ (red dots) fields.
ED numerics are performed on L=18 sites with periodic boundary conditions at $(\theta=0.52\pi,\phi=0.15\pi)$.
Both $h_{\text{I}}$ and $h_{\text{II}}$ fields are taken to be $10^{-4}$.
} 
\label{fig:D3order_A}
\end{figure*}

\subsubsection{The strong coupling limit for $|J|>|J_{c2}|$}
\label{sec:FM_from_LL4}

The above analysis completes the discussion for the strong coupling limit in the region $|J|<|J_{c1}(\Delta)|$.
Next we identify the order in the parameter region $|J|>|J_{c2}(\Delta)|$.
When $\cos(4\sqrt{\pi}\phi)$ becomes relevant, $\phi_n$ orders at $(2n+1)\sqrt{\pi} /4$ or $n\sqrt{\pi} /2$ depending on the sign of the coupling constant.
According to the bosonization formula in Eq. (\ref{eq:abelian_bosonize2}),
the system develops a N\'eel ordering along $z^\prime$-direction.
Performing the inverse of the transformation in Eq. (\ref{eq:rotate_6rot}),
it corresponds to a N\'eel-$\hat{n}_a$ order in the six-sublattice rotated frame,
where $\hat{n}_a$ is the unit vector along the $(111)$-direction.
As discussed in Sec. \ref{sec:RG_KG_H}, by taking into account the distortions of the quantization axes in the bosonization formula and performing the inverse of the six-sublattice rotation, 
the N\'eel-$\hat{n}_a$ ordering corresponds to an FM spin order in the original frame. 
Thus we conclude that the system should transit from the ``LL4" phase to the ``FM" phase by increasing the magnitude of $J$ where  $J<0$.

We can also make a tree-level estimation on the value of $J_{c2}$.
As can be seen from Eq. (\ref{eq:Hxxz}),
the anisotropy of the $H_{XXZ}+\Delta H^{(2)}$ Hamiltonian in Eq. (\ref{eq:Hxxz}) becomes easy-axis when $J<-2\Gamma$.
Neglecting the effects of $\Delta H$ in Eq. (\ref{eq:Hprime_tilde}),
the critical value $J_{c2}(\Delta)$ is determined to be  $J_{c2}(\Delta)\equiv-2\Gamma$.
At $\Delta=0$, this gives the point $(\theta_{c2}=0.804\pi,\phi=0.25\pi)$ in the phase diagram.
Of course, the value of $J_{c2}(\Delta)$ must be shifted due to the effects of $\Delta H$ and higher order effects of $\Delta H^{(2)}$.

\subsubsection{Phase diagram around the FM2 point}

Based on the above analytic analysis, we propose the following phase diagram which applies at least in a neighborhood of the FM2 point:
\begin{eqnarray}
\text{``$D_3$-breaking I, II"},& \theta<\theta_{c1}(\phi);\nn\\
\text{``LL4"},&  \theta_{c1}(\phi)<\theta<\theta_{c2}(\phi);\nn\\
\text{``FM"}, & \theta>\theta_{c2}(\phi),
\end{eqnarray}
where we have expressed $J$ and $\Delta$ in terms of $\theta$ and $\phi$,
and the ranges of $\theta$ all refer to the corresponding value of $\phi$.
In the ``LL4" phase, the quantization axis for the longitudinal fluctuation is along the $(111)$-direction in the six-sublattice rotated frame which becomes staggered in the original frame with possible site-dependent distortions as discussed in Sec. \ref{sec:RG_KG_H}.
However, the staggered sign in the definition of the coordinates in Eq. (\ref{eq:rotate_6rot}) indicates an FM-type quasi-long range order in the six-sublattice rotated frame. 
Hence, no oscillation is drawn in the cartoon plot of the ``LL4" phase in Fig. \ref{fig:LL4_quantization}.

\subsection{Numerical results}
\label{sec:LL4_numerics}

\begin{figure*}[htbp]
\includegraphics[width=12.0cm]{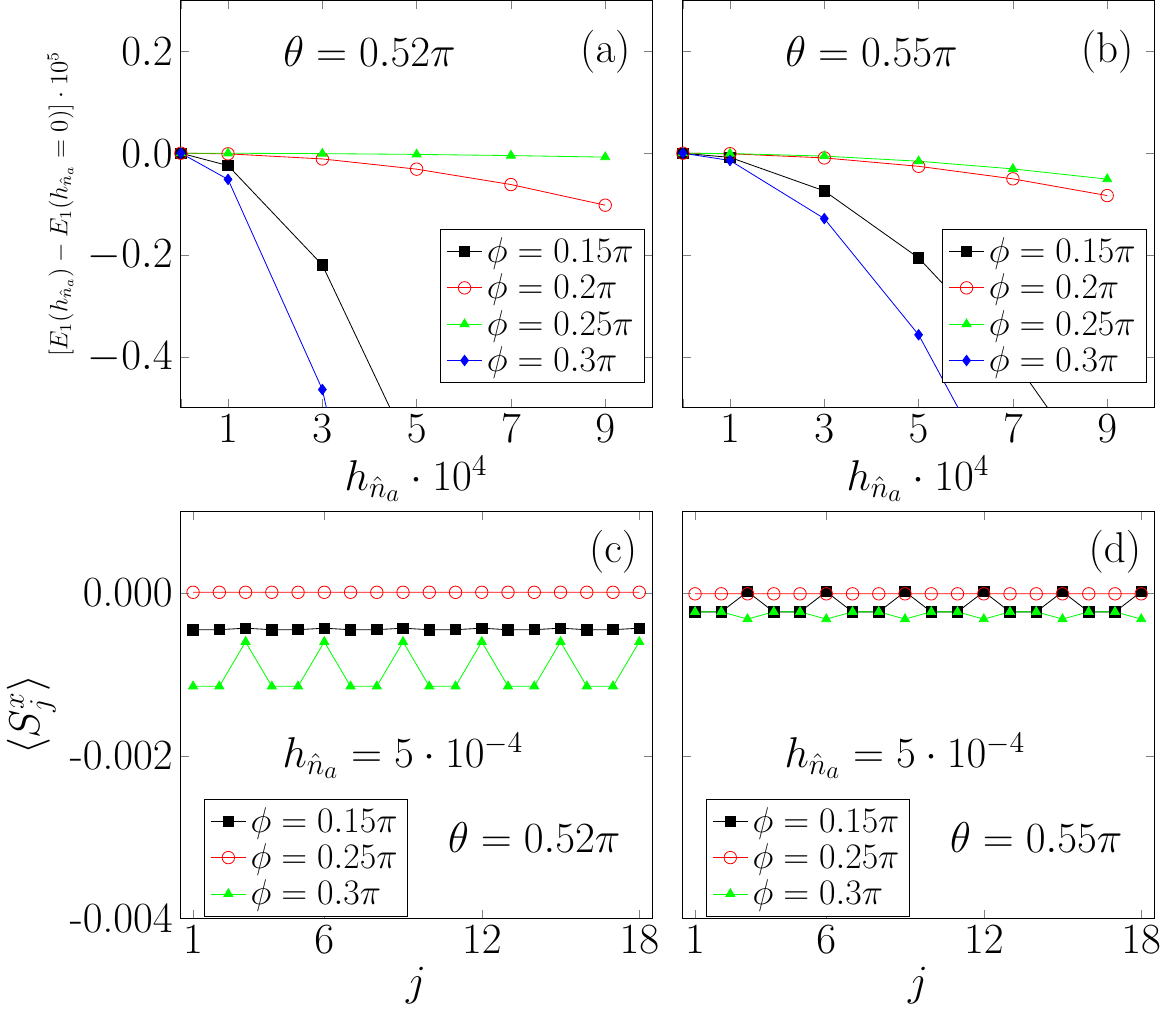}
\caption{$\Delta E$ ($=E(h_{\hat{n}_a})-E(h_{\hat{n}_a}=0)$) vs $h_{\hat{n}_a}$ for (a) $\theta=0.52\pi$, (b) $\theta=0.55\pi$ at several fixed values of  $\phi$;
and $\left< S_j^x \right>$ vs $j$ for (c) $\theta=0.52\pi$, (d) $\theta=0.55\pi$ at several fixed values of $\phi$.
The magnetic field $h_{\hat{n}_a}$ is taken along the $(111)$-direction with a magnitude $h_{\hat{n}_a}=5\times 10^{-4}$.
ED numerics are performed on $L=18$ sites with periodic boundary conditions. 
} 
\label{fig:D3order}
\end{figure*}

\subsubsection{The ``LL4" phase}

The numerical studies on the Luttinger parameter reveal that the ``LL4" phase is rather narrow as shown in Fig. \ref{fig:LL_K}. 
While the phase transition line between the ``LL4" and ``FM" phases can be clearly identified at the far end of the ``LL4 peninsula",
it cannot be accurately determined close to the equator since the Luttinger liquid behaviors in the ``LL1" phase percolate into the ``FM" phase in finite size systems as discussed in Sec. \ref{sec:LL1_K_value}.
Thus the segment of the phase transition line between the ``LL4" and ``FM" phases in the vicinity of the equator is plotted as a dashed rather than solid line in Fig. \ref{fig:LL_K}. 
Here we note an interesting observation. 
According to the discussion in Sec. \ref{sec:LL4_LL_analysis} and Sec. \ref{sec:FM_from_LL4},
a rough estimation of the range of the ``LL4" phase at $\phi=0.25\pi$ is $0.51\pi<\theta<0.80\pi$.
However, as can be seen from Fig. \ref{fig:LL_K}, the actual range greatly shrinks compared with the above estimation.
An explanation of why high order terms have such a huge effect is worth further considerations.  

\subsubsection{The ``$D_3$-breaking I, II" phases}

Next we numerically study the spin ordering in the ``$D_3$-breaking I, II" phases in Fig. \ref{fig:phase}.
We emphasize that the six-sublattice rotated frame is taken, not the frame after the further two-sublattice transformation defined in Eq. (\ref{eq:rotate_6rot}). 

As discussed in Sec. \ref{sec:numerics_dspiral}, a small field has to be applied to test the spin orders.
We consider two types of fields: $h_{\text{I}}$ along $(-1,0,1)$-direction, and $h_{\text{II}}$ along $(1,-1,1)$-direction. 
If the system is in the ``$D_3$-breaking I" phase, then the field $h_{\text{I}}$ is able to polarize the system such that the spins are aligned according to the pattern in Eq. (\ref{eq:D3_order_6rot});
on the other hand, if the system is in the ``$D_3$-breaking II" phase, then an $h_{\text{II}}$-field will polarize the spins into the pattern in Eq. (\ref{eq:D3_order2_6rot}). 

However, we note a difficulty of this method.
In fact, the above method is not able to distinguish between the ``$D_3$-breaking I, II" phases.
Suppose the system is in the ``$D_3$-breaking I" phase and subject to an $h_{\text{II}}$-field,
then as can be seen from Fig. \ref{fig:Spins_6rot_D3I},
the energies of two red circles located at $(0,-1,1)$ and $(1,-1,0)$ 
are lowered by the largest amount among the six solid red circles.
Thus the system will end up with a linear combination of the two states represented by the red circles at $(0,-1,1)$ and $(1,-1,0)$.
We will demonstrate that the spin expectation values in the state of such linear combination will exhibit the same pattern as the one in Eq. (\ref{eq:D3_order2_6rot}).
Thus, the ``$D_3$-breaking I" phase will respond to the  $h_{\text{II}}$ field in the same way as the ``$D_3$-breaking II" phase.

To see the above point, recall that it is the unbroken symmetry group $\left<TR_II\right>$ of the state corresponding to the $(1,-1,1)$ vertex that 
determines the spin alignment pattern in Eq. (\ref{eq:D3_order2_6rot}). 
Denote $\Psi_{\vec{n}}$ to be the state corresponding to the solid circle located at position $\vec{n}$ in Fig. \ref{fig:Spins_6rot_D3I}.
Then it is straightforward to verify that 
\begin{eqnarray}
TR_II\Psi_{(0,-1,1)}&=&\Psi_{(1,-1,0)},\nn\\
TR_II\Psi_{(1,-1,0)}&=&\Psi_{(0,-1,1)}.
\end{eqnarray}
Thus the linear combination 
\begin{eqnarray}
\Psi_{\pm}=\frac{1}{\sqrt{2}}[\Psi_{(0,-1,1)}\pm\Psi_{(1,-1,0)}]
\label{eq:Psi}
\end{eqnarray}
is invariant under $TR_II$ up to an overall sign.
As a result, the spin expectation values in the state in Eq. (\ref{eq:Psi})
will exhibit exactly the same pattern as Eq. (\ref{eq:D3_order2_6rot}) (which applies to the ``$D_3$-breaking II" phase) even though the system is within the ``$D_3$-breaking I" phase.
Despite  such difficulty, we note that the applications of the $h_{\text{I}}$ and $h_{\text{II}}$ fields are still useful since they are able to test the existence of the ``$D_3$-breaking I, II" orders, although not able to distinguish between the two.

Fig. \ref{fig:D3order_A} shows the numerically measured expectation values of $S_j^\alpha$ ($\alpha=x,y,z$) under $h_{\text{I}}$-, $h_{\text{II}}$-fields (both equal to $10^{-4}$)  at a representative point $(\theta=0.52\pi,\phi=0.15\pi)$ within the ``$D_3$-breaking" phase in Fig. \ref{fig:phase}.
ED numerics are performed on L=18 system with periodic boundary conditions.
As can be seen from Fig. \ref{fig:D3order}, the patterns of $\langle S_j^\alpha \rangle$ are consistent with Eq. (\ref{eq:D3_order_6rot}) (Eq. (\ref{eq:D3_order2_6rot})) under the $h_{\text{I}}$- ($h_{\text{II}}$-) field shown as the black (red) data points.
The magnitudes of both spin orders are huge (about $10^3$ times larger than the applied fields),
indicating the existence of the two types of classical orders.

To determine in which phase the system resides, we further study the response of the system to a small field $h_{\hat{n}_a}$ along the $(111)$-direction.
As can be inferred from Fig. \ref{fig:Spins_6rot_D3I},
the ``$D_3$-breaking I" phase does not respond to $h_{\hat{n}_a}$, since the six solid red circles are perpendicular to the $(111)$-direction.
On the other hand,
 the ``$D_3$-breaking II" phase should respond, as the field $h_{\hat{n}_a}$ lowers the energies of the three states corresponding to the three solid blue circles located at $(1,1,-1)$, $(0,1,1)$ and $(1,-1,1)$.

Fig. \ref{fig:D3order} (a,b) shows the energy change $\Delta E= E(h_{\hat{n}_a})-E(h_{\hat{n}_a}=0)$ as a function of $h_{\hat{n}_a}$ at several representative points in the negative $J$ region.
Clearly, the system responds significantly at some of the points, whereas the response is nearly negligible at others. 
Based on the results in Fig. \ref{fig:D3order} (a,b), we conclude that the points 
$(\theta=0.52\pi,\phi=0.2\pi,0.24\pi)$ and $(\theta=0.55\pi,\phi=0.15\pi,0.2\pi,0.25\pi)$ are within the ``$D_3$-breaking I" phase,
whereas the points
$(\theta=0.52\pi,\phi=0.15\pi,0.3\pi)$ and $(\theta=0.55\pi,\phi=0.3\pi)$
are in the ``$D_3$-breaking II" phase.
Notice that the range of the ``$D_3$-breaking I" phase expands by increasing $\theta$,
which is consistent with the classical phase diagram as shown in Fig. \ref{fig:phase}.

As a further check, Fig. \ref{fig:D3order} (c,d) displays the response of $\left<S_j^x\right>$ to $h_{\hat{n}_a}=5\times 10^{-4}$ at several different points in the negative $J$ region.
As is clear from Fig. \ref{fig:D3order} (c,d),
the point $(\theta=0.52\pi,\phi=0.24\pi)$ nearly has no response to the field,
hence should locate within the ``$D_3$-breaking I" phase.
On the other hand, the response at the points $(\theta=0.52\pi, \phi=0.15\pi,0.3\pi)$
are huge, and they should be within the ``$D_3$-breaking II" phase.

\section{The ``FM" phase}
\label{sec:FM_phase}

\begin{figure}
\includegraphics[width=8cm]{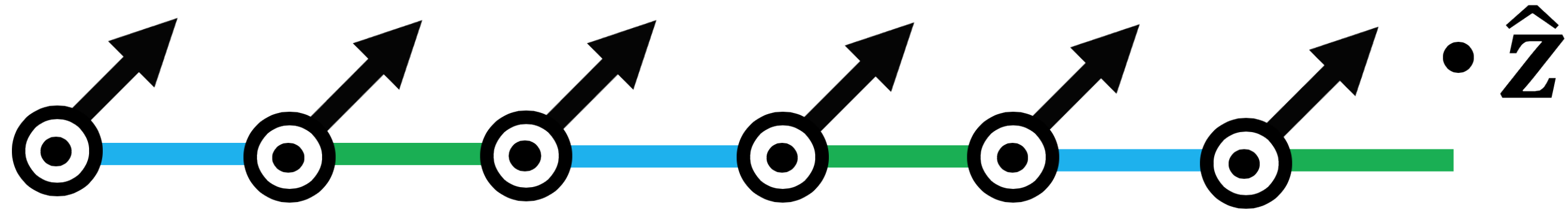}
\caption{Spin alignments in the FM phase within the original frame.
The $z$-direction in spin space is chose to be perpendicular to the plane,
and the $x$-direction is along the chain to the right.
} 
\label{fig:Spin_align_FM}
\end{figure}

In Sec. \ref{sec:RG_KG_H}, Sec. \ref{sec:sym_LL3} and Sec. \ref{sec:FM_from_LL4}, we have inferred the ``FM" phase based on 
RG analysis in three different regions, i.e. in the region close to the ``Emergent SU(2)$_1$" line on the equator of Fig. \ref{fig:phase}, the region close to the ``LL3" phase, and the region close to the ``LL4" phase.
The spin alignments in the ``FM" phase are shown to be
\begin{eqnarray}
\vec{S}_i=(a,a,b)^T.
\label{eq:FM_aab}
\end{eqnarray}
A plot of the spin ordering in Eq. (\ref{eq:FM_aab}) is displayed in Fig. \ref{fig:Spin_align_FM}.

\begin{figure*}[htbp]
\includegraphics[width=18cm]{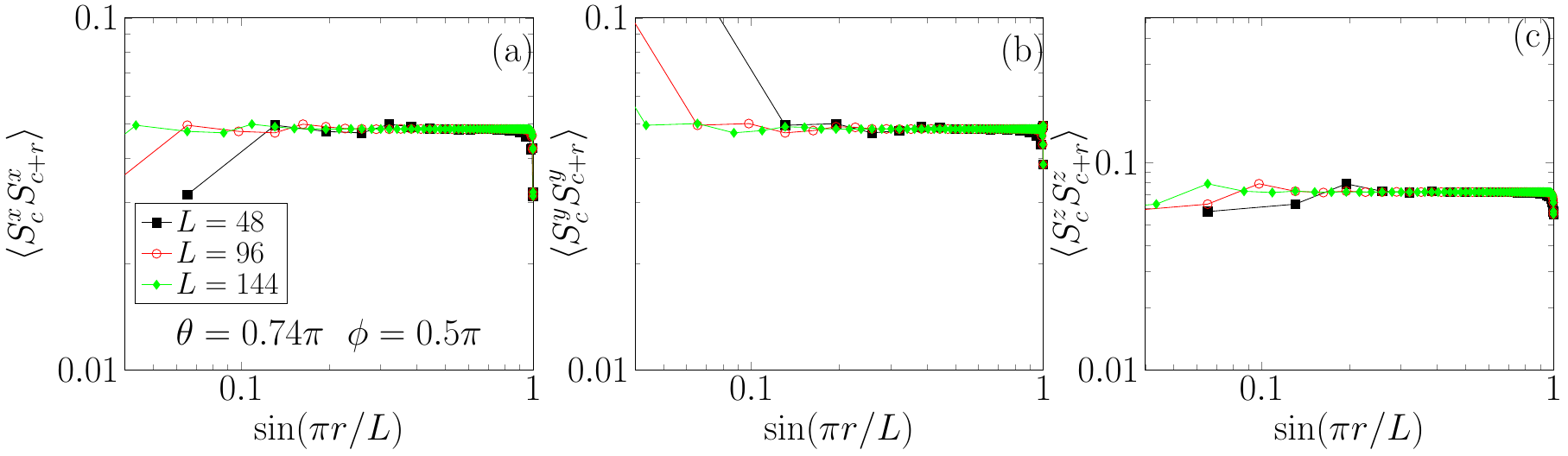}
\caption{(a) $\langle S_{1}^xS_{1+L}^x \rangle$, (b) $\langle S_{1}^yS_{1+L}^y \rangle$ and (c) $\langle S_{1}^zS_{1+r}^z \rangle$ vs $\sin(\pi r/L)$ at a representative point $(\theta=0.74\pi,\phi=0.5\pi)$ in the ``FM" phase.
DMRG numerics are performed on three system sizes $L=48,96,144$ with open boundary conditions.
} 
\label{fig:FM_corr}
\end{figure*}

Numerics have provided evidence for the FM order in Eq. (\ref{eq:FM_aab}).
Fig. \ref{fig:FM_corr} shows the correlation functions $\langle S_1^\alpha S_{1+r}^\alpha\rangle$ ($\alpha=x,y,z$) at a representative point $(\theta=0.74\pi,\phi=0.5\pi)$ in the ``FM" phase.
DMRG numerics are performed on three system sizes $L=48,96,144$ with open boundary conditions.
As can be checked from Fig. \ref{fig:FM_corr}, the numerical results are consistent with Eq. (\ref{eq:FM_aab}).
The extracted values of $a^2,b^2$ are $a^2\simeq 0.0484$, $b^2\simeq 0.0722$. 
We have checked several other points in the ``FM" phase, and they all exhibit an FM order given by Eq. (\ref{eq:FM_aab}).

\section{Conclusions}
\label{sec:summary}

In summary, we have studied  the phase diagram of the spin-1/2 Kitaev-Heisenberg-Gamma chain.
There are nine phases in total, including four Luttinger liquid phases, an FM phase, a N\'eel phase, a narrow ``d-Spiral" phase in which the spins align in a distorted-spiral pattern, and ``$D_3$-breaking I, II" phases.
Good agreements are reached between analytic and numerical calculations for all the nine phases,
though there is a narrow region close to the AFM Kitaev point on the equator where the nature of the phase diagram remains unclear and it is worth future investigations. 
Our comprehensive study of the phase diagram of the 1D generalized Kitaev model provides a road-map to the exotic physics in Kitaev materials. 

{\it Acknowledgments}
We thank H.-Y. Kee for interesting remarks and helpful discussions.
WY and IA acknowledge support from NSERC Discovery Grant 04033-2016.
AN acknowledges computational resources and services provided by Compute Canada and
Advanced Research Computing at the University of British Columbia.
AN is supported by the Canada First Research Excellence Fund.

\let\oldaddcontentsline\addcontentsline
\renewcommand{\addcontentsline}[3]{}




\let\addcontentsline\oldaddcontentsline


\begin{widetext}
\clearpage

\centerline{ {\Large \bf Supplementary Materials} }

\appendix

\section{The Hamiltonians in different frames}
\label{app:Ham}

In this section, we spell out the terms in the Hamiltonians in different frames. 
In general, we write the Hamiltonian $H$ as $H=\sum_{j=1}^L H_{j,j+1}$ where $H_{j,j+1}$ is the term on the bond between the sites $j$ and $j+1$.   
The forms of $H_{j,j+1}$ will be written explicitly. 

In the unrotated frame, the form of $H_{j,j+1}$ has a two-site periodicity. 
We have
\begin{eqnarray}
H_{2n+1,2n+2}&=&K S_{2n+1}^x S_{2n+2}^x +\Gamma (S_{2n+1}^y S_{2n+2}^z+S_{2n+1}^z S_{2n+2}^y)+J\vec{S}_{2n+1}\cdot \vec{S}_{2n+2}, \nn\\
H_{2n+2,2n+3}&=&K S_{2n+2}^y S_{2n+3}^y +\Gamma (S_{2n+2}^z S_{2n+3}^x+S_{2n+2}^x S_{2n+3}^z)+J\vec{S}_{2n+2}\cdot \vec{S}_{2n+3}.
\end{eqnarray}

In the six-sublattice rotated frame, the form of $H^\prime_{j,j+1}$ has a three-site periodicity. 
We have
\begin{eqnarray}
H^\prime_{3n+1,3n+2}&=&-KS_{3n+1}^xS_{3n+2}^x-\Gamma (S_{3n+1}^yS_{3n+2}^y+S_{3n+1}^zS_{3n+2}^z)-J(S_{3n+1}^xS_{3n+2}^x+S_{3n+1}^yS_{3n+2}^z+S_{3n+1}^zS_{3n+2}^y),\nn\\
H^\prime_{3n+2,3n+3}&=&-KS_{3n+2}^zS_{3n+3}^z-\Gamma (S_{3n+2}^xS_{3n+3}^x+S_{3n+2}^yS_{3n+3}^y)-J(S_{3n+2}^zS_{3n+3}^z+S_{3n+2}^xS_{3n+3}^y+S_{3n+2}^yS_{3n+3}^x),\nn\\
H^\prime_{3n+3,3n+4}&=&-KS_{3n+3}^yS_{3n+4}^y-\Gamma (S_{3n+3}^zS_{3n+4}^z+S_{3n+3}^xS_{3n+4}^x)-J(S_{3n+3}^yS_{3n+4}^y+S_{3n+3}^zS_{3n+4}^x+S_{3n+3}^xS_{3n+4}^z).\nn\\
\end{eqnarray}

In the four-sublattice rotated frame, the form of $H^{\prime\prime}_{j,j+1}$ has a four-site periodicity. 
We have
\begin{eqnarray}
H^{\prime\prime}_{4n+1,4n+2}&=& (K+2J)S_{4n+1}^xS_{4n+2}^x-J\vec{S}_{4n+1}\cdot\vec{S}_{4n+2}+\Gamma (S^y_{4n+1}S^z_{4n+2}+S^z_{4n+1}S^y_{4n+2}), \nn\\
H^{\prime\prime}_{4n+2,4n+3}&=& (K+2J)S_{4n+2}^yS_{4n+3}^y-J\vec{S}_{4n+2}\cdot\vec{S}_{4n+3}+\Gamma (S^z_{4n+2}S^x_{4n+3}+S^x_{4n+2}S^z_{4n+3}), \nn\\
H^{\prime\prime}_{4n+3,4n+4}&=& (K+2J)S_{4n+3}^xS_{4n+4}^x-J\vec{S}_{4n+3}\cdot\vec{S}_{4n+4}-\Gamma (S^y_{4n+3}S^z_{4n+4}+S^z_{4n+3}S^y_{4n+4}), \nn\\
H^{\prime\prime}_{4n+4,4n+5}&=& (K+2J)S_{4n+4}^yS_{4n+5}^y-J\vec{S}_{4n+4}\cdot\vec{S}_{4n+5}-\Gamma (S^z_{4n+4}S^x_{4n+5}+S^x_{4n+4}S^z_{4n+5}).
\end{eqnarray}

In the literature, another $\Gamma^\prime$-term sometimes is also considered 
\cite{Rau2014}, 
which is defined as the following in the original frame
\begin{eqnarray}
H_{\Gamma^\prime}=\sum_{<ij>\in\gamma\,\text{bond}}\Gamma^\prime (S_i^\alpha S_j^\gamma+S_i^\gamma S_j^\alpha+S_i^\beta S_j^\gamma+S_i^\gamma S_j^\beta).
\label{eq:gammaprime}
\end{eqnarray}
Since $\Gamma^\prime$ is much smaller than the other three couplings 
\cite{Rau2014}, 
the  $\Gamma^\prime$ term is neglected in this work.

\section{Transformation rules of the SU(2)$_1$ WZW fields}
\label{app:transform_WZW}

In this section, we work out the symmetry transformations of the WZW primary field and its descendent fields.
The strategy is to first work within the fermion representation, 
then use the nonabelian bosonization formula to translate them into the WZW language.

We first summarize the transformation properties of the primary and descendent fields in the SU(2)$_1$ WZW theory.
The derivations are left for the subsequent subsections.
The scaling fields in the SU(2)$_1$ WZW theory are known to be either $(J_L^{\alpha_1}...J_L^{\alpha_n})(w)(J_R^{\beta_1}...J_R^{\beta_n})(\bar{w})$ or $(J_L^{\alpha_1}...J_L^{\alpha_n}J_R^{\beta_1}...J_R^{\beta_n}g)(w,\bar{w})$   \cite{DiFrancesco1997},
in which $(AB)(w)$ in the holomorphic sector is the $O((z-w)^0)$ term in the operator product expansion (OPE) of $A(z)B(w)$ as a Laurent series in $z-w$ around the point $w$,
and definitions are similar for the anti-holomorphic sector.
The transformation laws of $g$ and $\vec{J_L},\vec{J}_R$ under time reversal $T$, spatial translation $T_a$, inversion $I$ and spin rotation $R\in SO(3)$ are summarized as
\begin{eqnarray}
T: &\epsilon(x)\rightarrow \epsilon(x), &\vec{N}(x)\rightarrow -\vec{N}(x),\nn\\
&\vec{J}_L(x)\rightarrow -\vec{J}_R(x), &\vec{J}_R(x)\rightarrow -\vec{J}_L(x), 
\label{eq:transformT}
\end{eqnarray}
\begin{eqnarray}
T_a: &\epsilon(x)\rightarrow -\epsilon(x), &\vec{N}(x)\rightarrow -\vec{N}(x),\nn\\
&\vec{J}_L(x)\rightarrow \vec{J}_L(x), &\vec{J}_R(x)\rightarrow \vec{J}_R(x), 
\label{eq:transformTa}
\end{eqnarray}
\begin{eqnarray}
I: & \epsilon(x)\rightarrow -\epsilon(-x), &\vec{N}(x)\rightarrow \vec{N}(-x),\nn\\
&\vec{J}_L(x)\rightarrow \vec{J}_R(-x), &\vec{J}_R(x)\rightarrow \vec{J}_L(-x), 
\label{eq:transformI}
\end{eqnarray}
\begin{eqnarray}
R: &\epsilon(x)\rightarrow \epsilon(x), &N^\alpha(x)\rightarrow R^{\alpha}_{\,\,\beta}N^\beta(x),\nn\\
&J^\alpha_L(x)\rightarrow R^{\alpha}_{\,\,\beta} J^\beta_L(x), &J^\alpha_R(x)\rightarrow R^{\alpha}_{\,\,\beta}J^\beta_R(x), 
\label{eq:transformR}
\end{eqnarray}
in which $x$ is the spatial coordinate; $R^{\alpha}_{\,\,\beta}$ ($\alpha,\beta=x,y,z$) is the matrix element of the $3\times 3$ rotation matrix $R$;
$\epsilon(x)=\text{tr}g(x)$ is the dimer order parameter; and $\vec{N}(x)=i\text{tr}(g(x)\vec{\sigma})$ is the N\'eel order parameter \cite{Affleck1988}.

\subsection{Translation by one lattice site}

In fermion language,
\begin{eqnarray}
\psi(x)=e^{ik_F x} \psi_L(x)+e^{-ik_F x} \psi_R(x),
\end{eqnarray}
and
\begin{eqnarray}
T_a \psi(x) T_a^{-1} &=&\psi(x+a) =ie^{ik_Fx} \psi_L(x+a)-i e^{-ik_Fx} \psi_R(x+a),
\end{eqnarray}
where $\psi_\lambda=(\psi_{\lambda\uparrow},\psi_{\lambda\downarrow})^T$ ($\lambda=L,R$).
Comparing 
\begin{eqnarray}
T_a \psi(x) T_a^{-1}=e^{ik_Fx}T_a \psi_L(x) T_a^{-1} +e^{-ik_Fx}T_a \psi_R(x) T_a^{-1} 
\end{eqnarray}
with the above expressions, we obtain
\begin{eqnarray}
T_a \psi_L(x) T_a^{-1}&=&i\psi_L(x+a),\nn\\
T_a \psi_R(x) T_a^{-1}&=&-i\psi_R(x+a).
\end{eqnarray}
The bosonization rule is  \cite{Affleck1988}
\begin{eqnarray}
\psi_{L\alpha} (x)\psi^\dagger_{R\beta} (x) \sim g_{\alpha\beta} (x) e^{i\sqrt{2\pi}\phi(x)},
\label{eq:nonabelian_bosonization}
\end{eqnarray}
in which $\alpha,\beta=\uparrow,\downarrow$,
and $\phi(x)$ is the charge boson.
This gives
\begin{eqnarray}
T_a g(x) T_a^{-1} =-g(x+a).
\end{eqnarray}

\subsection{Spatial inversion}

In fermion language,
\begin{eqnarray}
I \psi(x)I &=&\psi(-x)=e^{ik_Fx} \psi_R(-x)+e^{-ik_Fx} \psi_L(-x).
\end{eqnarray}
Hence
\begin{eqnarray}
I\psi_L(x)I^{-1}&=&\psi_R(-x),\nn\\
I\psi_R(x)I^{-1}&=&\psi_L-x).
\end{eqnarray}
From this, we obtain,
\begin{eqnarray}
I\psi_{L\alpha}(x) \psi_{R\beta}^\dagger(x)I^{-1}&=&\psi_{R\alpha}(-x) \psi_{L\beta}^\dagger(-x)=[ \psi_{L\beta}(-x) \psi^\dagger_{R\alpha}(-x)]^\dagger.
\end{eqnarray}
Using the bosonization formula, we have
\begin{eqnarray}
I g_{\alpha\beta} (x) e^{-i\sqrt{2\pi} \phi(x)} I^{-1} &=& g^*_{\beta\alpha} (-x) e^{i\sqrt{2\pi} \phi(-x)}= g^{-1}_{\alpha\beta} (-x) e^{i\sqrt{2\pi} \phi(-x)}.
\end{eqnarray}
When the charge boson is gapped out, $e^{-i\sqrt{2\pi} \phi(x)}$ is purely imaginary,
hence $e^{i\sqrt{2\pi} \phi(x)}=-e^{-i\sqrt{2\pi} \phi(x)}$.
This yields
\begin{eqnarray}
I g(x) I^{-1} = -g^{-1}(-x).
\end{eqnarray}

\subsection{Time reversal}

In fermion language,
\begin{eqnarray}
T\psi(x)T^{-1} = e^{-ik_Fx} [\psi_L^\dagger(x)i\sigma_2]^\dagger+e^{ik_Fx} [\psi_R^\dagger(x)i\sigma_2]^\dagger,
\end{eqnarray}
hence
\begin{eqnarray}
T\psi_L(x)T^{-1} &=&-i\sigma_2 \psi_R(x)\nn\\
T\psi_R(x)T^{-1} &=&-i\sigma_2 \psi_L(x).
\end{eqnarray}
Then
\begin{eqnarray}
T\psi_L\psi_R^\dagger T^{-1}= -i\sigma_2 \psi_R\psi_L^\dagger i\sigma_2.
\end{eqnarray}
This gives
\begin{eqnarray}
T g(x) e^{-i\sqrt{2\pi} \phi(x)}T^{-1} =-i\sigma_2 g^\dagger(x) e^{i\sqrt{2\pi}\phi(x)} i\sigma_2.
\end{eqnarray}
By canceling the charge boson, we obtain
\begin{eqnarray}
Tg(x)T^{-1}=-i\sigma_2 g^\dagger(x)i\sigma_2.
\end{eqnarray}

\subsection{Spin rotation}

The transformation law under spin rotation $R$ is
\begin{eqnarray}
Rg(x)R^{-1}=U(R) g(x)U^{-1}(R),
\end{eqnarray}
in which the $2\times 2$ matrix $U$ is the SU(2) representation of the rotation $R$.\\

In summary, the transformation rules are
\begin{eqnarray}
T_a&:&  g(x)\rightarrow -g(x+a)\nn\\
I&:&  g(x)\rightarrow -g^{-1}(x)\nn\\
T&:& g(x)\rightarrow  -i\sigma_2 g^{-1}(x)i\sigma_2\nn\\
R&:& g(x)\rightarrow  U(R) g(x) U^{-1}(R).
\end{eqnarray}

Sometimes, it is more convenient to use the following combinations of $g$
\begin{eqnarray}
\epsilon=\text{tr} (g), \vec{N}=i\text{tr} (g\vec{\sigma}).
\end{eqnarray}
(Note: the insertion of $i$ is to make $\vec{N}$ hermitian.)
Since $g=\cos(\frac{\theta}{2})I_2+i\sin(\frac{\theta}{2})\vec{\sigma}\cdot \hat{n}$
where $I_2$ is the $2\times 2$ identity matrix,
we obtain
\begin{eqnarray}
\text{tr} (g\vec{\sigma}) = -\text{tr} (g^{-1}\vec{\sigma}),\,\text{tr} (g)=\text{tr} (g^{-1}).
\label{eq:tr_relation}
\end{eqnarray}
Using Eq. (\ref{eq:tr_relation}), we arrive at Eqs. (\ref{eq:transformT},\ref{eq:transformTa},\ref{eq:transformI},\ref{eq:transformR}).
We note that these transformation laws are consistent with the physical meaning of $\epsilon$ and $\vec{N}$ as dimer and N\'eel order fields.

\section{Redundancies in the descendent fields in the SU(2)$_1$ WZW model}
\label{app:quadratic_WZW}

In this section, we show that the existence of null fields in the SU(2)$_1$ WZW model leads to redundancies in the descendent fields.
We focus on several cases that are relevant for our purposes,
including the quadratic WZW current terms,
and the dimension-3/2 fields $J_L^\alpha N^\beta+J_L^\beta N^\alpha$ where $\alpha\neq \beta$.
The holomorphic sector is taken as an example,
and the antiholomorphic sector can be treated in a similar manner.


First let's consider the quadratic WZW current terms.
We compute $(J_L^a J_L^b)$ by working out the $O(1)$ terms in the OPE $J_L^a (x+a)J_L^b(x)$.
In what follows, all higher order terms in $a$ will be neglected.

The free fermion contraction rule is \cite{DiFrancesco1997}
\begin{eqnarray}
\left< \psi_L^\dagger(z) \psi_L(w)  \right>=-\left< \psi_L(w) \psi^\dagger_L(z)  \right>=\frac{1}{z-w}.
\end{eqnarray}
The OPE $J_L^a (x+a)J_L^b(x)$ can be calculated as
\begin{flalign}
&J_L^a (x+a)J_L^b(x) \nn\\
&= \frac{1}{4} \sigma^a_{\alpha\beta} \sigma^b_{\gamma\delta} \psi_{L\alpha}^\dagger(x+a) \psi_{L\beta} (x+a) \psi^\dagger_{L\gamma} (x) \psi_{L\delta} (x)\nn\\
&= \frac{1}{4} \sigma^a_{\alpha\beta} \sigma^b_{\gamma\delta}
\big( 
:\psi^\dagger_{L\alpha} (x+a) \psi_{L\beta}(x+a)\psi^\dagger_{L\gamma}(x)\psi_{L\delta}(x) :
+\text{$
\contraction{:\psi^\dagger_{L\alpha} (x+a)}{\psi}{{}_{L\beta}(x+a)}{\psi}
:\psi^\dagger_{L\alpha} (x+a) \psi_{L\beta}(x+a)\psi^\dagger_{L\gamma}(x)\psi_{L\delta}(x) :
$}\nn\\
&~~~~~ +\text{$
\contraction{:}{\psi}{{}^\dagger_{L\alpha} (x+a) \psi_{L\beta}(x+a)\psi^\dagger_{L\gamma}(x) }{\psi}
:\psi^\dagger_{L\alpha} (x+a) \psi_{L\beta}(x+a)\psi^\dagger_{L\gamma}(x)\psi_{L\delta}(x) :
$}
 +\text{$
\contraction[2ex]{:}{\psi}{{}^\dagger_{L\alpha} (x+a) \psi_{L\beta}(x+a)\psi^\dagger_{R\gamma}(x) }{\psi}
\contraction{:\psi^\dagger_{L\alpha} (x+a)}{\psi}{{}_{L\beta}(x+a)}{\psi}
:\psi^\dagger_{L\alpha} (x+a) \psi_{L\beta}(x+a)\psi^\dagger_{L\gamma}(x)\psi_{L\delta}(x) :
$}
\big)\nn\\
&=\frac{1}{4} \big(\sigma^a_{\alpha\beta} \sigma^b_{\gamma\delta}
:\psi^\dagger_{L\alpha}  \psi_{L\beta}\psi^\dagger_{L\gamma}\psi_{L\delta} :
+\frac{1}{a}  :\psi_L^\dagger(x+a) (\delta_{ab}+\epsilon^{abc} \sigma^c) \psi_L(x):
-\frac{1}{a} :\psi_L^\dagger(x) (\delta_{ab}-\epsilon^{abc} \sigma^c) \psi_L(x+a)  :
-\delta_{ab}\frac{1}{2a^2}
\big).
\label{eq:Quadr_f}
\end{flalign}

If $a=b$, then Eq. (\ref{eq:Quadr_f}) gives
\begin{flalign}
J_L^a (x+a)J_L^a(x) &= 
\frac{1}{4} \big(\sigma^a_{\alpha\beta} \sigma^{a}_{-\alpha,-\beta}
:\psi^\dagger_{L\alpha}  \psi_{L\beta}\psi^\dagger_{L\,-\alpha}\psi_{L,-\beta} :
+  :\partial_x\psi_L^\dagger\cdot \psi_L: - :\psi_L^\dagger\partial_x \psi_L :
\big)\nn\\
&=-\frac{1}{2} :n_{L\uparrow} n_{L\downarrow}: +\frac{1}{4} (:\partial_x\psi_L^\dagger\cdot \psi_L: - :\psi_L^\dagger\partial_x \psi_L :)-\frac{1}{2a^2},
\label{eq:JaJa}
\end{flalign}
where $n_{L\alpha}=:\psi^\dagger_{L\alpha}\psi_{L\alpha}:$ ($\alpha=\uparrow,\downarrow$).
Hence all the three $(J_L^a J_L^a)$ ($a=x,y,z$) are equal given by Eq. (\ref{eq:JaJa}).
As a result, $(J_L^a J_L^a)=\frac{1}{3}(\vec{J}_L\cdot \vec{J}_L)$. 

If $a\neq b$, the first term in the last line of Eq. (\ref{eq:Quadr_f}) vanishes.
Taking the $O(1)$ part of Eq. (\ref{eq:Quadr_f}), we obtain
\begin{flalign}
(J_L^a J_L^b)=\epsilon^{abc} \partial_x :\psi_L^\dagger \sigma^c \psi_L:=\epsilon^{abc}\partial_x J_L^c.
\end{flalign}
Using $z=\tau+ix$, we also have
\begin{eqnarray}
(J_L^a J_L^b)=i\epsilon^{abc}\partial_z J_L^c.
\end{eqnarray}

Next we consider the dimension-3/2 fields $J_L^c N^d+J_L^d N^c$ ($c\neq d$).
Due to the nonabelian bosonization formula Eq. (\ref{eq:nonabelian_bosonization}),
we consider the $O(1)$ part of the fermionic OPE 
\begin{eqnarray}
:\psi_{L\alpha}^\dagger(x+a) \sigma^c_{\alpha\beta} \psi_{L\beta}(x+a)::\psi_{R\gamma}^\dagger(x) (-i)\sigma^d_{\gamma\delta} \psi_{L\delta}(x):+
:\psi_{L\alpha}^\dagger(x+a) \sigma^d_{\alpha\beta} \psi_{L\beta}(x+a)::\psi_{R\gamma}^\dagger(x) (-i)\sigma^c_{\gamma\delta} \psi_{L\delta}(x):,
\label{eq:JN_NJ}
\end{eqnarray}
which differs from the $J_L^c N^d+J_L^d N^c$ only by an overall factor $e^{i\sqrt{2\pi}\phi}$ in the charge sector.
Using the Wick theorem and the fermionic contraction rules, Eq. (\ref{eq:JN_NJ}) is equal to
\begin{flalign}
&(-i)\big[:\psi_{L\alpha}^\dagger(x+a) \sigma^c_{\alpha\beta} \psi_{L\beta}(x+a)\psi_{R\gamma}^\dagger(x) \sigma^d_{\gamma\delta} \psi_{L\delta}(x):+
:\psi_{L\alpha}^\dagger(x+a) \sigma^d_{\alpha\beta} \psi_{L\beta}(x+a)\psi_{R\gamma}^\dagger(x) \sigma^c_{\gamma\delta} \psi_{L\delta}(x):\nn\\
&+
\text{$
\contraction{:}{\psi}{{}_{L\alpha}^\dagger(x+a) \sigma^c_{\alpha\beta} \psi_{L\beta}(x+a)\psi_{R\gamma}^\dagger(x) \sigma^d_{\gamma\delta}}{\psi}
:\psi_{L\alpha}^\dagger(x+a) \sigma^c_{\alpha\beta} \psi_{L\beta}(x+a)\psi_{R\gamma}^\dagger(x) \sigma^d_{\gamma\delta} \psi_{L\delta}(x):
$}
+\text{$
\contraction{:}{\psi}{{}_{L\alpha}^\dagger(x+a) \sigma^d_{\alpha\beta} \psi_{L\beta}(x+a)\psi_{R\gamma}^\dagger(x) \sigma^c_{\gamma\delta}}{\psi}
:\psi_{L\alpha}^\dagger(x+a) \sigma^d_{\alpha\beta} \psi_{L\beta}(x+a)\psi_{R\gamma}^\dagger(x) \sigma^c_{\gamma\delta} \psi_{L\delta}(x):
$} \big].
\end{flalign}
If we only keep the O(1) terms, then it becomes
\begin{flalign}
&(-i)\big[\sigma^c_{\alpha\beta} \sigma^d_{\gamma\delta}
 :\psi_{L\alpha}^\dagger(x)  \psi_{L\beta}(x)\psi_{R\gamma}^\dagger(x) \psi_{L\delta}(x):
 +\frac{1}{2\pi i} \epsilon^{dce} \sigma^e_{\gamma\beta} :\partial_x\psi_{L\beta}(x)\cdot \psi_{R\gamma}^\dagger(x):\nn\\
 &+\sigma^d_{\alpha\beta} \sigma^c_{\gamma\delta}
 :\psi_{L\alpha}^\dagger(x)  \psi_{L\beta}(x)\psi_{R\gamma}^\dagger(x) \psi_{L\delta}(x):
 +\frac{1}{2\pi i} \epsilon^{cde} \sigma^e_{\gamma\beta} :\partial_x\psi_{L\beta}(x)\cdot \psi_{R\gamma}^\dagger(x):\big]\nn\\
 =&(-i) (\sigma^c_{\alpha\beta} \sigma^d_{\gamma\delta}+\sigma^d_{\alpha\beta} \sigma^c_{\gamma\delta}) 
 :\psi_{L\alpha}^\dagger(x)  \psi_{L\beta}(x)\psi_{R\gamma}^\dagger(x) \psi_{L\delta}(x):.
 \label{eq:JN_NJ_b}
\end{flalign}
Note that due to Fermi statistics, we must have $\beta\neq \delta$.

For our purpose, let's take $c=x,d=y$ in accordance with Eq. (\ref{eq:dim_3o2_LL3}).
Then the coefficient in Eq. (\ref{eq:JN_NJ_b}) becomes
\begin{eqnarray}
\sigma^x_{-\beta,\beta} \sigma^y_{\beta,-\beta}+\sigma^y_{-\beta,\beta} \sigma^x_{\beta,-\beta}
\end{eqnarray}
since both $\sigma^x$ and $\sigma^y$ are off-diagonal.
Take $\beta=\uparrow$ as an example.
Then it is clear that the coefficient vanishes.
The same conclusion holds for $\beta=\downarrow$, and for the antiholomorphic term $J_R^c N^d+J_R^d N^c$ where $c=x,d=y$.


\section{Operator algebra of the SU(2)$_1$ WZW model}
\label{app:derivation_OPE}

\subsection{The SU(2)$_1$ affine algebra}
In this subsection, we fix the normalization conventions for the SU(2)$_1$ affine algebra.
The commutation relations of the affine generators can be obtained from 
the Ward identities \cite{DiFrancesco1997}.
Alternatively, they can be determined from the 1D Dirac fermions, which we briefly review in this subsection.


We work with radial ordering \cite{DiFrancesco1997}.
The OPE of the free fermion fields are
\begin{eqnarray}
\left< \psi_L^\dagger(z) \psi_L(w)  \right>=-\left< \psi_L(w) \psi^\dagger_L(z)  \right>=\frac{1}{z-w},\nn\\
\left< \psi_R^\dagger(\bar{z}) \psi_R(\bar{w})  \right>=-\left< \psi_R(\bar{w}) \psi^\dagger_R(\bar{z})  \right>=\frac{1}{\bar{z}-\bar{w}}.
\label{eq:contraction_f}
\end{eqnarray}
The WZW current operators are
\begin{eqnarray}
J^a&=&:\psi_L^\dagger \frac{1}{2}\sigma^a \psi_L:\nn\\
\bar{J}^a&=&:\psi_R^\dagger \frac{1}{2}\sigma^a \psi_R:,
\label{eq:fermion_JLJR}
\end{eqnarray}
in which $a=x,y,z$.
The anomalous constant term of the OPE $J^a(z)J^b(w)$ (similar for $\bar{J}$)
can be obtained from the current-current correlation function \cite{Gogolin1998}.
The result of the OPE is
\begin{eqnarray}
J^a(z)J^b(w) \sim \frac{\delta_{ab}}{2(z-w)^2}+\sum_c \frac{i\epsilon_{abc}J^c(w)}{z-w},
\label{eq:JOPE}
\end{eqnarray}
in which $\epsilon_{abc}$ is the rank-3 totally antisymmetric tensor. 

Define the affine generator as
\begin{eqnarray}
J_n^a=\frac{1}{2\pi i} \oint dz z^n J^a(z).
\end{eqnarray}
It is straightforward to obtain the following commutation relations from Eq. (\ref{eq:JOPE}) \cite{DiFrancesco1997},
\begin{eqnarray}
[J_n^a,J_m^b]=\sum_c i\epsilon_{abc} J_{n+m}^c+\frac{1}{2} n\delta_{ab} \delta_{n+m,0}.
\label{eq:affine_algebra}
\end{eqnarray}
Note the $1/2$ factor in the anomalous term in Eq. (\ref{eq:affine_algebra}), which is a consequence of the choice of normalization in Eq. (\ref{eq:fermion_JLJR}).

\subsection{Operator algebra of the SU(2)$_1$ WZW model}

In this part, we use the affine symmetry to determine the OPE $g_{\alpha\beta}(z,\bar{z}) g_{\gamma \delta}(w,\bar{w})$  up to quadratic order of the current operators.
Throughout this section, repeated indices indicate summation unless otherwise stated.

First, let's figure out what kinds of primary fields appear in the OPE.
Notice that $\mathbbm{1}$ and $g$ are the only two primary fields in the SU(2)$_1$ WZW model.
Thus the components of $g_{\alpha\beta}(z,\bar{z}) g_{\gamma \delta}(w,\bar{w})$ on these 
two primary fields are given by the vacuum expectation values $\langle g_{\alpha\beta}(z,\bar{z}) g_{\gamma \delta}(w,\bar{w})\rangle$
and $\langle  g_{\alpha\beta}(z,\bar{z}) g_{\gamma \delta}(w,\bar{w}) g^{-1}_{\lambda\mu}(0,0) \rangle$, respectively.
However, the three-point function $\langle  g_{\alpha\beta} g_{\gamma \delta} g^{-1}_{\lambda\mu} \rangle$ vanishes due to the chiral rotation symmetry.
In fact, consider the holomorphic sector,
then $\langle g_{\alpha\beta} g_{\gamma \delta} g^{-1}_{\lambda\mu} \rangle$
is spin-1/2 $\otimes$ spin-1/2 $\otimes$ spin-1/2,
which does not contain an SU(2) singlet.
On the other hand, the vacuum is SU(2) invariant, so the expectation value is zero.
As a result, the OPE $g_{\alpha\beta}(z,\bar{z}) g_{\gamma \delta}(w,\bar{w})$ only contains the primary field $\mathbbm{1}$.

In the case of the Virasoro algebra, once the three-point functions of the primary fields are known, the full OPE can be obtained by using the conformal symmetry and the Virasoro algebra \cite{DiFrancesco1997}. 
This is also true for the affine algebra.
Thus we conclude that the OPE $g_{\alpha\beta}(z,\bar{z}) g_{\gamma \delta}(w,\bar{w})$
has no components on all affine descendent fields of the primary field g.
Hence it only contains descendent fields of $\mathbbm{1}$,
which means that  $g_{\alpha\beta}(z,\bar{z}) g_{\gamma \delta}(w,\bar{w})$ can be written in terms of affine current operators.

We write 
\begin{eqnarray}
g_{\alpha\beta}(z,\bar{z}) g_{\gamma \delta}(0,0)=(\sum_{\{K\}} \beta^{\{K\}}_{\alpha\gamma} z^{-\frac{1}{2}+K})\cdot (\sum_{\{K\}} \bar{\beta}^{\{\bar{K}\}}_{\alpha\gamma} \bar{z}^{-\frac{1}{2}+\bar{K}}) \mathbbm{1}^{\{K\},\{\bar{K}\}},
\end{eqnarray}
in which 
\begin{eqnarray}
\mathbbm{1}^{K}=(J_{-n_1}^{a_1}...J_{-n_k}^{a_k}\mathbbm{1}),~~~
\mathbbm{1}^{\bar{K}}=(\bar{J}_{-\bar{n}_1}^{\bar{a}_1}...\bar{J}_{-\bar{n}_k}^{\bar{a}_k}\mathbbm{1}),
\label{eq:1_decendents}
\end{eqnarray}
\begin{eqnarray}
\{K\}=\left\{ \begin{array}{cccc}
n_1&n_2&...&n_k\\
a_1&a_2&...&a_k
\end{array}\right\},~~~
\{\bar{K}\}=\left\{ \begin{array}{cccc}
\bar{n}_1&\bar{n}_2&...&\bar{n}_k\\
\bar{a}_1&\bar{a}_2&...&\bar{a}_k
\end{array}\right\},
\end{eqnarray}
\begin{eqnarray}
K=\sum_{j=1}^k n_j, \,\bar{K}=\sum_{j=1}^k \bar{n}_j,
\end{eqnarray}
and for some field $\phi$,
$(J_{-n}^a\phi) (z)$ is defined as
\begin{eqnarray}
(J_{-n}^a\phi) (z)=\frac{1}{2\pi i} \oint _z dw (w-z)^n J^a(w)\phi(z).
\end{eqnarray}
In Eq. (\ref{eq:1_decendents}), $\phi(z)$ is taken to be the identity field $\mathbbm{1}$.

Acting with the OPE on the vacuum $\ket{\Omega}$, we have
\begin{eqnarray}
g_{\alpha\beta} (z,\bar{z}) \ket{g_{\gamma\delta}} = 
\frac{1}{z^{1/2} \bar{z}^{1/2}} (\sum_{\{K\}} \beta^{\{K\}}_{\alpha\gamma} z^{-\frac{1}{2}+K})\cdot (\sum_{\{K\}} \bar{\beta}^{\{\bar{K}\}}_{\alpha\gamma} \bar{z}^{-\frac{1}{2}+\bar{K}}) \mathbbm{1}^{\{K\},\{\bar{K}\}} \ket{\Omega},
\end{eqnarray}
in which $\ket{g_{\gamma\delta}}>=g_{\gamma\delta}(0,0)\ket{\Omega}$.
In what follows, we only keep the holomorphic part.
The full expression is a product of the holomorphic part and the anti-holomorphic part.

Define
\begin{eqnarray}
\ket{N}_{\alpha\gamma}=\sum_{K+N} \beta^{K}_{\alpha\gamma} \mathbbm{1}^{\{K\}}\ket{\Omega},
\end{eqnarray}
then we have
\begin{eqnarray}
g_{\alpha\beta}(z,\bar{z}) \ket{g_{\gamma\delta}} = \frac{1}{z^{1/2}} \sum_N \ket{N}_{\alpha\gamma}.
\end{eqnarray}
We are going to show that all the coefficients of $\beta^{K}_{\alpha\gamma}$ 
can be determined from $\beta^{0}_{\alpha\gamma}=\beta_{\alpha\gamma}$.

We now act $J_n^a$ on $g_{\alpha\beta} (z,\bar{z}) \ket{g_{\gamma\delta}}$ with $n>0$.
Then
\begin{eqnarray}
J_n^a g_{\alpha\beta} (z,\bar{z}) \ket{g_{\gamma\delta}} = \frac{1}{z^{1/2}} \sum_N z^N J_n^a \ket{N}_{\alpha\gamma}.
\end{eqnarray}
On the other hand, since $J_n^a$ ($n>0$) annihilates $\ket{g_{\gamma\delta}}$ \cite{DiFrancesco1997},
we obtain
\begin{eqnarray}
J_n^a g_{\alpha\beta} (z,\bar{z}) \ket{g_{\gamma\delta}} = [J_n^a,g_{\alpha\beta} (z,\bar{z})]\ket{g_{\gamma\delta}}.
\label{eq:Jna}
\end{eqnarray}
Using the OPE 
\begin{eqnarray}
J^a(w) g_{\alpha\beta} (z,\bar{z}) \sim (-)\frac{\frac{1}{2} \sigma^a_{\alpha\alpha^\prime} g_{\alpha^\prime\beta}(z,\bar{z})}{w-z},
\end{eqnarray}
the commutator can be calculated as
\begin{eqnarray}
[J_n^a,g_{\alpha\beta} (z,\bar{z})]&=&\frac{1}{2\pi i} \oint_{|w|=|z|+0} dw w^n J^a(w)g_{\alpha\beta} (z,\bar{z})-\frac{1}{2\pi i} \oint_{|w|=|z|-0} dw w^n J^a(w)g_{\alpha\beta} (z,\bar{z})\nn\\
&=&\frac{1}{2\pi i} \oint_{z} dw w^n J^a(w)g_{\alpha\beta} (z,\bar{z})\nn\\
&=& -z^n (\frac{1}{2}\sigma^a)_{\alpha\alpha^\prime}g_{\alpha^\prime\beta}(z,\bar{z}).
\end{eqnarray}
Hence
\begin{eqnarray}
[J_n^a,g_{\alpha\beta}(z,\bar{z})] \ket{g_{\gamma\delta}} &=& -z^n (\frac{1}{2} \sigma^a)_{\alpha\alpha^\prime} g_{\alpha^\prime\beta} (z,\bar{z}) \ket{g_{\gamma\delta}}\nn\\
&=&-\frac{1}{z^{1/2}} \sum_N z^{N+n} (\frac{1}{2}\sigma^a)_{\alpha\alpha^\prime}\ket{N}_{\alpha^\prime\gamma}.\nn\\
\end{eqnarray}

Then Eq. (\ref{eq:Jna}) gives
\begin{eqnarray}
 \frac{1}{z^{1/2}} \sum_N z^{N+n} J_n^a \ket{N+n}_{\alpha\gamma} =
\frac{1}{z^{1/2}}   \sum_N z^{N+n}(-\frac{1}{2}\sigma^a)_{\alpha\alpha^\prime} \ket{N}_{\alpha^\prime\gamma}.
\end{eqnarray}
Thus we obtain
\begin{eqnarray}
J_n^a\ket{N+n}_{\alpha\gamma} = -(\frac{1}{2}\sigma^a)_{\alpha\alpha^\prime} \ket{N}_{\alpha^\prime\gamma}.
\label{eq:key_eq}
\end{eqnarray}
Eq. (\ref{eq:key_eq}) determines all $\beta^{\{K\}}_{\alpha\gamma}$ from $\beta_{\alpha\gamma}$.
Recall that $\beta_{\alpha\gamma}=\epsilon_{\alpha\gamma}$ which is easily obtained from the CG coefficients to form an SU(2) singlet.

We also note that similar treatment can be performed for the anti-holomorphic sector,
and the result is
\begin{eqnarray}
\bar{J}_n^a \ket{N+n}_{\alpha\beta} = \ket{N}_{\delta\beta^\prime} (\frac{1}{2} \sigma^a)_{\beta^\prime\beta}.
\end{eqnarray}

\subsection{First order expansion of the OPE}

Let's first consider the holomorphic sector.
Take $N=0,n=1$, then
\begin{eqnarray}
J_n^a\ket{1}_{\alpha\gamma} &=& (\frac{1}{2} \sigma^a)_{\alpha\alpha^\prime}\ket{0}_{\alpha^\prime\gamma}=(\frac{1}{2} \sigma^a)_{\alpha\alpha^\prime} \epsilon_{\alpha^\prime \gamma}\ket{0}=(\frac{1}{2} \sigma^a i\sigma^2)_{\alpha\gamma} \ket{\Omega}. 
\end{eqnarray}
Using 
\begin{eqnarray}
\ket{1}_{\alpha\gamma}=(\beta_{\alpha\gamma}^{1x} J_{-1}^x +\beta_{\alpha\gamma}^{1y} J_{-1}^y+\beta_{\alpha\gamma}^{1z} J_{-1}^z)\ket{\Omega},
\end{eqnarray}
(in which the superscript $1a$ is $\left\{\begin{array}{c} 1\\a \end{array} \right\}$ for short)
and the affine algebra Eq. (\ref{eq:affine_algebra}), we have 
\begin{eqnarray}
J_1^a\ket{1}_{\alpha\gamma} = \frac{1}{2} \beta^{1a}_{\alpha\gamma}\ket{\Omega}.
\end{eqnarray}
Thus it is clear that
\begin{eqnarray}
\beta^{1a}_{\alpha\gamma}\ket{\Omega}=-(\sigma^a i\sigma^2)_{\alpha\gamma}. 
\end{eqnarray}
Then to lowest order,
the holomorphic part of the OPE $g_{\alpha\beta} (z,\bar{z}) g_{\gamma\delta} (w,\bar{z})$
is given by
\begin{flalign}
&g_{\alpha\beta} (z,\bar{z}) g_{\gamma\delta} (w,\bar{z})=\frac{1}{z^{1/2}} \big[ \epsilon_{\alpha\gamma} - \sum_{a=x,y,z} (\sigma^a i\sigma^2)_{\alpha\gamma} (J_{-1}^a \mathbbm{1}) (w)
+...  \big]. 
\end{flalign}

Notice that in the above expression
\begin{eqnarray}
(J_{-1}^a \mathbbm{1}) (w)=\frac{1}{2\pi i} \oint_w dz \frac{1}{z-w} J^a(z)=J^a(w)
\end{eqnarray}
is just the WZW current at the point $w$, hence
\begin{flalign}
&g_{\alpha\beta} (z,\bar{z}) g_{\gamma\delta} (w,\bar{z})=\frac{1}{z^{1/2}} \big[ \epsilon_{\alpha\gamma} - \sum_{a=x,y,z} (\sigma^a i\sigma^2)_{\alpha\gamma} J^a  (w)
+...  \big]. 
\end{flalign}
Also note that we only know
\begin{eqnarray}
\langle g_{\alpha\beta}(z,\bar{z}) g_{\gamma\delta} (w,\bar{w})\rangle
\propto\epsilon_{\alpha\gamma} \epsilon_{\beta\delta}\frac{1}{|z-w|}.
\label{eq:app_norm_gg}
\end{eqnarray}
In the above calculation of the OPE, 
we have fixed the coefficient of this vacuum expectation value to be $1$,
which amounts to a multiplicative rescaling of the primary field $g$.

Proceeding exactly similarly for the anti-holomorphic part, we obtain
\begin{eqnarray}
\bar{\beta}^{1a}_{\delta\beta} = -(i\sigma^2\sigma^a)_{\delta\beta}.
\end{eqnarray}

Combining the holomorphic and anti-holomorphic parts together, the result up to first order in the WZW currents is
\begin{eqnarray}
g_{\alpha\beta}(z,\bar{z}) g_{\gamma\delta} (w,\bar{w}) 
= \frac{1}{|z-w|} \big[ \epsilon_{\alpha\gamma} \epsilon_{\beta\delta} -\epsilon_{\beta\delta} 
(z-w)\sum_a (\sigma^ai\sigma^2)_{\alpha\gamma} J^a(w) -\epsilon_{\alpha\gamma} (\bar{z}-\bar{w}) \sum_b (i\sigma^2\sigma^b)_{\delta\beta} \bar{J}^b(\bar{w})+... \big].
\end{eqnarray}

\subsection{Second order expansion of the OPE}

We first consider the holomorphic sector.
We write
\begin{eqnarray}
\ket{2}_{\alpha\gamma} = \big[ \sum_a \gamma^a_{\alpha\gamma} (J_{-2}^a\mathbbm{1}) +\sum_{a,b} \beta^{ab}_{\alpha\gamma} (J_{-1}^aJ_{-1}^b\mathbbm{1}) \big]\ket{\Omega},
\label{eq:OPE2}
\end{eqnarray}
in which $\gamma^a$ and $\beta^{ab}$ are $\beta^{\{K_1\}}$ and $\beta^{\{K_2\}}$
where $\{K_1\}=\left\{ \begin{array}{c}1\\a\end{array} \right\}$, 
and $\{K_2\}=\left\{ \begin{array}{cc}1&1\\a&b\end{array} \right\}$.
We also note that in Eq. (\ref{eq:OPE2}),
\begin{eqnarray}
(J_{-2}^a\mathbbm{1})=\frac{1}{2\pi i} \oint_0 dz z^{-2} J^a(z)=\partial_z J^a(0),
\end{eqnarray}
and 
\begin{eqnarray}
(J_{-1}^aJ_{-1}^b\mathbbm{1})&=&\frac{1}{2\pi i} \oint_0 dz \frac{1}{z} J^a(z) (J_{-1}^b\mathbbm{1}) (0)= \frac{1}{2\pi i} \oint_0 dz \frac{1}{z} J^a(z) J^b(0)=(J^aJ^b)(0),
\end{eqnarray}
in which $(J^aJ^b)(w)$ means the zeroth order term in the expansion of the OPE $J^a(z)J^b(w)$ in powers of $z-w$.

First we take $N=0,n=2$ in Eq. (\ref{eq:key_eq}). Then
\begin{eqnarray}
J_2^c\ket{2}_{\alpha\gamma} &=& -(\frac{1}{2} \sigma^c)_{\alpha\alpha^\prime} \ket{0}_{\alpha^\prime\gamma}= -(\frac{1}{2} \sigma^c i\sigma^2)_{\alpha\gamma} \ket{\Omega}.
\end{eqnarray}
On the other hand, we can use the affine commutation relations to calculate $J_2^c\ket{2}_{\alpha\gamma}$ as
\begin{flalign}
J_2^c\big( \gamma_{\alpha\gamma}^aJ_{-2}^a\ket{\Omega} +\beta^{ab}_{\alpha\gamma} J_{-1}^aJ_{-1}^b\ket{\Omega} \big)
&=\gamma_{\alpha\gamma}^a \delta_{ac} \ket{\Omega} + \beta_{\alpha\gamma}^{ab} 
(J_{-1}^aJ_2^c+i\epsilon^{cad}J_1^d) J_{-1}^b\ket{\Omega}\nn\\
&=\gamma_{\alpha\gamma}^c \ket{\Omega} + \beta_{\alpha\gamma}^{ab} i\epsilon^{cad}\frac{1}{2} \delta_{ab}\ket{\Omega}\nn\\
&=(\gamma_{\alpha\gamma}^c + \frac{1}{2} i\epsilon^{cad} \beta_{\alpha\gamma}^{ab})
 \ket{\Omega}.
\end{flalign}
This gives
\begin{eqnarray}
\gamma_{\alpha\gamma}^c + \frac{1}{2} i\epsilon^{cad} \beta_{\alpha\gamma}^{ab}=-(\frac{1}{2} \sigma^ci\sigma^2)_{\alpha\gamma},
\end{eqnarray}
or as a matrix identity
\begin{eqnarray}
\gamma^c + \frac{1}{2} i\epsilon^{cad} \beta^{ab}=-\frac{1}{2} \sigma^ci\sigma^2.
\end{eqnarray}

Next we take $N=1,n=1$.
Then
\begin{eqnarray}
J_1^c\ket{2}_{\alpha\gamma}&=&-\frac{1}{2}\sigma^c_{\alpha\alpha^\prime}\ket{1}_{\alpha^\prime\gamma}\nn\\
&=&-\frac{1}{2}\sigma^c_{\alpha\alpha^\prime} (-\sigma^\mu i\sigma^2)_{\alpha^\prime\gamma} J_{-1}^\mu \ket{\Omega}\nn\\
&=&(\frac{1}{2} \sigma^c\sigma^\mu i\sigma^2)_{\alpha\gamma} J_{-1}^\mu \ket{\Omega}
\end{eqnarray}
On the other hand, using the affine algebra we get
\begin{flalign}
J_1^c(\gamma_{\alpha\gamma}^a J_{-2}^a +\beta_{\alpha\gamma}^{ab} J_{-1}^aJ_{-1}^b )\ket{\Omega}
&=\big[  \gamma_{\alpha\gamma}^a i\epsilon^{ca\mu} J_{-1}^\mu +   \beta_{\alpha\gamma}^{ab} (J_{-1}^a J_{1}^c +i\epsilon^{cad} J_0^d+\frac{1}{2} \delta_{ac}  ) J_{-1}^b
\big]\ket{\Omega}\nn\\
&=\big[ i\epsilon^{ca\mu} \gamma_{\alpha\gamma}^a +\frac{1}{2} (\beta_{\alpha\gamma}^{\mu c} + \beta_{\alpha\gamma}^{ c\mu}) -\epsilon^{cad}\epsilon^{db\mu} \beta^{ab}
\big] J_{-1}^\mu \ket{\Omega}\nn\\
&=\big[ i\epsilon^{ca\mu} \gamma_{\alpha\gamma}^a +\frac{1}{2} (\beta_{\alpha\gamma}^{\mu c} - \beta_{\alpha\gamma}^{ c\mu}) +\delta_{c\mu} \beta^{aa}_{\alpha\gamma}
\big] J_{-1}^\mu \ket{\Omega}.
\end{flalign}
Thus we get the equation
\begin{flalign}
&\frac{1}{2} (\beta^{ab}-\beta^{ba}) +\delta_{ab} \beta^{dd} -i\epsilon^{abc} \gamma^c=\frac{1}{2} \sigma^a\sigma^b i\sigma^2= \frac{1}{2} (\delta_{ab} + i\epsilon^{abc}\sigma^c) i\sigma^2.
\end{flalign}

In conclusion, the equations determining $\gamma^a$ and $\beta^{ab}$ are
\begin{eqnarray}
\gamma^c+\frac{1}{2} i \epsilon^{abc} \beta^{ab} &=& -\frac{1}{2} \sigma^c i\sigma^2,\nn\\
\frac{1}{2} (\beta^{ab}-\beta^{ba}) +\delta_{ab} \beta^{dd} -i\epsilon^{abc} \gamma^c&=& 
\frac{1}{2} \delta_{ab} i\sigma^2+ \frac{1}{2}i\epsilon^{abc}\sigma^c i\sigma^2.
 \end{eqnarray}
The first equation can be simplified by multiplying $-i\epsilon^{cde}$ on both sides.
We then arrive at 
\begin{eqnarray}
\frac{1}{2} (\beta^{ab}-\beta^{ba}) -i\epsilon^{abc}\gamma^c&=& \frac{1}{2} i \epsilon^{abc}\sigma^c i\sigma^2,\nn\\
\frac{1}{2} (\beta^{ab}-\beta^{ba}) +\delta_{ab} \beta^{dd} -i\epsilon^{abc} \gamma^c&=& 
\frac{1}{2} \delta_{ab} i\sigma^2+\frac{1}{2} i\epsilon^{abc}\sigma^c i\sigma^2.
\label{eq:OPE2_main}
 \end{eqnarray}

Plugging the first equation in Eq. (\ref{eq:OPE2_main}) into the second,
we obtain
\begin{eqnarray}
\delta_{ab} \beta^{dd} = \frac{1}{2} \delta_{ab} i\sigma^2.
\end{eqnarray}
We have three comments here.

1. We can only solve the sum $\sum_d \beta^{dd}$ as $\frac{1}{2} i \sigma^2$, and the individual $\beta^{xx},\beta^{yy},\beta^{zz}$ cannot be determined individually.
However, according to Sec. \ref{app:quadratic_WZW},
$(J^aJ^a)$ are all equal for $a=x,y,z$.
Hence we can well take
\begin{eqnarray}
\beta^{xx}=\beta^{yy}=\beta^{zz}=\frac{1}{6} i\sigma^2.
\end{eqnarray}

2. The information  about the off-diagonal symmetric part of the tensor $\beta^{ab}$ is missing,
i.e., there is no information about $\frac{1}{2} (\beta^{ab}+\beta^{ba})$ ($a\neq b$).
However, according to Sec. \ref{app:quadratic_WZW},
$(J^aJ^b)+(J^bJ^a)=0$ when $a=b$.
Thus this is not a problem.

3.We only know a combination of the anti-symmetric part of $\beta^{ab}$ and $\gamma^c$ as given by the first equation in Eqs. (\ref{eq:OPE2_main}),
and we are not able to solve them individually. 
However, according to Sec. \ref{app:quadratic_WZW},
\begin{eqnarray}
(J^aJ^b)-(J^bJ^a)=i\epsilon^{abc} \partial_z J^c\, (a\neq b).
\end{eqnarray}
Thus we can drop $(J_{-2}^c\mathbbm{1})$ and absorb it into the antisymmetric combination $i\epsilon^{abc} (J_{-1}^aJ_{-1}^b\mathbbm{1})$.

In summary, Eq. (\ref{eq:OPE2_main}) can be simplified as
\begin{eqnarray}
\frac{1}{2} (\beta^{ab}-\beta^{ba}) &=& \frac{1}{2} i \epsilon^{abc} \sigma^ci\sigma^2,\nn\\
 \sum_d \beta^{dd}&=&\frac{1}{2} i\sigma^2.
\end{eqnarray}
The solution is clearly,
\begin{eqnarray}
\beta^{ab}=\frac{1}{6} \delta_{ab} i\sigma^2+\frac{1}{2} i\epsilon^{abc} \sigma^ci\sigma^2.
\end{eqnarray}

By doing exactly similar treatment to the anti-holomorphic sector, we obtain
\begin{eqnarray}
\bar{\beta}^{ab}=-\frac{1}{6} \delta_{ab} i\sigma^2+\frac{1}{2} i \epsilon^{abc} i\sigma^2 \sigma^c.
\end{eqnarray}

In summary, we are able to write down the OPE of $g$ up to quadratic order in the WZW currents, as
\begin{eqnarray}
g_{\alpha\beta} (z,\bar{z})g_{\gamma\delta} (w,\bar{w}) &=
\frac{1}{|z-w|} \big[
\epsilon_{\alpha\gamma} -(z-w) \sum_a (\sigma^a i\sigma^2)_{\alpha\gamma} J^a(w)+
(z-w)^2 \sum_{ab} (\frac{1}{6} \delta_{ab}i\sigma^2+\frac{1}{2} i\epsilon^{abc} \sigma^ci\sigma^2 )_{\alpha\gamma} (J^aJ^b)(w)+...
 \big]\nn\\
 &\times
 \big[
\epsilon_{\beta\delta} -(\bar{z}-\bar{w}) \sum_a ( i\sigma^2 \sigma^a)_{\delta\beta} \bar{J}^a(w)+
(\bar{z}-\bar{w})^2 \sum_{ab} (-\frac{1}{6} \delta_{ab}i\sigma^2+\frac{1}{2} i\epsilon^{abc} i\sigma^2\sigma^c )_{\delta\beta} (\bar{J}^a\bar{J}^b)(\bar{w})+...
 \big],\nn\\
\end{eqnarray}
i.e.,
\begin{flalign}
g_{\alpha\beta} (z,\bar{z})g_{\gamma\delta} (w,\bar{w})
&=\frac{1}{|z-w|} \big[ \epsilon_{\alpha\gamma} \epsilon_{\beta\delta}
-(z-w)  \epsilon_{\beta\delta} (\sigma^ai\sigma^2)_{\alpha\gamma} J^a(w)\nn\\
&-(\bar{z}-\bar{w}) \epsilon_{\alpha\gamma} (i\sigma^2\sigma^a)_{\delta\beta} \bar{J}^a(\bar{w})+ |z-w|^2 (\sigma^ai\sigma^2)_{\alpha\gamma}  (i\sigma^2\sigma^a)_{\delta\beta} J^a(w) \bar{J}^b(\bar{w})\nn\\
&+(z-w)^2 \epsilon_{\beta\delta} (\frac{1}{6} \delta_{ab}\epsilon_{\alpha\gamma} +\frac{1}{2} i\epsilon^{abc} (\sigma^ci\sigma^2)_{\alpha\gamma}   ) (J^aJ^b) (w)\nn\\
&+(\bar{z}-\bar{w})^2 \epsilon_{\alpha\gamma} (\frac{1}{6} \delta_{ab}\epsilon_{\beta\delta} +\frac{1}{2} i\epsilon^{abc} (i\sigma^2\sigma^c)_{\delta\beta}   ) (\bar{J}^a\bar{J}^b) (w)+...
\big]
\label{eq:OPE_gg}
\end{flalign}.

Using Eq. (\ref{eq:OPE_gg}), we obtain the OPE between the Neel order fields,
\begin{eqnarray}
&\text{tr} (g(z,\bar{z}) \sigma^\lambda)\text{tr} (g(w,\bar{w}) \sigma^\mu)=
\frac{1}{|z-w|} \big(
-2\delta_{\lambda\mu} -2(z-w) i\epsilon^{\lambda\mu \alpha} J_L^\alpha(w)-2(\bar{z}-\bar{w}) i\epsilon^{\lambda\mu \alpha} J_R^\alpha(\bar{w})\nn\\
&+2|z-w|^2 [J_L^\lambda(w)J_R^\mu(\bar{w}) + J_L^\mu(w)J_R^\lambda(\bar{w}) -\delta_{\lambda\mu}
\vec{J}_L(w)\cdot \vec{J}_R(\bar{w})]\nn\\
&+(z-w)^2 [-J_L^\lambda(w)J_L^\mu(w) + J_L^\mu(w)J_L^\lambda(w) -\frac{1}{3}\delta_{\lambda\mu}
\vec{J}_L(w)\cdot \vec{J}_L(w)]\nn\\
&+(\bar{z}-\bar{w})^2 [-J_R^\lambda(\bar{w})J_R^\mu(\bar{w}) + J_R^\mu(\bar{w})J_R^\lambda(\bar{w}) -\frac{1}{3}\delta_{\lambda\mu}
\vec{J}_R(\bar{w})\cdot \vec{J}_R(\bar{w})]+...
\big),
\label{eq:OPENeel}
\end{eqnarray}
in which $\lambda,\mu=x,y,z$.
We note that the normalization in Eq. (\ref{eq:OPENeel}) is chosen as Eq. (\ref{eq:app_norm_gg}), which is different from the one used in the main text which is
\begin{eqnarray}
\left< g_{\alpha\beta}(z,\bar{z}) g_{\gamma\delta} (w,\bar{w})\right>
=\epsilon_{\alpha\gamma} \epsilon_{\beta\delta}\frac{2\pi a}{|z-w|}.
\label{eq:maintext_gg_corr}
\end{eqnarray}

On the other hand,  the 1D spin-1/2 free Dirac fermion can be 
decomposed into a U(1) charge boson and an SU(2) WZW boson
according to the nonabelian bosonization theory.
Then the OPE calculated in the bosonic language must coincide 
with the fermionic calcuation.
By properly taking into account the U(1) charge sector, we have verified that the fermionic approach gives the same result as Eq. (\ref{eq:OPENeel}).

\section{Symmetry groups in different frames}

\subsection{Symmetry group in the original frame}
\label{app:sym_group_neel}

\subsubsection{$\Gamma=0$}

In this subsection, we discuss the symmetry group within the original frame.
We first consider the case of $\Gamma=0$, i.e., the case of a Kitaev-Heisenberg chain.

For the Kitaev-Heisenberg chain, in addition to the symmetry operations 
$\{T,T_{2a},T_aI, R(\hat{z},\frac{\pi}{2})T_a,R(\hat{y},\pi)\}$
there are other symmetry transformations, including $R(\hat{x},\pi),R(\hat{z},\pi),R(\hat{n}_N,\pi)T_a,R(\frac{1}{2}(1,1,0)^T,\pi)T_a,R(\hat{z},-\frac{\pi}{2})T_a $.
However, they can all be generated by the above five operations:
\begin{eqnarray}
R(z,\pi)&=&(R(\hat{z},\frac{\pi}{2})T_a)^2\cdot T_{2a}^{-1},\nn\\
R(\hat{x},\pi)&=&R(\hat{y},\pi)\cdot R(\hat{z},\pi),\nn\\
R(\hat{n}_N,\pi)T_a&=&R(\hat{x},\pi)\cdot R(\hat{z},\frac{\pi}{2})T_a,\nn\\
R(\frac{1}{2}(1,1,0)^T,\pi)T_a&=&R(\hat{x},\pi)\cdot R(\hat{n}_N,\pi)T_a\cdot R(\hat{x},\pi),\nn\\
R(\hat{z},-\frac{\pi}{2})T_a&=&R(\hat{z},\pi)\cdot R(\hat{z},\frac{\pi}{2})T_a.
\end{eqnarray}
Therefore, the symmetry group of the Kitaev-Heisenberg chain is
\begin{flalign}
G_0&=\mathopen{<}
T,T_{2a},T_aI,R(\hat{y},\pi),R(\hat{z},\frac{\pi}{2})T_a
\mathclose{>}.
\label{eq:group_G0_app}
\end{flalign}
The group structure of $G_0$ will be shown in Sec. \ref{app:fourrot_gamma_0} to be
$G_0\simeq D_{4d}\ltimes (\mathbb{Z}_2\ltimes 2\mathbb{Z})$.

\subsubsection{$\Gamma\neq 0$}

Next we turn on a nonzero $\Gamma$ and study the symmetry group of a general Kitaev-Heisenberg-Gamma chain.
The symmetry transformations are now
\begin{eqnarray}
1.&T &:  (S_i^x,S_i^y,S_i^z)\rightarrow (-S_{i}^x,-S_{i}^y,-S_{i}^z)\nn\\
2.& T_{2a}&:  (S_i^x,S_i^y,S_i^z)\rightarrow (S_{i+2}^x,S_{i+2}^y,S_{i+2}^z)\nn\\
3.&T_a I&: (S_i^x,S_i^y,S_i^z)\rightarrow (S_{-i+1}^x,S_{-i+1}^y,S_{-i+1}^z)\nn\\
4.&R(\hat{n}_N,\pi)T_a&: (S_i^x,S_i^y,S_i^z)\rightarrow (-S_{i+1}^y,-S_{i+1}^x,-S_{i+1}^z),
\label{eq:Sym_Neel_KHG}
\end{eqnarray}
where in particular, with the presence of the cross terms $S_i^\alpha S_{i+1}^\beta$ ($\alpha\neq\beta$) in the Hamiltonian, the operations $R(\hat{\alpha},\pi)$ ($\alpha=x,y,z$) are no longer symmetries.
We thus conclude that the symmetry group $G_N$ is 
\begin{eqnarray}
G_N=\mathopen{<}T,T_{2a},T_aI,R(\hat{n}_N,\pi)T_a
\mathclose{>}.
\label{eq:GN}
\end{eqnarray}
By a similar analysis with the $G_0$ case discussed in the main text, the group structure of $G_N$ when $\Gamma\neq 0$ is
\begin{eqnarray}
G_N
=(\mathbb{Z}_2\times \mathbb{Z}_2)\ltimes (\mathbb{Z}_2\ltimes 2\mathbb{Z}),
\label{eq:group_structure_Neel}
\end{eqnarray}
in which from left to right, $\mathbb{Z}_2\times \mathbb{Z}_2=\mathopen{<} T\mathclose{>}\times \mathopen{<}R(\hat{n}_N,\pi)T_a \mathclose{>}/\mathopen{<}T_{2a}\mathclose{>}$, $\mathbb{Z}_2=\mathopen{<}T_a I\mathclose{>}$,
and $2\mathbb{Z}=\mathopen{<}T_{2a}\mathclose{>}$.

On the other hand, in the main text we have discussed the same model but in the six-sublattice rotated frame. 
The group structure in the six-subalttice rotated frame has been worked out to be $ G_1\cong D_{3d}\ltimes 3\mathbb{Z}$.
Notice that $G_1$ and $G_N$
must be isomorphic since they are essentially the same group up to a unitary transformation.
An explicit verification of this isomorphism is included in Appendix \ref{app:isomorphism}.

\subsection{Symmetry group in the four-sublattice rotated frame}
\label{app:GN_to_G3}

\subsubsection{$\Gamma=0$}
\label{app:fourrot_gamma_0}

To determine the group structure of $G_0$,
first  notice that $N_{0s}=\mathopen{<}T_{2a},T_aI\mathclose{>}=\mathopen{<}T_aI
\mathclose{>}\ltimes \mathopen{<}T_{2a}
\mathclose{>}$ is a normal subgroup of $G_0$,
hence $G_0$ can be written as
\begin{eqnarray}
G_0=G_{0s}\ltimes N_{0s},
\end{eqnarray}
in which 
\begin{eqnarray}
G_{0s}=\mathopen{<}T,R(\hat{z},\frac{\pi}{2})T_a,R(\hat{y},\pi)
\mathclose{>}/ \mathopen{<} T_{2a}\mathclose{>}.
\end{eqnarray}
Then by defining $a=R(\hat{z},\frac{\pi}{2})T_a$, $b=R(\hat{y},\pi)$,
it is straightforward to verify that 
$a^4=b^2=(ab)^2=e\mod  T_{2a}$.
Since the generator-relation representation for the group $D_n$ is
\begin{eqnarray}
D_n=\mathopen{<} \alpha,\beta| \alpha^n=\beta^2=(\alpha\beta)^2=e \mathclose{>},
\label{eq:generator_Dn}
\end{eqnarray}
and the relations in Eq. (\ref{eq:generator_Dn}) are satisfied for the two generators of $G_{0s}/\mathopen{<}T\mathclose{>}$,
we see that $G_{0s}/\mathopen{<}T\mathclose{>}$ must be a subgroup of $D_4$.
On the other hand, since the actions of 
\begin{eqnarray}
\{1,a,a^2,a^3,b,ab,a^2b,a^3b\}
\end{eqnarray} 
restricted within the spin space are 
\begin{eqnarray}
&\{1,R(\hat{z},\frac{\pi}{2}),R(\hat{z},\pi),R(\hat{z},-\frac{\pi}{2}),R(\hat{y},\pi),\nn\\
&R(\frac{1}{\sqrt{2}}(\hat{x}-\hat{y}),\pi),R(\hat{x},\pi),R(\frac{1}{\sqrt{2}}(\hat{x}+\hat{y}),\pi)\},
\end{eqnarray}
which are all distinct operations,
there must be at least eight distinct group  elements in $G_{0s}/\mathopen{<}T\mathclose{>}$.
But the order of $D_4$ is eight, hence we conclude that $G_{0s}/\mathopen{<}T\mathclose{>}\simeq D_4$, i.e., $G_{0s}\simeq D_{4d}$.
In conclusion, the group structure of $G_0$ is
\begin{eqnarray}
G_0\simeq D_{4d}\ltimes (\mathbb{Z}_2\ltimes 2\mathbb{Z}),
\label{eq:G0_group_structure}
\end{eqnarray}
in which $D_{4d}=G_{0s}$, $\mathbb{Z}_2=\mathopen{<}T_a I\mathclose{>}$,
and $2\mathbb{Z}=\mathopen{<}T_{2a} \mathclose{>}$.

Alternatively, we can write the group structure in another way.
Recall  that in the expression of $G_0$, $\mathbb{Z}_2\ltimes 2\mathbb{Z}$
represents $\mathopen{<}T_aI\mathclose{>}\ltimes \mathopen{<}T_{2a} \mathclose{>}$.
Distilling out $\mathopen{<}T_{4a} \mathclose{>}$, it can be rewritten as
\begin{eqnarray}
\mathopen{<}T_aI\mathclose{>}\ltimes \mathopen{<}T_{2a} \mathclose{>}=
(\mathopen{<}T_aI\mathclose{>}\times \mathopen{<} T_{2a} \mathclose{>})\ltimes \mathopen{<} T_{4a} \mathclose{>},
\end{eqnarray}
Therefore, the group structure of $G_0$ is 
\begin{eqnarray}
G_0=[(\mathbb{Z}_2\times \mathbb{Z}_2) \ltimes D_{4d}] \ltimes 4\mathbb{Z},
\label{eq:G0_d_sp}
\end{eqnarray}
in which from left to right, $\mathbb{Z}_2=\mathopen{<}T_aI\mathclose{>}$,
$\mathbb{Z}_2=\mathopen{<} T_{2a} \mathclose{>} \mod T_{4a} $, $D_{4d}=\mathopen{<}T,R(\hat{z},\frac{\pi}{2})T_a,R(\hat{y},\pi)T_a I
\mathclose{>} \mod T_{4a}$, and $4\mathbb{Z}=\mathopen{<} T_{4a} \mathclose{>}$.

\subsubsection{$\Gamma\neq 0$}

Recall that the symmetry group $G_N$ discussed in Sec. \ref{app:sym_group_neel} in the original frame is $G_N=\mathopen{<}T,T_aI,R(\hat{n}_N,\pi)T_a\mathclose{>}$.
In this section, we work out $G_3=U_4 G_NU_4^{-1}$,
which is the symmetry group in the four-sublattice rotated frame.
We take the inversion center of $I$ to be at site $2$ throughout this section. 

1) Clearly, $U_4TU_4^{-1}=T$.

2) For, $U_4 T_aIU_4^{-1}$, we have
\begin{flalign}
&(x_1,y_1,z_1)\xrightarrow{\text{$U_4^{-1}$}}(-x_1,y_1,-z_1)\xrightarrow{\text{$T_aI$}}(-x_4,y_4,-z_4)\xrightarrow{\text{$U_4$}}(-x_4,y_4,-z_4),\nn\\
&(x_2,y_2,z_2)\xrightarrow{\text{$U_4^{-1}$}}(-x_2,-y_2,z_2)\xrightarrow{\text{$T_aI$}}(-x_3,-y_3,z_3)\xrightarrow{\text{$U_4$}}(-x_3,y_3,-z_3),\nn\\
&(x_3,y_3,z_3)\xrightarrow{\text{$U_4^{-1}$}}(x_3,-y_3,-z_3)\xrightarrow{\text{$T_aI$}}(x_2,-y_2,-z_2)\xrightarrow{\text{$U_4$}}(-x_2,y_2,-z_2),\nn\\
&(x_4,y_4,z_4)\xrightarrow{\text{$U_4^{-1}$}}(x_4,y_4,z_4)\xrightarrow{\text{$T_aI$}}(x_1,y_1,z_1)\xrightarrow{\text{$U_4$}}(-x_1,y_1,-z_1),
\end{flalign}
which is exactly the operation $R(\hat{y},\pi)T_aI$.

3) For $U_4R(\hat{n}_N,\pi)T_aU_4^{-1}$, we have
\begin{flalign}
&(x_1,y_1,z_1)\xrightarrow{\text{$U_4^{-1}$}}(-x_1,y_1,-z_1)\xrightarrow{\text{$R(\hat{n},\pi)T_a$}}(y_2,-x_2,z_2)\xrightarrow{\text{$U_4$}}(-y_2,x_2,z_2),\nn\\
&(x_2,y_2,z_2)\xrightarrow{\text{$U_4^{-1}$}}(-x_2,-y_2,z_2)\xrightarrow{\text{$R(\hat{n},\pi)T_a$}}(y_3,x_3,-z_3)\xrightarrow{\text{$U_4$}}(-y_3,x_3,z_3),\nn\\
&(x_3,y_3,z_3)\xrightarrow{\text{$U_4^{-1}$}}(x_3,-y_3,-z_3)\xrightarrow{\text{$R(\hat{n},\pi)T_a$}}(-y_4,x_4,z_4)\xrightarrow{\text{$U_4$}}(-y_4,x_4,z_4),\nn\\
&(x_4,y_4,z_4)\xrightarrow{\text{$U_4^{-1}$}}(x_4,y_4,z_4)\xrightarrow{\text{$R(\hat{n},\pi)T_a$}}(-y_1,-x_1,-z_1)\xrightarrow{\text{$U_4$}}(-y_1,x_1,z_1),
\end{flalign}
which is exactly the operation $R(\hat{z},-\frac{\pi}{2})T_a$.

In summary, the symmetry group in the four-sublattice rotated frame is
\begin{eqnarray}
G_3=\mathopen{<} T,R(\hat{y},\pi)T_a I,R(\hat{z},-\frac{\pi}{2})T_a \mathclose{>},
\end{eqnarray}
which verifies the expression of $G_3$ given in the main text.

\subsection{Group isomorphism between $G_1$ and $G_N$}
\label{app:isomorphism}

We have seen that the symmetry group for the Kitaev-Heisenberg-Gamma chain is $G_1=\mathopen{<} T,R_a T_a,R_I I \mathclose{>}$ in the six-sublattice rotated frame,
and $G_N=\mathopen{<}T,T_a I,R(\hat{n}_N,\pi)T_a\mathclose{>}$ in the original frame.
These two groups should be related by the six-sublattice rotation $U_6$.
Furthermore, their group structures $G_1^\prime=G_1/\mathopen{<}T\mathclose{>}=D_{3}\ltimes 3\mathbb{Z}$ and 
$G_N^\prime=G_N/\mathopen{<}T\mathclose{>}= \mathbb{Z}_2\ltimes (\mathbb{Z}_2\ltimes 2\mathbb{Z})$ must be isomorphic.
In this section, we explicitly verify these.
For later convenience, we define the inversion operation $I$ to have the inversion center at site $5$.
Recall that in the main text, the inversion center is chosen at site zero for  $I$ in $G_N$.
However, this is not essential, since the symmetry operation $T_{2a}$  in $G_N$ is able to
shift the inversion center by any integer number of sites.

We first show that 
\begin{eqnarray}
U_6^{-1} R_aT_a U_6&=&R(\hat{n}_N,\pi)T_a\nn\\ 
U_6^{-1} R_I I U_6&=&T_{2a}^{-1}\cdot R(\hat{n}_N,\pi)T_a\cdot T_aI,
\label{eq:U6_G1}
\end{eqnarray}
thereby confirming the fact that $G_1$ and $G_N$ are indeed related by $U_6$.

The action of $U_6^{-1}\cdot R_aT_a\cdot U_6$ is given by
\begin{flalign}
&(
S_1^x,
S_1^y,
S_1^z
)
\xrightarrow{\text{$U_6$}}
(
S_1^x,
S_1^y,
S_1^z
)
\xrightarrow{\text{$R_aT_a$}}
(
S_2^z,
S_2^x,
S_2^y
)
\xrightarrow{\text{$U_6^{-1}$}}
(
-S_2^y,
-S_2^x,
-S_2^z
),\nn\\
&(
S_2^x,
S_2^y,
S_2^z
)
\xrightarrow{\text{$U_6$}}
(
-S_2^x,
-S_2^z,
-S_2^y
)
\xrightarrow{\text{$R_aT_a$}}
(
-S_3^z,
-S_3^y,
-S_3^x
)
\xrightarrow{\text{$U_6^{-1}$}}
(
-S_3^y,
-S_3^x,
-S_3^z
),\nn\\
&(
S_3^x,
S_3^y,
S_3^z
)
\xrightarrow{\text{$U_6$}}
(
S_3^y,
S_3^z,
S_3^x
)
\xrightarrow{\text{$R_aT_a$}}
(
S_4^x,
S_4^y,
S_4^z
)
\xrightarrow{\text{$U_6^{-1}$}}
(
-S_4^y,
-S_4^x,
-S_4^z
),\nn\\
&(
S_4^x,
S_4^y,
S_4^z
)
\xrightarrow{\text{$U_6$}}
(
-S_4^y,
-S_4^x,
-S_4^z
)
\xrightarrow{\text{$R_aT_a$}}
(
-S_5^x,
-S_5^z,
-S_5^y
)
\xrightarrow{\text{$U_6^{-1}$}}
(
-S_5^y,
-S_5^x,
-S_5^z
),\nn\\
&(
S_5^x,
S_5^y,
S_5^z
)
\xrightarrow{\text{$U_6$}}
(
S_5^z,
S_5^x,
S_5^y
)
\xrightarrow{\text{$R_aT_a$}}
(
S_6^y,
S_6^z,
S_6^x
)
\xrightarrow{\text{$U_6^{-1}$}}
(
-S_6^y,
-S_6^x,
-S_6^z
),\nn\\
&(
S_6^x,
S_6^y,
S_6^z
)
\xrightarrow{\text{$U_6$}}
(
-S_6^z,
-S_6^y,
-S_6^x
)
\xrightarrow{\text{$R_aT_a$}}
(
-S_7^y,
-S_7^x,
-S_2^z
)
\xrightarrow{\text{$U_6^{-1}$}}
(
-S_7^y,
-S_7^x,
-S_7^z
),
\end{flalign}
which is exactly $R(\hat{n}_N,\pi)T_a$.

The action of $U_6^{-1}\cdot R_I I \cdot U_6$ is given by
\begin{flalign}
&(
S_1^x,
S_1^y,
S_1^z
)
\xrightarrow{\text{$U_6$}}
(
S_1^x,
S_1^y,
S_1^z
)
\xrightarrow{\text{$R_II$}}
(
-S_9^z,
-S_9^y,
-S_9^x
)
\xrightarrow{\text{$U_6^{-1}$}}
(
-S_9^y,
-S_9^x,
-S_9^z
),\nn\\
&(
S_2^x,
S_2^y,
S_2^z
)
\xrightarrow{\text{$U_6$}}
(
-S_2^x,
-S_2^z,
-S_2^y
)
\xrightarrow{\text{$R_II$}}
(
S_8^z,
S_8^x,
S_8^y
)
\xrightarrow{\text{$U_6^{-1}$}}
(
-S_8^y,
-S_8^x,
-S_8^z
),\nn\\
&(
S_3^x,
S_3^y,
S_3^z
)
\xrightarrow{\text{$U_6$}}
(
S_3^y,
S_3^z,
S_3^x
)
\xrightarrow{\text{$R_II$}}
(
-S_7^y,
-S_7^x,
-S_7^z
)
\xrightarrow{\text{$U_6^{-1}$}}
(
-S_7^y,
-S_7^x,
-S_7^z
),\nn\\
&(
S_4^x,
S_4^y,
S_4^z
)
\xrightarrow{\text{$U_6$}}
(
-S_4^y,
-S_4^x,
-S_4^z
)
\xrightarrow{\text{$R_II$}}
(
S_6^y,
S_6^z,
S_6^x
)
\xrightarrow{\text{$U_6^{-1}$}}
(
-S_6^y,
-S_6^x,
-S_6^z
),\nn\\
&(
S_5^x,
S_5^y,
S_5^z
)
\xrightarrow{\text{$U_6$}}
(
S_5^z,
S_5^x,
S_5^y
)
\xrightarrow{\text{$R_II$}}
(
-S_5^x,
-S_5^z,
-S_5^y
)
\xrightarrow{\text{$U_6^{-1}$}}
(
-S_5^y,
-S_5^x,
-S_5^z
),\nn\\
&(
S_6^x,
S_6^y,
S_6^z
)
\xrightarrow{\text{$U_6$}}
(
-S_6^z,
-S_6^y,
-S_6^x
)
\xrightarrow{\text{$R_II$}}
(
S_4^x,
S_4^y,
S_4^z
)
\xrightarrow{\text{$U_6^{-1}$}}
(
-S_4^y,
-S_4^x,
-S_4^z
),
\end{flalign}
which is equal to $R(\hat{n}_N,\pi)I$.
Using $R(\hat{n}_N,\pi)I=T_{2a}^{-1}\cdot R(\hat{n}_N,\pi)T_a\cdot T_aI$,
we obtain the expression for $U_6^{-1} R_I I U_6$ in Eq. (\ref{eq:U6_G1}).

Next we show the equivalence of the two group structures,
i.e., $D_{3}\ltimes 3\mathbb{Z}\simeq \mathbb{Z}_2\ltimes (\mathbb{Z}_2\ltimes 2\mathbb{Z})$.

Denote $a=R_aT_a,b=R_I I$, then $G_1^\prime=\mathopen{<}a,b\mathclose{>}$.
It has been shown in the main text that $\mathopen{<}a,b\mathclose{>}=D_3\ltimes \mathopen{<}T_{3a}\mathclose{>}$.
Since $T_{3a}=(R_aT_a)^3$, we have $G_1^\prime=\mathopen{<}a,b\mathclose{>}\ltimes\mathopen{<}a^3\mathclose{>}$.
Recall the geometrical meaning of $D_3$ as the symmetry group of a regular triangle.
There are three rotations and three reflections,
in which the rotation group $C_3$ is a normal subgroup of $D_3$.
Thus $D_3$ can be written as 
\begin{eqnarray}
D_3\cong \mathbb{Z}_2\ltimes C_3,
\end{eqnarray}
in which $\mathbb{Z}_2$ is the group generated by a reflection.
For example, $ab$ is a refelection and $a^2=a^{-1}$ is a rotation.
Hence,
\begin{eqnarray}
D_3\cong (\mathopen{<}ab\mathclose{>}\ltimes \mathopen{<}a^2\mathclose{>})/\mathopen{<}a^3\mathclose{>}.
\end{eqnarray}
This gives 
\begin{eqnarray}
G_1^\prime=(\mathopen{<}ab\mathclose{>}\ltimes \mathopen{<}a^2\mathclose{>})\ltimes\mathopen{<}a^3\mathclose{>}.
\end{eqnarray}
On the other hand, $\mathopen{<}ab\mathclose{>}\ltimes \mathopen{<}a^2\mathclose{>}$
is also a normal subgroup of $G_1^\prime$.
This is because $T_{3a}R_IIT_{3a}^{-1}=(R_aT_a)^{-6}R_II$.
Therefore, $G_1^\prime$ can also be written as 
\begin{eqnarray}
G_1^\prime=\mathopen{<}a^3\mathclose{>}\ltimes(\mathopen{<}ab\mathclose{>}\ltimes \mathopen{<}a^2\mathclose{>}),
\end{eqnarray}
or alternatively,
\begin{eqnarray}
G_1^\prime=\mathopen{<}a\mathclose{>}\ltimes(\mathopen{<}ab\mathclose{>}\ltimes \mathopen{<}a^2\mathclose{>}).
\end{eqnarray}
Using
\begin{eqnarray}
U_6^{-1}R_aT_aU_6&=&R(\hat{n}_N,\pi)T_a\nn\\
U_6^{-1}a^2U_6&=&T_{2a}\nn\\
U_6^{-1}abU_6&=&T_aI,
\end{eqnarray}
we obtain
\begin{eqnarray}
U_6^{-1}G_1^\prime U_6=\mathopen{<}R(\hat{n}_N,\pi)\mathclose{>}\ltimes(\mathopen{<}T_aI\mathclose{>}\ltimes \mathopen{<}T_{2a}\mathclose{>}),
\end{eqnarray}
which is exactly $G_N^\prime$.

\section{Symmetry analysis of the low energy field theory}
\subsection{Symmetry analysis in the ``LL1" phase}
\label{app:sym_LL1}

In this section, we perform a symmetry analysis of the low energy field theory in the ``LL1" phase. 
The symmetry group of the system in the six-sublattice rotated frame has been worked out to be
$G_1=\mathopen{<}T, R_aT_a, R_I I\mathclose{>}\cong D_{3d}\ltimes 3\mathbb{Z}$.
We will exhaust all the symmetry allowed relevant and marginal terms in the low energy field theory.
Throughout this section, we work in the six-sublattice rotated frame unless otherwise stated.

All operators with scaling dimensions not greater than $2$ in the SU(2)$_1$ WZW model are listed as follows,
\begin{eqnarray}
\begin{array}{cc}
\text{Dimension} & \text{Operators}\\
1/2 & \epsilon,N^\alpha\\
1& J^\alpha_L,J^\alpha_R\\
3/2 & J^\alpha_L\epsilon,J^\alpha_R\epsilon, J_L^{\alpha} N^{\beta},J_R^\alpha N^\beta \\
2 & J^\alpha_L J^\beta_L, J^\alpha_R J^\beta_R,J^\alpha_L J^\beta_R.
\end{array}
\label{eq:low_terms}
\end{eqnarray}
Next we inspect which of these terms are compatible with the $D_{3d}\ltimes 3\mathbb{Z}$ symmetry.

1) The dimension $1/2$ operators are forbidden since they change sign under odd powers of $T_{a}$ (see Ref. \cite{Affleck1988}) and in the our case $T_{3a}$ is a symmetry of the system.

2) The dimension $3/2$ operators also change sign under $T_{3a}$.

3) To analyze the dimension $1$ operators, we first classify the current operators $\vec{J}_L, \vec{J}_R$
according to the group $\mathopen{<}R_aT_a,R_II\mathclose{>}/\mathopen{<}T_{3a}\mathclose{>}\cong D_3$. 
The three dimensional vector representation $\vec{J}_s$ ($s=1,-1$ corresponding to $L,R$) of $D_3$
 can be decomposed as $V_{1s}\oplus V_{2s}$,
in which 
\begin{eqnarray}
V_{1s}&=&\text{Span} \{ J^0_s\},\nn\\
V_{2s}&=&\text{Span} \{J^1_s,J^2_s\},
\end{eqnarray}
where ``Span$\{...\}$" denotes the $\mathbb{R}$-linear space spanned by the vectors in the bracket, 
and
\begin{eqnarray}
J^0_s&=& \frac{1}{\sqrt{3}} (J^x_s+J^y_s+J^z_s),\nn\\
J^1_s&=& \frac{1}{\sqrt{6}} (2J^x_s-J^y_s-J^z_s),\nn\\
J^2_s&=& \frac{1}{\sqrt{2}} (J^y_s-J^z_s).
\end{eqnarray}
The actions of $R_aT_a,R_I I$ in $V_{1s},V_{2s}$ can be worked out as
\begin{eqnarray}
(R_aT_a) J^0_s(x)(R_aT_a)^{-1}&=&J^0_s(x)\nn\\ 
(R_I I) J^0_s (x) (R_I I)^{-1}&=& -J^0_{-s}(-x),
\label{eq:act_V1}
\end{eqnarray}
and 
\begin{flalign}
&(R_aT_a) (J^1_s(x) \,J^2_s(x)) (R_aT_a)^{-1}= (J^1_s(x) \,J^2_s(x))A, \nn\\
&(R_I I) (J^1_s(x) \,J^2_s(x)) (R_I I)^{-1}= (J^1_{-s}(-x) \,J^2_{-s}(-x))B,
\label{eq:act_V2}
\end{flalign}
respectively,
in which $R_aT_a$ keeps the chiral sector while $R_II$ flips the chiral index, and
\begin{eqnarray}
A=\left(\begin{array}{cc}
-\frac{1}{2} & \frac{\sqrt{3}}{2}\\
-\frac{\sqrt{3}}{2}& -\frac{1}{2}
\end{array}\right),
~~B=\left(\begin{array}{cc}
\frac{1}{2} & \frac{\sqrt{3}}{2}\\
\frac{\sqrt{3}}{2}& -\frac{1}{2}
\end{array}\right).
\end{eqnarray}
Notice that in $V_{2s}$, both $R_aT_a$ and $R_I I$ correspond to orthogonal matrices in the basis of $\{J^1_s,J^2_s\}$.

According to Eq. (\ref{eq:transformT}), $\vec{J}_L-\vec{J}_R$ is invariant under the time reversal operation.
On the other hand, because of Eqs. (\ref{eq:act_V1},\ref{eq:act_V2}), the only combination compatible with the $D_3$ symmetry is $J_L^0-J_R^0$.
Thus we conclude that $J_L^0-J_R^0$ is allowed in the low energy theory.

4) Finally, we analyze the dimension 2 operators.
We consider the actions of the time reversal operation $T$ and the group $D_3=\mathopen{<}R_aT_a,R_II\mathclose{>}$ separately.
Since time reversal switches the chiral indices $L$ and $R$, the combinations $J_L^\alpha J_L^\beta\pm J_R^\alpha J_R^\beta,J_L^\alpha J_R^\beta\pm J_L^\beta J_R^\alpha$ are even (for $+$) and odd (for $-$) under the time reversal operation.

Next consider the constraints imposed by the $D_3$ group. 
Neglecting the chiral index, since $J^0$ is an $A_2$ representation of $D_3$,
$J^0J^0$ is apparently $D_3$-invariant.
Hence, the terms
\begin{eqnarray}
J^0_LJ^0_R,~J^0_LJ^0_L+J^0_RJ^0_R
\end{eqnarray}
are allowed by $D_{3d}=D_3\times\mathopen{<}T\mathclose{>}$.
On the other hand,  $\{J^1,J^2\}$ form an $E$ representation of the $D_3$ group.
Using the multiplication rule 
\begin{eqnarray}
E\otimes E=A_1\oplus A_2\oplus E,
\end{eqnarray}
we know that 
$J^1J^1+J^2J^2$ and $J^1J^2-J^2J^1$
are within the $A_1$ and $A_2$ representations, respectively. 
Out of the $A_1$ representation, the following two terms do not change sign under the time reversal operation,
\begin{eqnarray}
J_L^1J_R^1+J_L^2J_R^2,~J_L^1J_L^1+ J_L^2J_L^2+J_R^1J_R^1+J_R^2J_R^2,
\end{eqnarray}
hence allowed in the low energy theory.
As to the $A_2$ representation, an odd combination of the two chiral sectors must be formed to ensure the invariance under $R_II$,
but then it spoils the time reversal invariance.

In summary, all the independent symmetry allowed relevant and marginal terms can be taken as
\begin{eqnarray}
&J_L^0-J_R^0,\nn\\
&J^0_LJ^0_R,~ \vec{J}_L\cdot \vec{J}_R,\nn\\
&J_L^0J_L^0+J_R^0J_R^0, ~\vec{J}_L\cdot \vec{J}_L+\vec{J}_R\cdot \vec{J}_R,
\label{eq:allowed_LL1}
\end{eqnarray}
in which we have used the relation $\vec{J}_s\cdot\vec{J}_{s^\prime}=J^0_sJ^0_{s^\prime}+J^1_sJ^1_{s^\prime}+J^2_sJ^2_{s^\prime}$ to replace $\{ J^0_sJ^0_{s^\prime},J^1_sJ^1_{s^\prime}+J^2_sJ^2_{s^\prime}\}$ with 
$\{J^0_sJ^0_{s^\prime},\vec{J}_s\cdot\vec{J}_{s^\prime}\}$ as the two linearly independent terms.

We make a final remark that all the conclusions in this subsection are based on 
symmetry analysis, hence not dependent on the detailed forms of the microscopic Hamiltonian.
Thus the analysis applies also to the $KH\Gamma\Gamma^\prime$ chain where a small $\Gamma^\prime$ term  is included,
and it holds even when the terms beyond nearest neighbor couplings are taken into account.
Of course, the coupling constants in the low energy theory  will be renormalized by these additional terms.

\subsection{Symmetry analysis in the N\'eel phase}
\label{app:sym_Neel}

In this section, we perform a symmetry analysis of the low energy field theory in the ``N\'eel" phase. 
The symmetry group of the system in the original frame when $\Gamma\neq 0$ has been worked out to be
$G_N=\mathopen{<}T,T_a I,R(\hat{n}_N,\pi)T_a\mathclose{>}$.
We will exhaust all the symmetry allowed relevant and marginal terms in the low energy field theory.
Throughout this section, we work in the original frame unless otherwise stated.

We start from the AFM Heisenberg model whose low energy properties are described by the SU(2)$_1$ WZW model,
and treat the Kitaev and Gamma terms as small perturbations,
which applies to a neighborhood of the AFM1 point.
Here we will focus on the case of a nonzero $\Gamma$.

The symmetry allowed relevant and marginal terms can be identified in a similar manner as Sec. \ref{app:sym_LL1}.
The list of terms with dimensions not greater than two is the same as Eq. (\ref{eq:low_terms}).

1) For dimension $1/2$ operators, $\epsilon$ is forbidden by $R(\hat{z},\frac{\pi}{2})T_a$,
and $N^\alpha$ is forbidden by $T$.

2) Time reversal symmetry restricts the dimension $1$ operators to be the combinations 
$J_L^\alpha-J_R^\alpha$.
However,  this combination changes sign under $T_aI$.

3) The dimension $3/2$ operators are $J_L^\alpha\epsilon,J_R^\alpha\epsilon,J_L^\alpha N^\beta,J_R^\alpha N^\beta$.
Time reversal restricts them to be the combinations $(J_L^\alpha-J_R^\alpha)\epsilon$ and $(J_L^\alpha+J_R^\alpha)N^\beta$.
However, according to the transformation laws in Eqs. (\ref{eq:transformTa},\ref{eq:transformI}), both combinations change sign under $T_aI$, hence forbidden.

4) To discuss the dimension $2$ operators, let's first study the linear space $\text{span}\{J_s^x,J_s^y,J_s^z\}$ ($s=L,R$) as a representation of the group $\mathopen{<} R(\hat{n}_N,\pi)\mathclose{>} \cong\mathbb{Z}_2$,
where $\text{span}\{\cdot\cdot\cdot\}$ represents the $\mathbb{R}$-linear space spanned by the vectors within the bracket.
The three operators $J_s^x,J_s^y,J_s^z$ can be recombined into 
\begin{eqnarray}
J_s^x+J_s^y,J_s^x-J_s^y,J_s^z,
\end{eqnarray}
which have $R(\hat{n}_N,\pi)$ eigenvalues $-1,1,-1$, respectively.
Therefore, the quadratic terms invariant under $R(\hat{n}_N,\pi)T_a$
are given by 
\begin{eqnarray}
&(J_s^x-J_s^y)\cdot (J_{s^\prime}^x-J_{s^\prime}^y),~(J_s^x+J_s^y)\cdot (J_{s^\prime}^x+J_{s^\prime}^y),~
J_s^z\cdot J_{s^\prime}^z,~
(J_s^x+J_s^y)\cdot J_{s^\prime}^z.
\end{eqnarray}
Time reversal and $T_aI$ further require $L$ and $R$ to appear symmetrically.

In summary, in addition to the terms that are already in the SU(2) invariant AFM Heisenberg model,
the linearly independent nontrivial additional terms in the $KHG$ chain are
\begin{eqnarray}
&J_L^zJ_R^z,&
J_L^xJ_R^y+J_L^yJ_R^x,~
J_L^z(J_R^x+J_R^y)+(J_L^x+J_L^y)J_R^z.
\end{eqnarray}
We note that the term $J_L^xJ_R^y+J_L^yJ_R^x$ does not appear in the first order perturbation low energy Hamiltonian derived in the main text.
However, it can be generated upon the RG flow.
In fact, this term appears in the low energy Hamiltonian of the $KH\Gamma\Gamma^\prime$ chain.
At low energies, generically, all symmetry allowed terms will be generated,
and the $KH\Gamma$ and $KH\Gamma\Gamma^\prime$ chains contain the same set of terms
though with different coefficients.

\subsection{Symmetry analysis in the ``LL2" phase}
\label{subapp:Phase_KH}

In this section, we analyze the phase diagram of the Kitaev-Heisenberg chain based on a combination of symmetry and RG analysis.
We will perform a symmetry analysis of the low energy field theory in the ``LL2" phase. 
The symmetry group of the system in the original frame when $\Gamma= 0$ has been worked out to be
$G_0=\mathopen{<}T,T_{2a},T_aI, R(\hat{z},\frac{\pi}{2})T_a,R(\hat{y},\pi)\mathclose{>}$.
We will exhaust all the symmetry allowed relevant and marginal terms in the low energy field theory.
Throughout this section, we work in the original frame unless otherwise stated.

All the relevant and marginal terms are given in Eq. (\ref{eq:low_terms}).

1,2,3) The dimension $1/2,1,3/2$ terms are forbidden as analyzed in Sec. \ref{app:sym_Neel}.

4) Time reversal symmetry constraints the dimension $2$ operators to be the combinations $J_L^\alpha J_L^\beta+J_R^\alpha J_R^\beta$ and $J_L^\alpha J_R^\beta$.
The three dimensional linear space spanned by $\{J_s^x,J_s^y,J_s^z\}$ ($s=L,R$)
can be decomposed into $B_2\oplus E_1$, in which $B_2=\text{span}\{J_s^z\}$ and
$E_1=\text{span} \{J_s^x,J_s^y\}$.
Since both $B_2\otimes B_2$ and $E_1\otimes E_1$ contain an $A_1$ component,
the $D_{4d}$ invariant combinations are 
\begin{eqnarray}
&J_L^xJ_L^x+J_L^yJ_L^y+J_R^xJ_R^x+J_R^yJ_R^y,
J_L^zJ_L^z+J_R^zJ_R^z,\nn\\
&J_L^xJ_R^x+J_L^yJ_R^y,
J_L^zJ_R^z.
\end{eqnarray}
In summary, the symmetry allowed terms are
\begin{eqnarray}
&\vec{J}_L\cdot \vec{J}_L+\vec{J}_R\cdot \vec{J}_R,
J_L^zJ_L^z+J_R^z J_R^z,\nn\\
&\vec{J}_L\cdot \vec{J}_R, J_L^zJ_R^z.
\end{eqnarray}
Clearly, the low energy Hamiltonian is still of the XXZ type.

On the other hand,
we can do a first order perturbation treatment to the Kitaev term.
Explicit calculations show that (for details, see Sec. \ref{app:low_KH}),
\begin{flalign}
&H^K\rightarrow Ka \int dx \big[
-\frac{c^2}{2\pi^2 a^2} +\frac{3+2c^2}{6} (\vec{J}_L\cdot\vec{J}_L+\vec{J}_R\cdot\vec{J}_R)
-\frac{1}{2} (J_L^zJ_L^z+J_R^zJ_R^z)+\vec{J}_L\cdot \vec{J}_R-(1+2c^2)J_L^zJ_R^z
\big].
\label{eq:first_order_KH}
\end{flalign}
Hence, we see that all the symmetry allowed terms indeed appear within the low energy Hamiltonian.

It is then straightforward to determine the phase diagram.
The RG flow is again of the KT type.
The phase diagram is shown in Fig. \ref{fig:phase_KH}.
When $K>0$, i.e., the ``LL2" phase,
the system is described by the Luttinger liquid theory with an emergent U(1) 
symmetry at low energies.
On the other hand, when $K<0$, the system develops a N\'eel order along $z$-direction.

\begin{figure}[h]
\includegraphics[width=6.0cm]{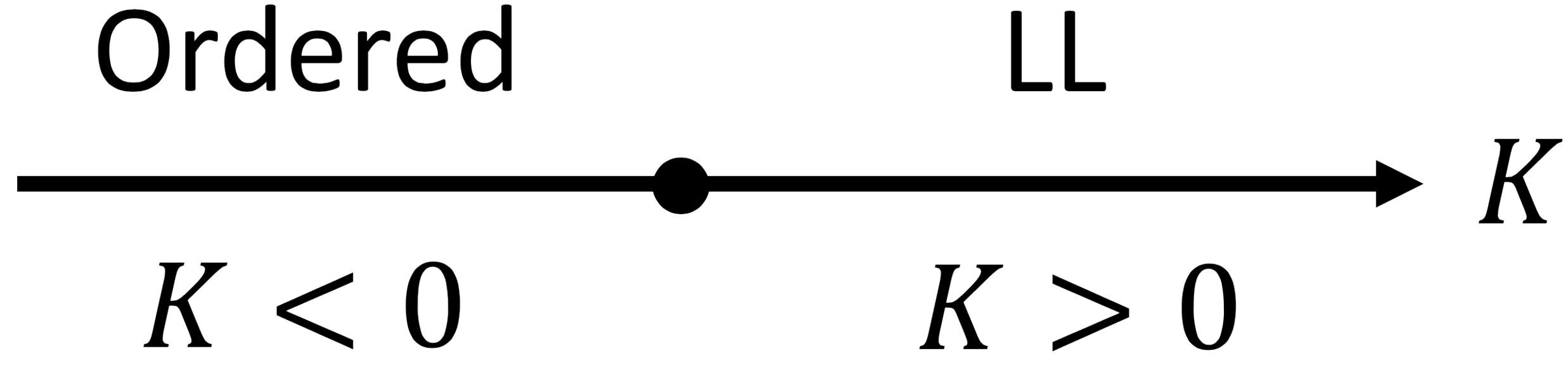}
\caption{Phase diagram for the Kitaev-Heisenberg chain,
in which ``LL" represents Luttinger liquid.
} \label{fig:phase_KH}
\end{figure}

\subsection{Symmetry analysis in the ``LL3" phase}
\label{app:sym_analysis_LL3}

In this section, we perform a symmetry analysis of the low energy field theory in the ``LL3" phase. 
The symmetry group of the system in the four-sublattice rotated frame when $\Gamma\neq 0$ has been worked out to be
$G_3=\mathopen{<} T,R(\hat{y},\pi)T_a I,R(\hat{z},-\frac{\pi}{2})T_a \mathclose{>}$.
We will exhaust all the symmetry allowed relevant and marginal terms in the low energy field theory.
Throughout this section, we work in the four-sublattice rotated frame unless otherwise stated.

The list of terms with dimensions not greater than two is the same as Eq. (\ref{eq:low_terms}).

1) The dimension $1/2$ operators $\epsilon$ and $\vec{N}$ change sign under $R(\hat{z},-\frac{\pi}{2})T_a$ and $T$, respectively. 
Hence both are forbidden. 

2) The dimension $1$ operators are restricted to the form $J_L^\alpha-J_R^\alpha$ ($\alpha=x,y,z$) due to time reversal symmetry,
in which only $J_L^z-J_R^z$ is allowed by $R(\hat{z},-\frac{\pi}{2})T_a$.
It can be seen that $J_L^z-J_R^z$ is invariant under $R(\hat{y},\pi)T_a I$.
Therefore, the term $J_L^z-J_R^z$ is allowed in the low energy theory.
 
2) The dimension $3/2$ operators are restricted to the forms 
$(J_L^\alpha-J_R^\alpha) \epsilon$ and $(J_L^\alpha+J_R^\alpha) N^\beta$
  due to time reversal symmetry.
The symmetry operation $R(\hat{z},-\frac{\pi}{2})T_a$ forbids $(J_L^\alpha-J_R^\alpha) \epsilon$.
This is because  $\epsilon $ changes sign under $T_a$,
and it is impossible for $J_L^\alpha-J_R^\alpha$ to change another sign under the rotation $R(\hat{z},-\frac{\pi}{2})$ since the $\pi/2$-rotation does not have the eigenvalue of $-1$.

However, it is possible for suitable combinations of $(J_L^\alpha+J_R^\alpha) N^\beta$ to change sign under $R(\hat{z},-\frac{\pi}{2})$.
In a vector representation $\{v_\alpha\}_{\alpha=x,y,z}$ of the SO(3) group,
the eigenvalues of $R(\hat{z},-\frac{\pi}{2})$ are $1,i,-i$ with eigenvectors $v_z,v_x+iv_y,v_x-iv_y$, respectively.
Hence, among the rank-2 tensors $(J_L^\alpha+J_R^\alpha) N^\beta$,
the following combinations change sign under $R(\hat{z},-\frac{\pi}{2})$
\begin{eqnarray}
&[(J_L^x+iJ_L^y)+(J_R^x+iJ_R^y)] (N^x+iN^y),~
[(J_L^x-iJ_L^y)+(J_R^x-iJ_R^y)] (N^x-iN^y).
\end{eqnarray}
They can be alternatively rewritten into real combinations as
 \begin{flalign}
&(J_L^x+J_R^x)N^x-(J_L^y+J_R^y)N^y, ~
(J_L^x+J_R^x)N^y+(J_L^y+J_R^y)N^x,
\label{eq:dim_3o2_LL3}
\end{flalign}
which are allowed by $R(\hat{z},-\frac{\pi}{2})T_a$.
 However, the term  $(J_L^x+J_R^x)N^x-(J_L^y+J_R^y)N^y$ changes sign under $R(\hat{y},\pi)T_a I$,
 and $(J_L^x+J_R^x)N^y+(J_L^y+J_R^y)N^x$ vanishes in the SU(2)$_1$ WZW model (see Sec. \ref{app:quadratic_WZW}).
 Hence, we conclude that there is no dimension-3/2 term in the low energy theory.
 
3)  For the dimension $2$ operators, let's focus on the left-right cross terms $J_L^\alpha J_R^\beta$.
The terms invariant under $R(\hat{z},-\frac{\pi}{2})T_a$ are
$J_L^zJ_R^z$, $J_L^xJ_R^x+J_L^yJ_R^y$, and $J_L^xJ_R^y-J_L^yJ_R^x$.
All these three terms are allowed by $R(\hat{y},\pi)T_aI$.
However, the third term is forbidden by time reversal symmetry. 

In summary, the nontrivial symmetry allowed terms with dimensions not greater than two are
\begin{eqnarray}
J_L^z-J_R^z,~J_L^zJ_R^z,
\end{eqnarray}
in which $J_L^z-J_R^z$ can be removed by a chiral rotation as discussed in the main text, 
and the system is again of an XXZ type at low energies.

\section{First order perturbation Hamiltonian by projection}

\subsection{Low energy Hamiltonian in the ``LL1" phase }
\label{app:deriv_proj}

In this section, we work in the six-sublattice rotated frame.
Expanding out explicitly, the Heisenberg term in the six-sublattic rotated frame is
\begin{eqnarray}
H^J_{12} &=& -J S_1^xS_2^x-J (S_1^yS_2^z+S_1^zS_2^y),\nn\\
H^J_{23} &=& -J S_2^zS_3^z-J (S_2^xS_3^y+S_2^yS_3^x),\nn\\
H^J_{34} &=& -J S_3^yS_4^y-J (S_3^zS_4^x+S_3^xS_4^z),
\end{eqnarray}
in which all the subscripts are written modulo $3$ for simplicity.
In the following, we project $H^J_{12}$. 
The other two terms can be treated similarly.

First consider the term $S_1^xS_2^x$.
Using
\begin{eqnarray}
S_1^x&=& D_1(J_L^x(x)+J_R^x(x))+C_1(-)^{\frac{x}{a}}\frac{1}{a} i \text{tr}(g(x)\sigma^x),\nn\\
S_2^x&=& D_1(J_L^x(x+a)+J_R^x(x+a))+C_1(-)^{\frac{x}{a}+1}\frac{1}{a} i\text{tr}(g(x+a)\sigma^x),
\end{eqnarray}
and the OPE formula for the N\'eel order fields,
we obtain
\begin{flalign}
&\sum_{\text{u.c.}}\frac{1}{a^2} S_1^xS_2^x\nn\\
&=\sum
D_1^2(J_L^xJ_L^x+J_L^xJ_L^x+2J_L^xJ_R^x)-C_1^2\frac{1}{a^2} (-1)
\big[-2+(2\pi a)^2 (4J_L^xJ_R^x-2\vec{J}_L\cdot\vec{J}_R+\frac{1}{3}\vec{J}_L\cdot\vec{J}_L+\frac{1}{3}\vec{J}_L\cdot\vec{J}_L)
\big]\nn\\
&= \sum-\frac{2C_1^2}{a^2}+(2D_1^2+16\pi^2 C_1^2) J_L^xJ_R^x-8\pi^2 C_1^2\vec{J}_L\cdot\vec{J}_R
+D_1^2(J_L^xJ_L^x+J_R^xJ_R^x)+\frac{4\pi^2C_1^2}{3} (\vec{J}_L\cdot \vec{J}_L+\vec{J}_R\cdot \vec{J}_R),
\end{flalign}
in which the summation is over the unit cells,
and the oscillating terms are dropped since they do not contribute.
Then it is straightforward to obtain
\begin{eqnarray}
\sum\frac{1}{a^2} (S_1^xS_2^x+S_2^zS_3^z+S_3^yS_4^y)=\sum-6C_1^2\frac{1}{a^2}
+(2D_1^2-8\pi^2C_1^2) \vec{J}_L\cdot\vec{J}_R+(D_1^2+4\pi^2 C_1^2) (\vec{J}_L\cdot \vec{J}_L+\vec{J}_R\cdot \vec{J}_R).
\end{eqnarray}
Summing over the unit cells and using $\sum_{\text{u.c.}}=\frac{1}{3a}\int dx$,
we obtain the following contribution to $H^J$,
\begin{eqnarray}
H^{J,(1)}=-\frac{1}{3}Ja\int dx \big[-6C_1^2\frac{1}{a^2}
+(2D_1^2-8\pi^2C_1^2) \vec{J}_L\cdot\vec{J}_R+(D_1^2+4\pi^2 C_1^2) (\vec{J}_L\cdot \vec{J}_L+\vec{J}_R\cdot \vec{J}_R)
\big].
\label{eq:proj_diag}
\end{eqnarray}

Next consider the cross term.
Using
\begin{eqnarray}
\frac{1}{a}S_1^y&=& D_1(J_L^y(x)+J_R^y(x))+C_1(-)^{\frac{x}{a}}\frac{1}{a} i\text{tr}(g(x)\sigma^y),\nn\\
\frac{1}{a}S_2^z&=& D_1(J_L^z(x+a)+J_R^z(x+a))+C_1(-)^{\frac{x}{a}+1}\frac{1}{a} i\text{tr}(g(x+a)\sigma^z),
\end{eqnarray}
we obtain
\begin{flalign}
&\sum\frac{1}{a^2} S_1^yS_2^z\nn\\
&=\sum D_1^2 (J_L^yJ_L^z+J_R^yJ_R^z+J_L^yJ_R^z+J_L^zJ_R^y)
-C_1^2  \frac{1}{a^2} (-1)
\big(-4\pi a (J_L^x-J_R^x)+
(2\pi a)^2 \big( 2J_L^{\{ y}J_R^{z\} }+J_L^{[y}J_L^{z]}+J_R^{[y}J_R^{z]}
\big)
\big]\nn\\
&=\sum -\frac{4\pi C_1^2}{a} (J_L^x-J_R^x)+ (D_1^2+8\pi^2 C_1^2) J_L^{\{ y}J_R^{z\} }
+\frac{1}{2}D_1^2(J_L^{\{ y}J_L^{z\} }+J_R^{\{ y}J_R^{z\} })
+(\frac{1}{2}D_1^2+4\pi^2 C_1^2) (J_L^{[ y}J_L^{z]}+J_R^{[y}J_R^{z] }).\nn\\
\label{eq:S1yS2z}
\end{flalign}
Similarly,
\begin{eqnarray}
&\sum\frac{1}{a^2} S_1^zS_2^y&= \sum\frac{4\pi C_2^2}{a} (J_L^x-J_R^x)
+(D_2^2+8\pi^2 C_2^2) J_L^{\{ y}J_R^{z\} }\nn\\
&&+\frac{1}{2}D_2^2(J_L^{\{ y}J_L^{z\} }+J_R^{\{ y}J_R^{z\} })
-(\frac{1}{2}D_2^2+4\pi^2 C_2^2) (J_L^{[ y}J_L^{z]}+J_R^{[ y}J_R^{z] }).
\label{eq:S1zS2y}
\end{eqnarray}
We note a subtlety here.
Notice that the spin operators commute when the two operators are on different sites.
However, Eq. (\ref{eq:S1yS2z},\ref{eq:S1zS2y}) do not satisfy this.
In fact, there is no reason to expect that such commutativity still holds when the operators are restricted to a low energy subspace. 
On the other hand, we do have an ambiguity in writing $S_i^\alpha S_j^\beta$ in terms of the low energy degrees of freedom, 
since $S_i^\alpha S_j^\beta$ and $ S_j^\beta S_i^\alpha$ are different.
To regularize such discrepancy, we use the symmetrized  expressions $\frac{1}{2}(S_i^\alpha S_j^\beta+S_j^\beta S_i^\alpha)$,
which removes the antisymmetric terms within Eq. (\ref{eq:S1yS2z},\ref{eq:S1zS2y}).
An alternative way is to keep using Eq. (\ref{eq:S1yS2z}),
but consider the sum of pairs $\frac{1}{2}(S_1^yS_2^z+S_3^y S_2^x)$,
which is ensured by the $R_I I$ symmetry operation.
Then apparently, combining $S_3^y S_2^x$ with $S_2^x S_3^y$ together lead to an effective symmetrization.
As a result, we obtain
\begin{flalign}
&\sum\frac{1}{a^2} (S_1^yS_2^z+S_1^zS_2^y)\nn\\
&=\sum-\frac{4\pi (C_1^2-C_2^2)}{a} (J_L^x-J_R^x)+[(D_1^2+D_2^2)+8\pi^2(C_1^2+ C_2^2)] J_L^{\{ y}J_R^{z\} }
+\frac{1}{2}(D_1^2+D_2^2)(J_L^{\{ y}J_L^{z\} }+J_R^{\{ y}J_R^{z\} }).\nn\\
\end{flalign}

Summing the three bonds up, we obtain
\begin{flalign}
&\sum\frac{1}{a^2} (S_1^yS_2^z+S_1^zS_2^y+S_2^xS_3^y+S_2^yS_3^x+S_3^zS_4^x+S_3^xS_4^z)
=\nn\\
&\sum-\frac{4\pi (C_1^2-C_2^2)}{a} \sqrt{3}(J_L^0-J_R^0)
+[(D_1^2+D_2^2)+8\pi^2(C_1^2+ C_2^2)] \frac{1}{2}(3J_L^0J_R^0-\vec{J}_L\cdot \vec{J}_R)\nn\\
&+\frac{1}{2}(D_1^2+D_2^2) 
\big[\frac{1}{2} (3J_L^0J_L^0-\vec{J}_L\cdot \vec{J}_L) +\frac{1}{2} (3J_R^0J_R^0-\vec{J}_L\cdot \vec{J}_R) \big].
\end{flalign}
in which 
$
J^{\{z}J^{y\}}+J^{\{y}J^{x\}}+J^{\{x}J^{z\}}=\frac{1}{2} (3 J^0J^0-\vec{J}\cdot\vec{J})
$
is used.
Hence, these terms contribute to $H^J$ the following expression,
\begin{flalign}
&H^{J,(2)}=-\frac{1}{3}Ja\int dx\big\{
-\frac{4\pi (C_1^2-C_2^2)}{a} \sqrt{3}(J_L^0-J_R^0)
+\frac{3}{2}[(D_1^2+D_2^2)+8\pi^2(C_1^2+ C_2^2)] J_L^0J_R^0\nn\\
&-\frac{1}{2}[(D_1^2+D_2^2)+8\pi^2(C_1^2+ C_2^2)] \vec{J}_L\cdot \vec{J}_R
+\frac{3}{4}(D_1^2+D_2^2) (J_L^0J_L^0+J_R^0J_R^0)
-\frac{1}{4}(D_1^2+D_2^2) (\vec{J}_L\cdot \vec{J}_L+\vec{J}_R\cdot \vec{J}_R)
\big\}.
\label{eq:proj_cross}
\end{flalign}
Combining Eqs. (\ref{eq:proj_diag},\ref{eq:proj_cross}) together,
we arrive at the low energy Hamiltonian close to the emergent SU(2)$_1$ line as discussed  in the main text, and the coefficients $u$'s are given by
\begin{eqnarray}
u_1&=&\frac{4\pi\sqrt{3}}{a}[(C_2)^2-(C_1)^2],\nn\\
u_2&=&\frac{3}{2} [(D_1)^2+(D_2)^2]+12\pi^2[(C_1)^2+(C_2)^2],\nn\\
u_3&=&\frac{1}{2} [3(D_1)^2-(D_2)^2]-4\pi^2[(C_1)^2-(C_2)^2],\nn\\
u_4&=&\frac{3}{4} [(D_1)^2+(D_2)^2],\nn\\
u_5&=&-\frac{1}{4} [(D_1)^2+(D_2)^2].
\label{eq:us}
\end{eqnarray}

\subsection{Low energy Hamiltonian in the ``LL2" phase}
\label{app:low_KH}

In this section, we work in the original frame. 
We will project the Kitaev term to the low energy subspace,
and confirm that indeed all the symmetry allowed terms will emerge.

The Kitaev term is 
\begin{eqnarray}
H^K=K\sum_i (S_{2i-1}^xS_{2i}^x+S_{2i}^yS_{2i+1}^y).
\end{eqnarray}
Using the nonabelian bosonization formula,
\begin{flalign}
\frac{1}{a} S_i^\alpha=J_L^\alpha+J_R^\alpha
+(-)^{i+1} \frac{c}{2\pi a} i\text{tr} (g\sigma^\alpha),
\label{eq:nonabel_H}
\end{flalign}
we have
\begin{flalign}
\frac{1}{a^2}\sum S_{2i-1}^xS_{2i}^x&=
\sum (J_L^xJ_L^x+J_R^xJ_R^x+2J_L^xJ_R^x)+(J_L^x(x)+J_R^x(x))  (-)\frac{c}{2\pi a} i\text{tr} g(x+a) \nn\\
&~~~+\frac{c}{2\pi a} i \text{tr} g(x) (J_L(x+a)+J_R(x+a))
 + \frac{c^2}{(2\pi a)^2} \text{tr} (g(x)\sigma^x) \text{tr} (g(x+a)\sigma^x).
\label{eq:low_SxSx}
\end{flalign}
Using the following OPE,
\begin{eqnarray}
\vec{J}_L(w) g(z,\bar{z})&=& -\frac{1}{2} \frac{\vec{\sigma} g(z,\bar{z}) }{w-z} +(\vec{J}_Lg) (z,\bar{z}),\nn\\
\vec{J}_R(w) g(z,\bar{z})&=& \frac{1}{2} \frac{ g(z,\bar{z})  \vec{\sigma}}{w-z} +(\vec{J}_Rg) (z,\bar{z}),
\end{eqnarray}
we obtain,
\begin{eqnarray}
J_L^x(x) \text{tr} (g(x+a)\sigma^x)&=&\frac{1}{4\pi i a} \text{tr}( g(x+a)) +\text{tr} ((J_L^xg)\sigma^x),\nn\\
J_R^x(x) \text{tr} (g(x+a)\sigma^x)&=&\frac{1}{4\pi i a} \text{tr}( g(x+a)) +\text{tr} ((J_R^xg)\sigma^x),\nn\\
\text{tr} (g(x)\sigma^x)J_L^x(x+a)&=&-\frac{1}{4\pi i a} \text{tr}( g(x)) +\text{tr} ((J_L^xg)\sigma^x),\nn\\
\text{tr} (g(x)\sigma^x)J_R^x(x+a)&=&-\frac{1}{4\pi i a} \text{tr}( g(x)) +\text{tr} ((J_R^xg)\sigma^x).
\end{eqnarray}
Therefore, the second and the third terms in Eq. (\ref{eq:low_SxSx}) sum up to
$
i\frac{c}{2\pi a} 
(-)\frac{1}{4\pi i a} \text{tr} g \times 4
$.
The fourth term in Eq. (\ref{eq:low_SxSx}) can be calculated using the OPEs Eqs. (\ref{eq:OPENeel}) as
\begin{eqnarray}
-\frac{2c^2}{(2\pi a)^2} +2c^2(J_L^xJ_R^x-J_L^yJ_R^y-J_L^zJ_R^z)+\frac{1}{3}c^2 (\vec{J}_L\cdot \vec{J}_L+\vec{J}_R\cdot \vec{J}_R).
\end{eqnarray}

In summary, 
\begin{flalign}
\sum S_{2i-1}^xS_{2i}^x=&\frac{1}{2} a \int dx
\big[
-\frac{2c^2}{(2\pi a)^2} -\frac{c}{2\pi^2 a^2} \text{tr} (g)
+J_L^xJ_L^x+J_R^xJ_R^x\nn\\
&+\frac{1}{3} c^2(\vec{J}_L\cdot \vec{J}_L+\vec{J}_R\cdot \vec{J}_R)
+(2+2c^2)J_L^xJ_R^x-2c^2J_L^yJ_R^y-2c^2J_L^zJ_R^z
\big].\nn\\
\end{flalign}
Similarly,
\begin{flalign}
\sum S_{2i}^yS_{2i+1}^y=&\frac{1}{2} a \int dx
\big[
-\frac{2c^2}{(2\pi a)^2} +\frac{c}{2\pi a^2} \text{tr} (g)
+J_L^yJ_L^y+J_R^yJ_R^y\nn\\
&+\frac{1}{3} c^2(\vec{J}_L\cdot \vec{J}_L+\vec{J}_R\cdot \vec{J}_R)
+(2+2c^2)J_L^yJ_R^y-2c^2J_L^xJ_R^x-2c^2J_L^zJ_R^z
\big].
\end{flalign}
Summing the two terms up, we arrive at the low energy Hamiltonian close to the AFM1 point along the circular boundary of Fig. 1 in the main text.

\subsection{Low energy Hamiltonian in the ``N\'eel" phase}
\label{app:low_northpole}

In this section, we derive the low energy Hamiltonian in the ``N
'eel" phase  close to the AFM1 point.
Throughout this section, we work in the original frame. 

The Gamma term is 
\begin{eqnarray}
H^\Gamma=\sum (S_{2i-1}^yS_{2i}^z+S_{2i-1}^zS_{2i}^y)+\sum (S_{2i-1}^xS_{2i}^z+S_{2i-1}^zS_{2i}^x).
\end{eqnarray}
Using 
\begin{flalign}
\frac{1}{a} S_i^\alpha=J_L^\alpha+J_R^\alpha
+(-)^{i+1} \frac{c}{2\pi a} i\text{tr} (g\sigma^\alpha),
\label{eq:nonabelian_Heisenberg}
\end{flalign}
 we obtain
\begin{flalign}
\sum \frac{1}{a^2}S_{2i-1}^\alpha S_{2i}^\beta& = 
\sum J_L^\alpha J_L^\beta+J_R^\alpha J_R^\beta +J_L^\alpha J_R^\beta+J_R^\alpha J_L^\beta
+ (J_L^\alpha(x)+J_R^\alpha(x)) (-)\frac{c}{2\pi a} i\text{tr} (g(x+a)\sigma^\beta) \nn\\
&+\frac{c}{2\pi a} i\text{tr} (g(x)\sigma^\alpha)  (J_L^\beta(x+a)+J_R^\beta(x+a))
+\frac{c^2}{(2\pi a)^2} \text{tr} (g(x)\sigma^\alpha)\text{tr} (g(x+a)\sigma^\beta), 
\end{flalign}
in which
\begin{flalign}
(J_L^\alpha(x)+J_R^\alpha(x)) (-)\frac{1}{2\pi a} i\text{tr} (g(x+a)\sigma^\beta)
&=-i\big\{
-\frac{1}{2\pi i a} \text{tr} \big[ (-\frac{1}{2} \sigma^\alpha) g\sigma^\beta \big]
+\frac{1}{(2\pi i a)^*} \text{tr} \big[  g\frac{1}{2} \sigma^\alpha\sigma^\beta \big]
+\text{tr}  \big[  (J_L^\alpha+J_R^\alpha )g \sigma^\beta  \big ]
\big\}\nn\\
&=-\frac{1}{2\pi a}i \text{tr} \big[  (J_L^\alpha g+J_R^\alpha g)\sigma^\beta \big]\nn\\
\frac{1}{2\pi a} i\text{tr} (g(x)\sigma^\alpha)  (J_L^\beta(x+a)+J_R^\beta(x+a))
&=\frac{1}{2\pi a}i \text{tr} \big[  (J_L^\beta g+J_R^\beta g)\sigma^\alpha \big]\nn\\
\frac{1}{(2\pi a)^2} \text{tr} (g(x)\sigma^\alpha)\text{tr} (g(x+a)\sigma^\beta)&=
-J_L^{[\alpha}J_L^{\beta]}-J_R^{[\alpha}J_R^{\beta]}
+2(J_L^\alpha J_R^\beta+J_L^\beta J_R^\alpha)
+\frac{1}{\pi a} \epsilon^{\alpha\beta\gamma} (-J_L^\gamma+J_R^\gamma).
\end{flalign}
Notice that in $H^\Gamma$ we need to symmetrize the indices $\alpha$ and $\beta$,
thus all the antisymmetric terms in $\alpha,\beta$ will drop off the expression. 
Finally, by summing up all the terms, we arrive at the low energy Hamiltonian in the ``N\'eel" phase close to the AFM1 point as discussed in the main text.

\section{Solution of the RG flow equations in the N\'eel phase}
\label{app:RG_flow_Neel}

Recall that there are three constants of the motions
\begin{eqnarray}
\lambda_x^2(l)-\lambda_y^2(l)&=& \frac{c^2g_c\Gamma a}{\sqrt{2}\pi^2 v^2}  \nn\\
\lambda_y^2(l)-\lambda_z^2(l)&=& -\frac{c^2g_c\Gamma a}{2\sqrt{2}\pi^2 v^2}  \nn\\
\lambda_z^2(l)-\lambda_x^2(l)&=&  -\frac{c^2g_c\Gamma a}{2\sqrt{2}\pi^2 v^2}. 
\label{eq:3hyperapp}
\end{eqnarray}
Define $E=\frac{c^2g_c\Gamma a}{2\sqrt{2}\pi^2 v^2}$.
Then according to Eq. (\ref{eq:3hyperapp}), we have
\begin{eqnarray}
\lambda_x=\sqrt{\lambda_y^2+2E^2},~\lambda_z=\sqrt{\lambda_y^2+E^2}.
\end{eqnarray}
The flow equation of $\lambda_y$ becomes
\begin{eqnarray}
\frac{d\lambda_y}{dl}=-\sqrt{\lambda_y^2+2E^2}\cdot \sqrt{\lambda_y^2+E^2},
\end{eqnarray}
which gives
\begin{eqnarray}
dl=-\frac{d\lambda_y}{\sqrt{(\lambda_y^2+2E^2)(\lambda_y^2+E^2)}}.
\end{eqnarray}
Notice that this is an elliptic function which can be solved exactly.
However, for our purpose to determine the asymptotic behaviors at $l\rightarrow \infty$,
there is no need to consider the accurate solutions,
and the information contained in  the constant of motions in Eq. (\ref{eq:3hyperapp}) is enough.

\begin{figure}[h]
\includegraphics[width=6.0cm]{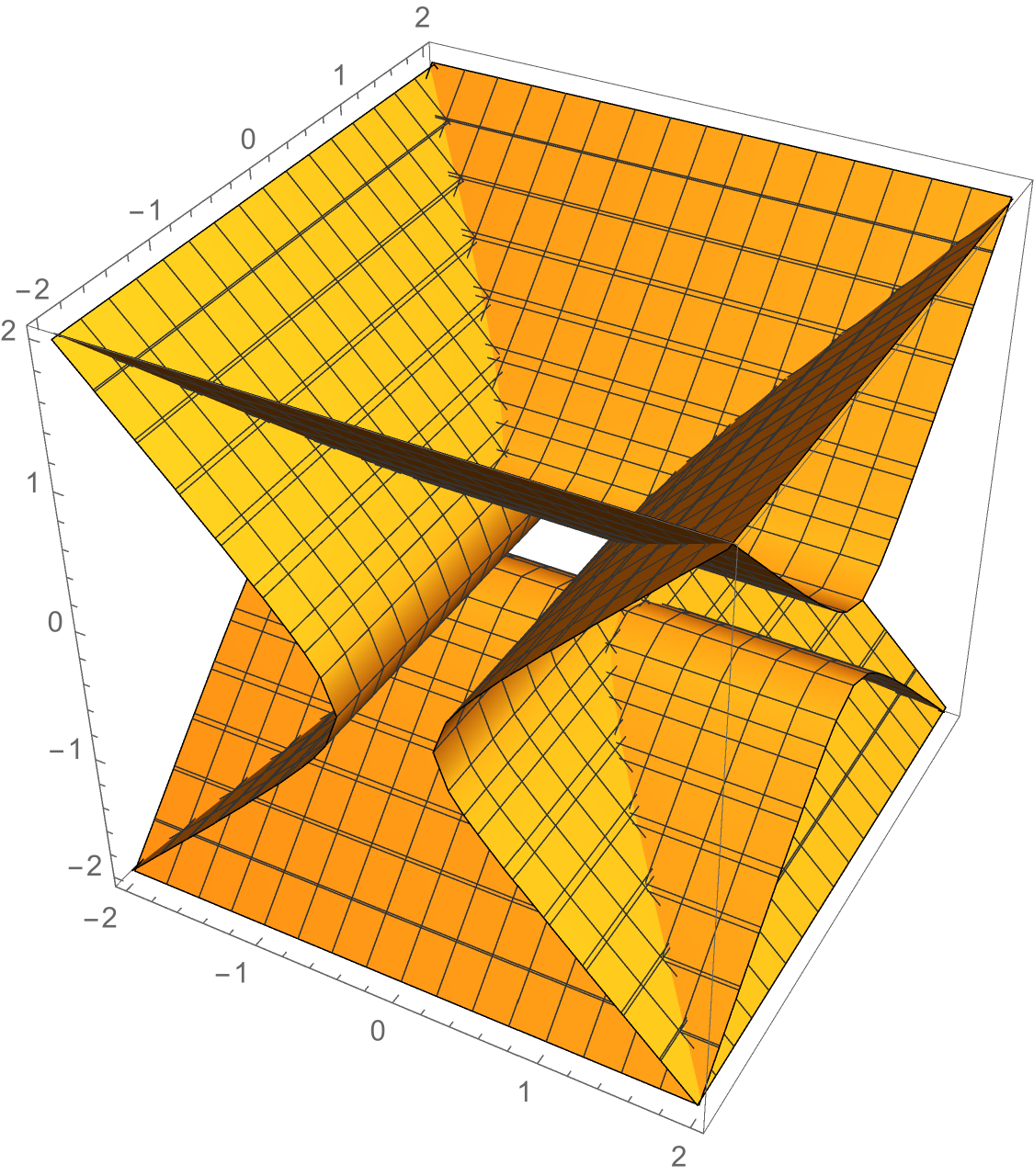}
\includegraphics[width=6.0cm]{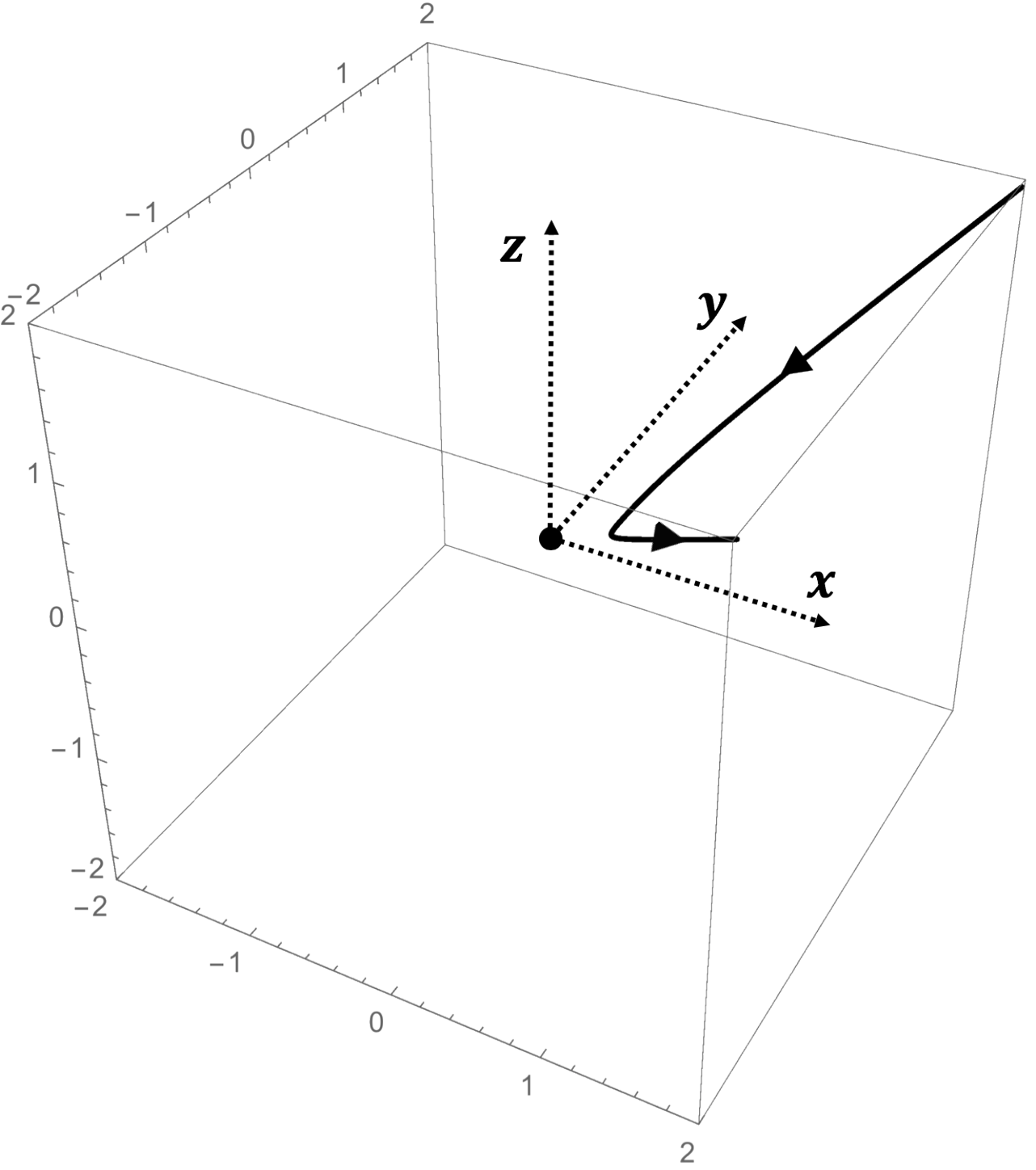}
\caption{Left: the combined plots of the two hyperbolic surfaces,
and right: the RG trajectory.
$E$ is taken as $0.1$ in plotting the figures.
} 
\label{fig:intersection}
\end{figure}

The RG trajectory is the intersection of the two surfaces determined by $x^2-z^2=E$ and $z^2-y^2=E$.
The left figure in Fig. \ref{fig:intersection} shows the two surfaces in a same plot. 
There are four intersection curves in total.
However, the initial conditions $\lambda_x,\lambda_y,\lambda_z$ determines the unique curve as shown in the right figure in Fig. \ref{fig:intersection}.

The direction of the flow can be determined from the flow equations and the intial conditions.
More precisely, the initial conditions imply that at $l=0$, $\frac{\lambda_\alpha}{dl}$ ($\alpha=x,y,z$) are all negative.
Therefore, $\lambda_x,\lambda_y,\lambda_z$ initially are all decreasing.
This determines the flow direction as shown by the arrows in Fig. \ref{fig:intersection}.
By tracing the flow to $l\rightarrow\infty$, we see that the final destiny of $\lambda_x,\lambda_y, \lambda_z$ is 
\begin{eqnarray}
&|\lambda_x|=|\lambda_y|=|\lambda_z|,\nn\\
&\lambda_x,\lambda_z\rightarrow+\infty,~\lambda_y\rightarrow-\infty.
\end{eqnarray}


\section{Classical analysis in the ``FM" phase}

We perform a classical analysis in the ``FM" phase, and show that the predicted spin orderings are consistent with the following pattern
\begin{eqnarray}
\vec{S}_i=(a,a,b)^T.
\label{eq:FM_aab}
\end{eqnarray}
Throughout this section, we work in the original frame unless otherwise stated.

Introducing the Lagrange multipliers $\{\lambda_i\}_{1\leq i \leq3}$ to impose the constraints $x_i^2+y_i^2+z_i^2=1$, 
the energy per unit cell becomes
\begin{eqnarray}
F_0=K^\prime(x_1x_2+y_1y_2)+2J^\prime (x_1x_2+y_1y_2+z_1z_2) 
+\Gamma^\prime (z_1(x_2+y_2)+z_2(x_1+y_1))
-\frac{1}{2}\sum_{i=1,2}\lambda_i(x_i^2+y_i^2+z_i^2-1),
\end{eqnarray}
in which 
\begin{eqnarray}
\Gamma^\prime=\Gamma S^2,~K^\prime=K S^2,~J^\prime=JS^2.
\end{eqnarray}
The saddle point equations read
\begin{eqnarray}
\frac{\partial F_0}{\partial x_1}&=&(K^\prime+2J^\prime) x_2+\Gamma^\prime z_2-\lambda_1 x_1=0\nn\\
\frac{\partial F_0}{\partial y_1}&=&(K^\prime+2J^\prime) y_2+\Gamma^\prime z_2-\lambda_1 y_1=0\nn\\
\frac{\partial F_0}{\partial z_1}&=&\Gamma^\prime (x_2+y_2)+2J^\prime z_2-\lambda_1 z_1=0\nn\\
\frac{\partial F_0}{\partial \lambda_1}&=&x_1^2+y_1^2+z_1^2-1=0,
\label{eq:saddle_FM1_1}
\end{eqnarray}
\begin{eqnarray}
\frac{\partial F_0}{\partial x_2}&=&(K^\prime+2J^\prime) x_1+\Gamma^\prime z_1-\lambda_2 x_2=0\nn\\
\frac{\partial F_0}{\partial y_2}&=&(K^\prime+2J^\prime) y_1+\Gamma^\prime z_1-\lambda_2 y_2=0\nn\\
\frac{\partial F_0}{\partial z_2}&=&\Gamma^\prime (x_1+y_1)+2J^\prime z_1-\lambda_2 z_2=0\nn\\
\frac{\partial F_0}{\partial \lambda_2}&=&x_2^2+y_2^2+z_2^2-1=0.
\label{eq:saddle_FM1_2}
\end{eqnarray}
Trying the ansatz 
\begin{flalign}
&x_1=y_1=x_2=y_2=a,\nn\\
&z_1=z_2=b,\nn\\
&\lambda_1=\lambda_2=\lambda,
\label{eq:trial_FM}
\end{flalign}
 Eqs. (\ref{eq:saddle_FM1_1},\ref{eq:saddle_FM1_2})  are reduced to
\begin{eqnarray}
(K^\prime+2J^\prime-\lambda)a+\Gamma^\prime b&=&0\nn\\
2\Gamma^\prime a +(2J^\prime-\lambda)b&=&0\nn\\
2a^2+b^2-1&=&0.
\label{eq:saddle_FM_red}
\end{eqnarray}
Since there are three variables $a,b,\lambda$ and an equal number of equations, 
a solution exists generically.
 
In what follows, we only discuss the special case $K^\prime=0$, $\Gamma^\prime>0$ for illustration.
Other cases can be solved exactly similarly. 
The solution of Eq. (\ref{eq:saddle_FM_red}) is
\begin{eqnarray}
\lambda=-2-\sqrt{2}\Gamma^\prime,~a=-\frac{1}{\sqrt{2}},~b=1.
\end{eqnarray}
To check if this is a minimum of the free energy,
the eigenvalues of the Hessian matrix can be calculated in a perturbative expansion over $\Gamma$.
We have determined the two lowest eigenvalues to be $\frac{1}{\sqrt{2}}\Gamma^\prime$ and $\sqrt{2}\Gamma^\prime$,
which are both positive when $\Gamma^\prime>0$.


\section{DMRG numerical results for the Luttinger parameters}
\label{app:DMRG_LL_K}

Recall that as discussed in the main text, for a finite size system with an open boundary condition, the energy density $\langle h(x)\rangle$ contains a uniform part $E_U(x)$ and a staggered part $E_A(x)$,
where
\begin{eqnarray}
E_A(x)\propto \frac{1}{[\frac{L}{\pi}\sin(\frac{\pi x}{L})]^\mathcal{K}},
\label{eq:LL_EA}
\end{eqnarray}
in which $x=j a$ ($j\gg 1$) is the distance measured from the boundary of the system.
The provides a method to accurately determine the Luttinger parameter $\mathcal{K}$.
We apply this method to the ``LL$i$" ($i=1,3,4$) phases.

\subsection{The ``LL1" phase}

\begin{figure}
\includegraphics[width=8.0cm]{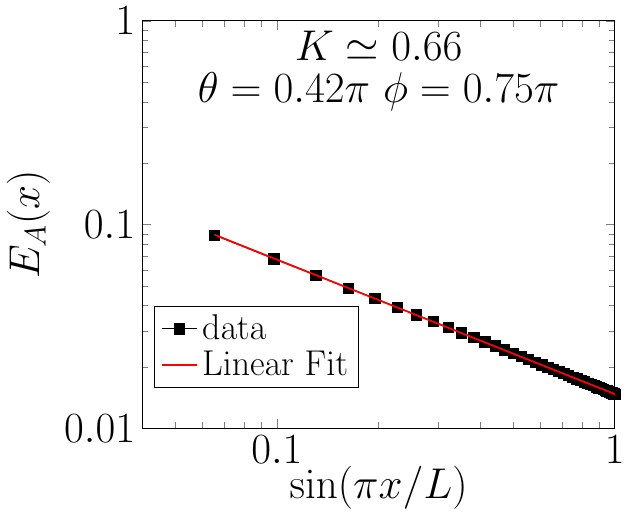}
\caption{$E_A(x)$ vs $\sin(\pi x/L)$ on a log-log scale,
in which $x$ is the distance measured from the boundary.
DMRG numerics are performed on an open system with $L=96$ sites at $(\theta=0.42\pi,\phi=0.75\pi)$.
The small solid black squares and the red line represent the numerical data and the linear fit, respectively, from which the Luttinger parameter is determined to be $\mathcal{K}\simeq 0.66$.
} \label{fig:LL1_LuttingerK}
\end{figure}

Fig. \ref{fig:LL1_LuttingerK} shows $E_A$ thus obtained vs $\sin(\pi x/L)$  on a log-log plot at a representative point $(\theta=0.42\pi,\phi=0.75\pi)$ in the ``LL1" phase,
where DMRG numerics are performed on a system of $L=96$ sites with an open boundary condition.
As can be seen from Fig. \ref{fig:LL1_LuttingerK}, an excellent linear fit can be obtained from which the Luttinger parameter is determined to be $\mathcal{K}\simeq 0.66$.

\subsection{The ``LL3" phase}

Fig. \ref{fig:K_fit_LL3} shows $E_A(x)$ vs $\sin(\pi x/L)$ on a log-log scale at a representative point in the ``LL3" phase $(\theta=0.64\pi,\phi=0.11\pi)$, where a good linear fit is obtained giving a Luttinger parameter equal to $0.591$.

\begin{figure}
\includegraphics[width=7.5cm]{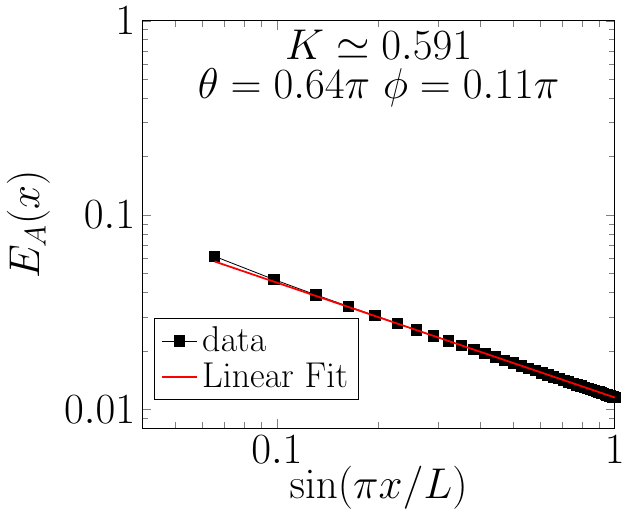}
\caption{$E_A(x)$ vs $\sin(\pi x/L)$ on a log-log scale,
in which $x$ is the distance measured from the boundary.
DMRG numerics are performed on an open system with $L=96$ sites at a representative point $(\theta=0.64\pi,\phi=0.11\pi)$ in the ``LL3" phase.
The small solid black squares and the red line represent the numerical data and the linear fit, respectively, from which the Luttinger parameter is determined to be $\mathcal{K}\simeq 0.591$.
} 
\label{fig:K_fit_LL3}
\end{figure}

\subsection{The ``LL4" phase}

The plots of $\log E_A(x)$ vs $\log \sin(\pi x/L)$ at a representative point in the ``LL4" phase close to the ``peninsular end" and a nearby point in the ``FM" phase are shown in Fig. \ref{fig:LL4_LL_K},
where DMRG numerics are performed on open systems of $L=96$ sites
and  $x$ is the distance measured from the boundary.
As can be seen from Fig. \ref{fig:LL4_LL_K}, while a good linear fit can be obtained for the point $(\theta=0.57\pi,\phi=0.30\pi)$ within the ``LL4" phase, no linear relation can be fitted for the point $(\theta=0.6\pi,\phi=0.30\pi)$ which is in the ``FM" phase.

\begin{figure}
\includegraphics[width=13.0cm]{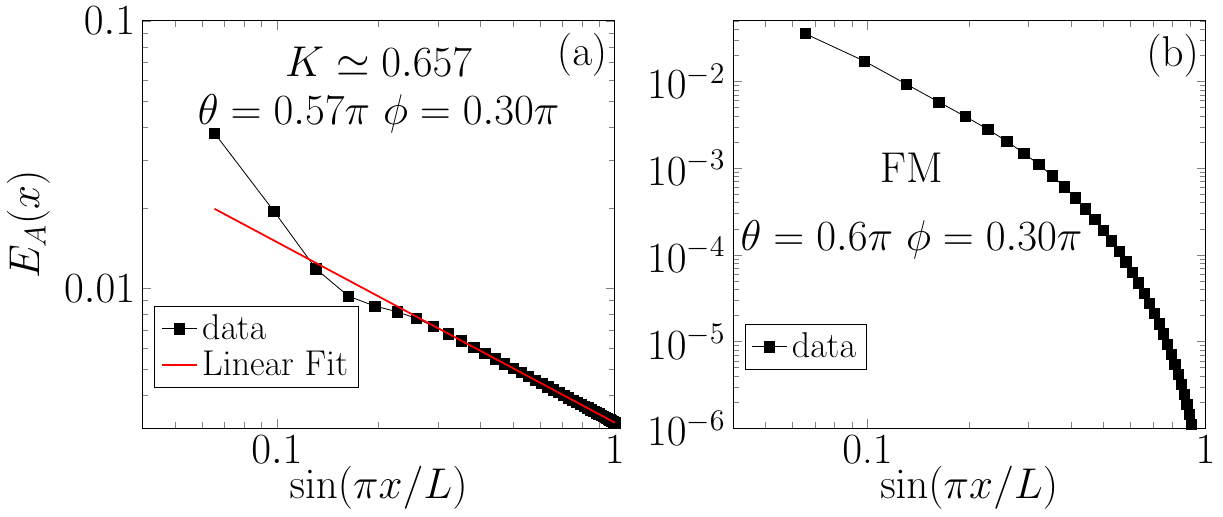}
\caption{$E_A(x)$ vs $\sin(\pi x/L)$ on a log-log scale at (a) $\theta=0.57\pi$, $\phi=0.30\pi$, and (b) $\theta=0.6\pi$, $\phi=0.30\pi$,
in which $x$ is the distance measured from the boundary.
DMRG calculations are performed on $L=96$ sites with open boundary conditions. 
} 
\label{fig:LL4_LL_K}
\end{figure}


\end{widetext}

\end{document}